\documentstyle[12pt]{article}   

\large
\makeatletter
\@addtoreset{equation}{section}
\makeatother

\newtheorem{Theorem}{Theorem}[section]
\newtheorem{Definition}{Definition}[section]
\newtheorem{Lemma}{Lemma}[section]
\newtheorem{Corollary}{Corollary}[section]

\def\be{\begin{equation}}
\def\ee{\end{equation}}
\def\ba{\begin{eqnarray}}
\def\ea{\end{eqnarray}}
\def\a{{\cal A}}
\def\ab{\overline{\a}}

\def\Nl{{\mathchoice
{\setbox0=\hbox{$\displaystyle\rm N$}\hbox{\hbox to0pt
{\kern0.4\wd0\vrule height0.9\ht0\hss}\box0}}
{\setbox0=\hbox{$\textstyle\rm N$}\hbox{\hbox to0pt
{\kern0.4\wd0\vrule height0.9\ht0\hss}\box0}}
{\setbox0=\hbox{$\scriptstyle\rm N$}\hbox{\hbox to0pt
{\kern0.4\wd0\vrule height0.9\ht0\hss}\box0}}
{\setbox0=\hbox{$\scriptscriptstyle\rm N$}\hbox{\hbox to0pt
{\kern0.4\wd0\vrule height0.9\ht0\hss}\box0}}}}
\def\Zl{{\mathchoice
{\setbox0=\hbox{$\displaystyle\rm Z$}\hbox{\hbox to0pt
{\kern0.4\wd0\vrule height0.9\ht0\hss}\box0}}
{\setbox0=\hbox{$\textstyle\rm Z$}\hbox{\hbox to0pt
{\kern0.4\wd0\vrule height0.9\ht0\hss}\box0}}
{\setbox0=\hbox{$\scriptstyle\rm Z$}\hbox{\hbox to0pt
{\kern0.4\wd0\vrule height0.9\ht0\hss}\box0}}
{\setbox0=\hbox{$\scriptscriptstyle\rm Z$}\hbox{\hbox to0pt
{\kern0.4\wd0\vrule height0.9\ht0\hss}\box0}}}}
\def\Ql{{\mathchoice
{\setbox0=\hbox{$\displaystyle\rm Q$}\hbox{\hbox to0pt
{\kern0.4\wd0\vrule height0.9\ht0\hss}\box0}}
{\setbox0=\hbox{$\textstyle\rm Q$}\hbox{\hbox to0pt
{\kern0.4\wd0\vrule height0.9\ht0\hss}\box0}}
{\setbox0=\hbox{$\scriptstyle\rm Q$}\hbox{\hbox to0pt
{\kern0.4\wd0\vrule height0.9\ht0\hss}\box0}}
{\setbox0=\hbox{$\scriptscriptstyle\rm Q$}\hbox{\hbox to0pt
{\kern0.4\wd0\vrule height0.9\ht0\hss}\box0}}}}
\def\Rl{{\mathchoice
{\setbox0=\hbox{$\displaystyle\rm R$}\hbox{\hbox to0pt
{\kern0.4\wd0\vrule height0.9\ht0\hss}\box0}}
{\setbox0=\hbox{$\textstyle\rm R$}\hbox{\hbox to0pt
{\kern0.4\wd0\vrule height0.9\ht0\hss}\box0}}
{\setbox0=\hbox{$\scriptstyle\rm R$}\hbox{\hbox to0pt
{\kern0.4\wd0\vrule height0.9\ht0\hss}\box0}}
{\setbox0=\hbox{$\scriptscriptstyle\rm R$}\hbox{\hbox to0pt
{\kern0.4\wd0\vrule height0.9\ht0\hss}\box0}}}}
\def\Co{{\mathchoice
{\setbox0=\hbox{$\displaystyle\rm C$}\hbox{\hbox to0pt
{\kern0.4\wd0\vrule height0.9\ht0\hss}\box0}}
{\setbox0=\hbox{$\textstyle\rm C$}\hbox{\hbox to0pt
{\kern0.4\wd0\vrule height0.9\ht0\hss}\box0}}
{\setbox0=\hbox{$\scriptstyle\rm C$}\hbox{\hbox to0pt
{\kern0.4\wd0\vrule height0.9\ht0\hss}\box0}}
{\setbox0=\hbox{$\scriptscriptstyle\rm C$}\hbox{\hbox to0pt
{\kern0.4\wd0\vrule height0.9\ht0\hss}\box0}}}}
\def\Hl{{\mathchoice
{\setbox0=\hbox{$\displaystyle\rm H$}\hbox{\hbox to0pt
{\kern0.4\wd0\vrule height0.9\ht0\hss}\box0}}
{\setbox0=\hbox{$\textstyle\rm H$}\hbox{\hbox to0pt
{\kern0.4\wd0\vrule height0.9\ht0\hss}\box0}}
{\setbox0=\hbox{$\scriptstyle\rm H$}\hbox{\hbox to0pt
{\kern0.4\wd0\vrule height0.9\ht0\hss}\box0}}
{\setbox0=\hbox{$\scriptscriptstyle\rm H$}\hbox{\hbox to0pt
{\kern0.4\wd0\vrule height0.9\ht0\hss}\box0}}}}
\def\Ol{{\mathchoice
{\setbox0=\hbox{$\displaystyle\rm O$}\hbox{\hbox to0pt
{\kern0.4\wd0\vrule height0.9\ht0\hss}\box0}}
{\setbox0=\hbox{$\textstyle\rm O$}\hbox{\hbox to0pt
{\kern0.4\wd0\vrule height0.9\ht0\hss}\box0}}
{\setbox0=\hbox{$\scriptstyle\rm O$}\hbox{\hbox to0pt
{\kern0.4\wd0\vrule height0.9\ht0\hss}\box0}}
{\setbox0=\hbox{$\scriptscriptstyle\rm O$}\hbox{\hbox to0pt
{\kern0.4\wd0\vrule height0.9\ht0\hss}\box0}}}}
\def\sh{\sinh}
\def\ch{\cosh}

\title{Gauge Field Theory Coherent States (GCS) : III.\\
Ehrenfest Theorems}
\author{T. Thiemann\thanks{thiemann@aei-potsdam.mpg.de},
O. Winkler\thanks{winkler@aei-potsdam.mpg.de} \\
MPI f. Gravitationsphysik, Albert-Einstein-Institut, \\
Am M\"uhlenberg 1, 14476 Golm near Potsdam, Germany}
\date{{\small Preprint AEI-2000-029}}   

\begin{document}

\maketitle

\begin{abstract}
In the preceding paper of this series of articles
we established peakedness properties of a family of coherent states 
that were introduced by Hall for any compact gauge group and were 
later generalized to gauge field theory by Ashtekar, Lewandowski,
Marolf, Mour\~ao and Thiemann.

In this paper we establish the ``Ehrenfest Property'' of these states
which are labelled by a point $(A,E)$, a connection and an electric field,
in the classical phase space.
By this we mean that \\
i) The expectation value of {\it all} elementary quantum 
operators $\hat{O}$ with respect to the coherent state with label $(A,E)$ is 
given to zeroth order in $\hbar$ by the value of the corresponding
classical function $O$ evaluated at the phase space point $(A,E)$ and\\
ii) The expectation value of the commutator between two elementary quantum 
operators $[\hat{O}_1,\hat{O}_2]/(i\hbar)$ divided by $i\hbar$ with 
respect to the coherent state with label $(A,E)$ is 
given to zeroth order in $\hbar$ by the value of the Poisson bracket 
between the corresponding classical functions $\{O_1,O_2\}$ 
evaluated at the phase space point $(A,E)$. 

These results can be extended to all polynomials of elementary operators and
to a certain non-polynomial function of the elementary operators associated
with the volume operator of quantum general relativity. It follows that
the infinitesimal quantum dynamics of quantum general relativity 
is  to zeroth order in $\hbar$ indeed given by classical general relativity.
\end{abstract}

\section{Introduction}
\label{s1}

Quantum General Relativity (QGR) has matured over the past decade to a 
mathematically well-defined theory of quantum gravity. 
In contrast to string theory, by definition GQR is a
manifestly background independent, diffeomorphism 
invariant and non-perturbative theory.
The obvious advantage is that one will never have to postulate the
existence of a non-perturbative extension of the theory,
which in string theory has been called the still unknown 
M(ystery)-Theory.

The disadvantage of a non-perturbative and background independent
formulation is, of course, that one is faced with new and interesting 
mathematical problems so that one cannot just go ahead and 
``start calculating scattering amplitudes'': 
As there is no background around which one could perturb, rather the full 
metric is fluctuating, one is not
doing quantum field theory on a spacetime but only on a differential
manifold. Once there is no (Minkowski) metric at our disposal, one loses
familiar notions such as causality, locality, Poincar\'e group 
and so forth, in other words, the theory is not a theory to which
the Wightman axioms apply. Therefore, one must build an entirely
new mathematical apparatus to treat the resulting quantum field theory 
which is {\it drastically different from the Fock space picture 
to which particle physicists are used to}.

As a consequence, the mathematical formulation of the theory was the main 
focus of research in the field over the past decade. The main 
achievements to date are the following (more or less in chronological 
order) : 
\begin{itemize}
\item[i)] {\it Kinematical Framework}\\
The starting 
point was the introduction of new field variables \cite{1} for the 
gravitational field which are better suited to a background  
independent formulation of the quantum theory than the ones employed
until that time. In its original version these variables were
complex valued, however, currently their real valued version,  
considered first in \cite{1a} for {\it classical} Euclidean gravity and 
later in \cite{1b} for {\it classical} Lorentzian gravity, is preferred 
because to date it seems that it is only with these variables that one can 
rigorously define the kinematics and dynamics of Euclidean or Lorentzian  
{\it quantum} gravity \cite{1c}. \\
These variables are coordinates for the infinite dimensional phase
space of an $SU(2)$ gauge theory subject to further constraints 
besides the Gauss law, that is, a connection and a canonically
conjugate electric field. As such, it is very natural to introduce
smeared functions of these variables, specifically Wilson loop and 
electric flux functions. (Notice that one does not need a metric 
to define these functions, that is, they are background independent).
This had been done for ordinary gauge fields already before in \cite{2} 
and was then reconsidered for gravity (see e.g. \cite{3}).\\
The next step was the choice of a representation of the canonical
commutation relations between the electric and magnetic degrees
of freedom. This involves the choice of a suitable space of 
distributional connections \cite{4} and a faithful measure thereon \cite{5}
which, as one can show \cite{6}, is $\sigma$-additive.
The proof that the resulting Hilbert space indeed solves the adjointness 
relations induced by the reality structure of the classical theory
as well as the canonical commutation relations induced by the symplectic 
structure of the classical theory can be found in \cite{7}.
Independently, a second representation, called the loop 
representation, of the canonical commutation
relations had been advocated (see e.g. \cite{8} and especially \cite{8a}
and references therein)
but both representations were shown to be unitarily equivalent in
\cite{9} (see also \cite{10} for a different method of proof).\\
This is then the first major achievement : The theory is based on
a rigorously defined kinematical framework.
\item[ii)] {\it Geometrical Operators}\\
The second major achievement concerns the spectra of positive 
semi-definite, self-adjoint geometrical
operators measuring lengths \cite{11}, areas \cite{12,13}
and volumes \cite{12,14,15,16,8} of curves, surfaces and regions
in spacetime. These spectra are pure point (discete) and imply a discrete
Planck scale structure. It should be pointed out that the discreteness
is, in contrast to other approaches to quantum gravity, not put in
by hand but it is a {\it prediction} !
\item[iii)] {\it Regularization- and Renormalization Techniques}\\
The third major achievement is that there is a new 
regularization and renormalization technique \cite{17,18}
for diffeomorphism covariant, density-one-valued operators at our disposal
which was successfully tested in model theories \cite{19}. This
technique can be applied, in particular, to the standard model
coupled to gravity \cite{20,21} and to the Poincar\'e generators at 
spatial infinity \cite{22}. In particular, it works for {\it Lorentzian}
gravity while all earlier proposals could at best work in the Euclidean 
context only (see, e.g. \cite{8a} and references therein). 
The algebra of important operators of the
resulting quantum field theories was shown to be consistent \cite{23}. 
Most surprisingly, these operators are {\it UV and IR finite} !
Notice that this result, at least as far as these operators are concerned, 
is stronger than the believed but unproved
finiteness of scattering amplitudes
order by order in perturbation theory of the five critical
string theories, in a sense we claim that the perturbation series converges.
The absence of the divergences that usually plague interacting quantum fields
propagating on a Minkowski background can be understood intuitively
from the diffeomorphism invariance of the theory : ``short and long distances
are gauge equivalent''. We will elaborate more on this point in future 
publications. 
\item[iv)] {\it Spin Foam Models}\\
After the construction of the densely defined Hamiltonian constraint
operator of \cite{17,18}, a formal, Euclidean functional integral was
constructed in \cite{23a} and gave rise to the so-called spin foam 
models   
(a spin foam is a history of a graph with faces as the history of edges)
\cite{23b}. Spin foam models are in close connection with causal
spin-network evolutions \cite{23c}, state sum models \cite{23d} and
topological quantum field theory, in particular BF theory \cite{23e}. To
date most results are at a formal level and for the Euclidean version of the
theory only but the programme is exciting since it may restore manifest
four-dimensional diffeomorphism invariance which in the Hamiltonian
formulation is somewhat hidden.
\item[v)]
Finally, the fifth major achievement is the existence of a rigorous and 
satisfactory framework \cite{24,25,26,27,28,29,30} for the quantum 
statistical description of black holes
which reproduces the Bekenstein-Hawking Entropy-Area relation and applies,
in particular, to physical Schwarzschild black holes while stringy black 
holes so far are under control only for extremal charged black holes.
\end{itemize}
Summarizing, the work of the past decade has now 
culminated in a promising starting point for a quantum theory of the 
gravitational field plus matter and the stage is set to pose and answer 
physical questions. 

The most basic and most important question that one should ask is :
{\it Does the theory have classical general relativity as its classical
limit ?} Notice that even if the answer is negative, the existence
of a consistent, interacting, diffeomorphism invariant quantum field theory 
in four dimensions is already a quite non-trivial result. However, we can 
claim to have a satisfactory quantum theory of Einstein's theory
only if the answer is positive. 

To settle this issue we have launched an attack based on coherent 
states which has culminated in a series of papers called 
``Gauge Field Theory Coherent States'' \cite{31,32,33,34,35,36}
and this paper is the third one of this collection (to be 
continued). It is closely connected with the companion paper
\cite{32}. In \cite{32} we established peakedness properties of the
coherent states of the heat kernel family introduced by Hall \cite{36a}
for arbitrary compact gauge groups which were later applied to
gauge field theories in \cite{36b}. The results of \cite{32} rest
on the explicit determination of the configuration space complexification
via the complexifier framework \cite{36c}. They reveal that the heat
kernel family more or less has all the properties that one would
like coherent states to have and that one is used to from the harmonic 
oscillator coherent states. In particular, these states $\psi^t_m$ are 
labelled by a point $m=(q,p)$ in the classical phase space and 1) are 
eigenstates of
certain annihilation operators, 2) are overcomplete, 3) saturate the 
unquenched Heisenberg uncertainty bound and 4) are peaked in the  
configuration representation at $x=q$, in the momentum representation at
$k=p$ and in the Bargmann-Segal representation at $z=q-ip\simeq m$. 
Here $t$ is 
a classicality parameter proportional to Planck's constant $\hbar$
and the peak is the sharper (the decay width $\propto \sqrt{t}$
the smaller) the smaller $t$ and resembles almost a Gaussian.

The properties listed ensure that normal ordered products of
creation and annihilation operators have exactly the expectation value, 
with respect to $\psi^t_m$,
given by the product of the associated classical functions evaluated at 
the phase space point $m$ {\it without any quantum corrections}.
However, to establish that this expectation value property also holds with 
respect to the elementary operators in terms of which important operators 
of quantum gauge field theory, such as Hamiltonians, are formulated  
is not granted a priori. The problem arises because the creation 
and annihilation operators of \cite{32} are not polynomial functions 
of the elementary operators which in turn is directly related to the 
kinematically non-linear nature of the theory.. 
Therefore, the framework of \cite{36d} to prove Ehrenfest  
theorems and the determination of the classical limit by using the 
harmonic oscillator coherent states does not extend to our case since
the methods of \cite{36d} crucially rest on the assumption 
that the basic operators are {\it linear combinations} of creation and  
annihilation operators.

The present paper is devoted to filling this gap. As in \cite{32}
all the proofs will be carried out for the case of rank one gauge groups,
that is $G=SU(2),U(1)$, only but by the arguments given in \cite{32}
they should readily extend to the case of an arbitrary compact gauge
group which we leave for future work \cite{36e}. 
With this restriction, the main result of the present article is that
the Ehrenfest property, to zeroth and first order, indeed holds for our 
coherent states. In other words,
the expectation values of polynomials of the elementary operators
as well as of an important operator, associated with the volume 
operator of quantum general relativity mentioned above, which is not a
polynomial (not even analytical !) function of the elementary operators,
reproduce, to zeroth order in $t$, the values of the correponding classical 
functions at the phase space point given by the coherent state. Moreover, 
the expectation values of commutators divided by $it$ reproduces the  
corresponding Poisson bracket, to zeroth order in $t$, at the given
phase space point. These results imply that the quantum dynamics
of the operators constructed in \cite{17,18,20}, as expected, reproduce
the infinitesimal classical dynamics of general relativity \cite{36f},
putting the worries raised in \cite{36g} ad acta.\\
\\
The architecture of the present article is as follows :\\

Section two summarizes the classical and quantum kinematical framework
for diffeomorphism invariant quantum gauge field theories.

In section three, after a brief review of the heat kernel coherent
states, we prove the above mentioned Ehrenfest theorems for the case
of the gauge-variant coherent states for the gauge group $G=SU(2)$. 
As stated already in \cite{32}, we are mostly interested in gauge-variant
coherent states because a) only those manage to verify the satisfaction
of a consistent quantum constraint algebra and b) the expectation
values of gauge -- and diffeomorphism invariant operators are gauge --
and diffeomorphism invariant since both gauge groups are represented 
unitarily on the Hilbert space.

Finally, in appendix A we repeat the analysis of section three for 
the gauge group $G=U(1)$. As in \cite{32}, the Abelian nature of $U(1)$
shortens all the proofs given for $SU(2)$ by an order of magnitude
and the reader is urged to study first the appendix before delving
into the technically much harder section three.

\section{Kinematical Structure of Diffeomorphism Invariant Quantum
Gauge Theories}
\label{s2}

In this section we will recall the main ingredients of the mathematical
formulation of (Lorentzian) diffeomorphism invariant classical and quantum 
field theories of 
connections with local degrees of freedom in any dimension and for
any compact gauge group. See \cite{7,37} and references therein
for more details. In this section we will take all quantities to be 
dimensionless. More about dimensionful constants will be said in section
\ref{s3}.

\subsection{Classical Theory}
\label{s2.1}

Let $G$ be a compact gauge group, $\Sigma$ a $D-$dimensional manifold 
admitting a principal $G-$bundle with connection over $\Sigma$.
Let us denote the pull-back to $\Sigma$ of the connection 
by local sections by $A_a^i$
where $a,b,c,..=1,..,D$ denote tensorial indices and $i,j,k,..=1,..,
\dim(G)$ denote indices for the Lie algebra of $G$. 
Likewise, consider a density-one vector bundle of electric fields, whose
pull-back to $\Sigma$ by local sections (their Hodge dual is a 
$D-1$ form) is a Lie algebra valued 
vector density of weight one. We will denote the set of generators
of the rank $N-1$ Lie algebra of $G$ by $\tau_i$ which are normalized
according to $\mbox{tr}(\tau_i\tau_j)=-N\delta_{ij}$ and 
$[\tau_i,\tau_j]=2f_{ij}\;^k\tau_k$ defines the structure constants 
of $Lie(G)$. 

Let $F^a_i$ be a Lie algebra valued vector density test field of weight one 
and let $f_a^i$ be a Lie algebra valued covector test field. 
We consider the smeared quantities
\be \label{2.1}
F(A):=\int_\Sigma d^Dx F^a_i A_a^i\mbox{ and } 
E(f):=\int_\Sigma d^Dx E^a_i f_a^i 
\ee
While both of them are diffeomorphism covariant, only the latter is gauge 
covariant, one reason to introduce the singular smearing functions 
discussed below.
The choice of the space of pairs of test fields $(F,f)\in{\cal S}$ 
depends on the boundary conditions on
the space of connections and electric fields which in turn depends on the 
topology of $\Sigma$ and will not be specified in what follows. 

The set
of all pairs of smooth functions $(A,E)$ on $\Sigma$ such that (\ref{2.1}) is 
well defined for any $(F,f)\in {\cal S}$ defines an infinite dimensional
set $M$. We define a topology on $M$ through the following globally
defined metric :
\ba \label{2.2}
&& d_{\rho,\sigma}[(A,E),(A',E')] \\
&:=& \sqrt{-\frac{1}{N}\int_\Sigma d^Dx 
[\sqrt{\det(\rho)} \rho^{ab} \mbox{tr}([A_a-A'_a][A_b-A'_b])+
\frac{[\sigma_{ab} \mbox{tr}([E^a-E^{a\prime}][E^b-E^{b\prime}])}
{\sqrt{\det(\sigma)}}]} \nonumber
\ea
where $\rho_{ab},\sigma_{ab}$ are fiducial metrics on $\Sigma$ of 
everywhere Euclidean signature. Their fall-off behaviour has to be suited
to the boundary conditions of the fields $A,E$ at spatial infinity.
Notice that the metric (\ref{2.2}) on $M$ is gauge invariant. It can be used   
in the usual way to equip $M$ with the structure of a smooth,
infinite dimensional differential
manifold modelled on a Banach (in fact Hilbert) space $\cal E$
where ${\cal S}\times {\cal S}\subset {\cal E}$. (It is the 
weighted Sobolev space $H_{0,\rho}^2\times H_{0,\sigma^{-1}}^2$ in the 
notation of \cite{38}). 

Finally, we equip $M$ with the structure of an infinite dimensional 
symplectic manifold through the following strong (in the sense of 
\cite{39})
symplectic structure 
\be \label{2.3}
\Omega((f,F),(f',F'))_m:=\int_\Sigma d^Dx [F^a_i f^{i\prime}_a
-F^{a\prime}_i f^i_a](x)
\ee
for any $(f,F),(f',F')\in {\cal E}$. We have abused the notation by 
identifying the tangent space to $M$ at $m$ with $\cal E$. To prove 
that $\Omega$ is a strong symplectic structure one uses standard 
Banach space techniques. Computing the Hamiltonian vector fields
(with respect to $\Omega$) of the functions $E(f),F(A)$ we obtain the
following elementary Poisson brackets
\be \label{2.4}
\{E(f),E(f')\}=\{F(A),F'(A)\}=0,\;\{E(f),A(F)\}=F(f)
\ee
As a first step towards quantization of the symplectic manifold
$(M,\Omega)$ one must choose a polariztion. As usual in gauge theories,
we will use a particular real polarization, specifically connections as 
the configuration variables and electric fields 
as canonically conjugate momenta. As a second step one must decide
on a complete set of coordinates of $M$ which are to become the elementary
quantum operators. The analysis just outlined suggests to use the 
coordinates $E(f),F(A)$. However, the well-known immediate problem is that 
these coordinates are not gauge covariant. Thus, we proceed as follows :

Let $\Gamma^\omega_0$ be the set
of all piecewise analytic, finite, oriented graphs $\gamma$ embedded into 
$\Sigma$ 
and denote by $E(\gamma)$ and $V(\gamma)$ respectively its sets of oriented
edges $e$ and vertices $v$ respectively. Here finite means that 
$E(\gamma)$ is a finite set. (One can extend the framework to 
$\Gamma^\omega_0$, the restriction to webs of the set of
piecewise smooth graphs \cite{40,41} but the description becomes more 
complicated and we refrain from doing this here). 
It is possible to consider the set $\Gamma^\omega_\sigma$ of piecewise 
analytic, infinite graphs with
an additional regularity property \cite{35} but for the purpose of this
paper it will be sufficient to stick to $\Gamma^\omega_0$. The subscript
$_0$ as usual denotes ``of compact support'' while $_\sigma$ denotes
``$\sigma$-finite''.

We denote by $h_e(A)$ the holonomy
of $A$ along $e$ and say that a function $f$ on $\a$ is cylindrical with 
respect to $\gamma$ if there exists a function $f_\gamma$ on 
$G^{|E(\gamma)|}$ such that $f=p_\gamma^\ast f_\gamma=f\circ p_\gamma$ 
where $p_\gamma(A)=\{h_e(A)\}_{e\in E(\gamma)}$. 
Holonomies are invariant under
reparameterizations of the edge and in this article we assume that
the edges are always analyticity preserving diffeomorphic images from 
$[0,1]$ to a
one-dimensional submanifold of $\Sigma$. Gauge transformations are functions
$g:\;\Sigma\mapsto G;\;x\mapsto g(x)$ and they act on
holonomies as $h_e\mapsto g(e(0))h_e g(e(1))$. 

Next, given a graph $\gamma$ we choose a polyhedronal decomposition
$P_\gamma$ of $\Sigma$ dual to $\gamma$. The precise definition
of a dual polyhedronal decomposition can be found in \cite{37} but
for the purposes of the present paper it is sufficient to know that
$P_\gamma$ assigns to each edge $e$ of $\gamma$ an open ``face''
$S_e$ (a polyhedron of codimension one embedded into $\Sigma$) with 
the following properties :\\ 
(1) the surfaces $S_e$ are mutually non-intersecting,\\ 
(2) only the edge $e$ intersects $S_e$, the intersection is transversal
and consists only of one point which is an interiour point of both
$e$ and $S_e$,\\
(3) $S_e$ carries the orientation which agrees with the orientation 
of $e$.\\
Furthermore, we choose a system $\Pi_\gamma$ of paths $\rho_e(x) \subset
S_e,\; x\in S_e,\; e\in E(\gamma)$ connecting the intersection point 
$p_e=e\cap S_e$ with $x$. The paths vary smoothly with
$x$ and the triples $(\gamma,P_\gamma,\Pi_\gamma)$ have the property
that if $\gamma,\gamma'$ are diffeomorphic, so
are $P_\gamma,P_{\gamma'}$ and $\Pi_\gamma,\Pi_{\gamma'}$, see
\cite{37} for details.

With these structures we define the following function on $(M,\Omega)$
\be \label{2.5}
P^e_i(A,E):=-\frac{1}{N}
\mbox{tr}(\tau_i h_e(0,1/2)[\int_{S_e} h_{\rho_e(x)} \ast E(x) 
h_{\rho_e(x)}^{-1}] h_e(0,1/2)^{-1})
\ee
where $h_e(s,t)$ denotes the holonomy of $A$ along $e$ between the 
parameter values $s<t$, $\ast$ denotes the Hodge dual, that is,
of $\ast E$ is a $(D-1)-$form on $\Sigma$, $E^a:=E^a_i\tau_i$ and
we have chosen a parameterization of $e$ such that $p_e=e(1/2)$.

Notice that in contrast to similar variables used earlier in the literature
the function $P^e_i$ is {\it gauge covariant}. Namely, under gauge 
transformations it transforms as $P^e\mapsto g(e(0)) P^e g(e(0))^{-1}$,
the price to pay being that $P^e$ depends on both $A$ and $E$ and not 
only on $E$. The idea is therefore to use the variables $h_e,P^e_i$
for all possible graphs $\gamma$ as the coordinates of $M$.

The problem with the functions $h_e(A)$ and $P^e_i(A,E)$ on $M$ is that 
they are not differentiable on $M$, that is, $Dh_e, DP^e_i$ are nowhere  
bounded operators on $\cal E$ as one can easily see. The reason for this is,
of course, that these are functions on $M$ which are not properly smeared 
with functions from $\cal S$, rather they are smeared with distributional
test functions with support on $e$ or $S_e$ respectively. Nevertheless
one would like to base the quantization of the theory on these functions 
as basic variables because of their gauge and diffeomorphism covariance.
Indeed, under diffeomorphisms $h_e\mapsto h_{\varphi^{-1}(e)},
P^e_i\mapsto P^{\varphi^{-1}(e)}_i$ where the latter notation means that
$P^{\varphi^{-1}(e)}_e$ is labelled by 
$\varphi^{-1}(S_e),\varphi^{-1}(\Pi_\gamma)$. We proceed as follows. 
\begin{Definition} \label{def2.1}
By $\bar{M}_\gamma$ we denote the direct product 
$[G\times Lie(G)]^{|E(\gamma)|}$. 
The subset of $\bar{M}_\gamma$ of pairs 
$(h_e(A),P^e_i(A,E))_{e\in E(\gamma)}$ as 
$(A,E)$ varies over $M$ will be denoted by $(\bar{M}_\gamma)_{|M}$. We have a 
corresponding pull-back map $p_\gamma:\;M\mapsto \bar{M}_\gamma$ which
maps $M$ onto $(\bar{M}_{\gamma})_{|M}$.
\end{Definition}
Notice that the set $(\bar{M}_\gamma)_{|M}$ is in general a proper subset of 
$\bar{M}_\gamma$,
depending on the boundary conditions on $(A,E)$, the topology of $\Sigma$ 
and the ``size'' of $e,S_e$. For instance, in the limit of $e,S_e\to  
e\cap S_e$ but holding the number of edges fixed, $(\bar{M}_\gamma)_{|M}$  
will consist of only one point in $M_\gamma$. This follows from the 
smoothness of the $(A,E)$. 

We equip a subset $M_\gamma$ of $\bar{M}_\gamma$ with the structure of a 
differentiable manifold modelled
on the Banach space ${\cal E}_\gamma=\Rl^{2\dim(G)|E(\gamma)|}$
by using the natural direct product manifold structure of
$[G\times Lie(G)]^{|E(\gamma)|}$. While $\bar{M}_\gamma$ is a kind of 
distributional phase space, $M_\gamma$ satisfies suitable regularity 
properties similar to $M$.

In order to proceed and to give $M_\gamma$ a symplectic structure 
{\it derived from $(M,\Omega)$} one must 
regularize the elementary functions $h_e, P^e_i$ by writing them as limits  
(in which the regulator vanishes) of functions which can be expressed  
in terms of the $F(A),E(f)$. Then one can compute their Poisson
brackets with respect to the symplectic structure $\Omega$ at finite
regulator and then take the limit pointwise on $M$. The result is the 
following  
well-defined strong symplectic structure $\Omega_\gamma$ on $M_\gamma$. 
\ba \label{2.6}
\{h_e,h_{e'}\}_\gamma &=& 0\nonumber\\
\{P^e_i,h_{e'}\}_\gamma &=&
\delta^e_{e'} \frac{\tau_i}{2}h_e\nonumber\\
\{P^e_i,P^{e'}_j\}_\gamma &=&
-\delta^{ee'}f_{ij}\;^k P^e_k
\ea
Since $\Omega_\gamma$ is obviously block diagonal, each block standing
for one copy of $G\times Lie(G)$, to check that $\Omega_\gamma$ is 
non-degenerate and closed reduces to doing it for each factor together
with an appeal to well-known Hilbert space techniques to establish that
$\Omega_\gamma$ is a surjection of ${\cal E}_\gamma$.
This is done in \cite{37} where it is shown that each copy is isomorphic
with the cotangent bundle $T^\ast G$ equipped with the symplectic structure
(\ref{2.6}) (choose $e=e'$ and delete the label $e$). \\
\\
Now that we have managed to assign to each graph $\gamma$ a symplectic
manifold $(M_\gamma,\Omega_\gamma)$ we can quantize it by using geometric
quantization. This can be done in a well-defined way because the relations
(\ref{2.6}) show that the corresponding operators are non-distributional.
This is therefore a clean starting point for the regularization of any 
operator
of quantum gauge field theory which can always be written in terms 
of the $\hat{h}_e,\hat{P}^e,\;e\in E(\gamma)$ if we apply this operator to
a function which depends only on the $h_e,\; e\in E(\gamma)$. 

The question is what $(M_\gamma,\Omega_\gamma)$ has to do with $(M,\Omega)$.
In \cite{37} it is shown that there exists a partial order $\prec$ on the 
set $\cal L$ of triples $l=(\gamma,P_\gamma,\Pi_\gamma)$. 
In particular, $\gamma\prec\gamma'$ means $\gamma\subset\gamma'$
and $\cal L$ is a directed set so that one can form
a generalized projective limit $M_\infty$ of the $M_\gamma$ (we abuse 
notation in 
displaying the dependence of $M_\gamma$ on $\gamma$ only rather than on
$l$). For this one verifies that the family 
of symplectic structures $\Omega_\gamma$ is self-consistent
in the sense that if 
$(\gamma,P_\gamma,\Pi_\gamma)\prec (\gamma',P_{\gamma'},\Pi_{\gamma'})$ 
then $p_{\gamma'\gamma}^\ast\{f,g\}_\gamma
=\{p_{\gamma'\gamma}^\ast f,p_{\gamma'\gamma}^\ast g\}_{\gamma'}$
for any $f,g\in C^\infty(M_\gamma)$ and 
$p_{\gamma'\gamma}:\;M_{\gamma'}\mapsto M_\gamma$ is a system
of natural projections, more precisely, of (non-invertible) 
symplectomorphisms. 

Now, via the maps $p_\gamma$ of definition \ref{def2.1} we can identify
$M$ with a subset of $M_\infty$. Moreover, in \cite{37} it is shown that
there is a generalized projective sequence $(\gamma_n,P_{\gamma_n},
\Pi_{\gamma_n})$
such that $\lim_{n\to\infty}p_{\gamma_n}^\ast\Omega_{\gamma_n}=\Omega$
pointwise in $M$. This displays $(M,\Omega)$ as embedded into a projective
generalized limit of the $(M_\gamma,\Omega_\gamma)$, intuitively speaking, 
as $\gamma$ 
fills all of $\Sigma$, we recover $(M,\Omega)$ from the 
$(M_\gamma,\Omega_\gamma)$. Of course, this works with $\Gamma^\omega_0$ 
only if $\Sigma$ is compact, otherwise we need the extension to 
$\Gamma^\omega_\sigma$.

It follows that quantization of $(M,\Omega)$, and conversely taking the 
classical limit, can be studied purely in terms of $(M_\gamma,\Omega_\gamma)$
for {\it all} $\gamma$. The quantum kinematical framework is
given in the next subsection.

\subsection{Quantum Theory}
\label{s2.2}

Let us denote the set of all smooth connections by $\a$. This is our
classical configuration space and we will choose for its coordinates the
holonomies $h_e(A),\;e\in\gamma,\;\gamma\in\Gamma^\omega_0$. 
$\a$ is naturally equipped with a metric topology induced by (\ref{2.2}). 

Recall the notion of a function cylindrical over a graph from the 
previous subsection.
A particularly useful set of cylindrical functions are the so-called 
spin-netwok functions \cite{42,43,9}. A spin-network function is 
labelled by a graph $\gamma$, a set of non-trivial irreducible 
representations 
$\vec{\pi}=\{\pi_e\}_{e\in E(\gamma)}$ (choose from each equivalence 
class of equivalent
representations once and for all a fixed representant), one for each 
edge of $\gamma$, and a set $\vec{c}=\{c_v\}_{v\in V(\gamma)}$ of
contraction matrices, one for each vertex of $\gamma$, which 
contract the indices of the tensor product 
$\otimes_{e\in E(\gamma)} \pi_e(h_e)$ in such a way that the resulting
function is gauge invariant. We denote spin-network functions as
$T_I$ where $I=\{\gamma,\vec{\pi},\vec{c}\}$ is a compound label.
One can show that these functions are linearly independent.
>From now on we denote by $\tilde{\Phi}_\gamma$ finite linear combinations of
spin-network functions over $\gamma$, by $\Phi_\gamma$ the finite linear 
combinations of elements from any possible $\tilde{\Phi}_{\gamma'},\;
\gamma'\subset\gamma$ a subgraph of $\gamma$  
and by $\Phi$ the finite linear 
combinations of spin-network functions over an arbitrary finite collection 
of graphs. Clearly $\tilde{\Phi}_{\gamma}$ is a subspace of 
$\Phi_\gamma$.  To express this distinction we will say that functions 
in $\tilde{\Phi}_\gamma$ are labelled by the ``coloured graphs'' $\gamma$
while functions in $\Phi_\gamma$ are labelled simply by graphs $\gamma$
where we abuse notation by using the same symbol $\gamma$.

The set $\Phi$ of finite linear combinations of spin-network functions 
forms an Abelian $^\ast$ algebra 
of functions on $\a$. By completing it with respect to the sup-norm 
topology it 
becomes an Abelian C$^\ast$ algebra (here the compactness of $G$ is 
crucial). The spectrum $\ab$ of this algebra, 
that is, the set of all algebraic homomorphisms ${\cal B}\mapsto\Co$
is called the quantum configuration space. This space is equipped with
the Gel'fand topology, that is, the space of continuous functions
$C^0(\ab)$
on $\ab$ is given by the Gel'fand transforms of elements of $\cal B$.
Recall that the Gel'fand transform is given by $\tilde{f}(\bar{A}):=
\bar{A}(f)\;\forall \bar{A}\in \ab$. It is a general result that $\ab$ with 
this topology is a compact Hausdorff space. Obviously, the elements of
$\a$ are contained in $\ab$ and one can show that $\a$ is even dense
\cite{44}. Generic elements of $\ab$ are, however, distributional.

The idea is now to construct a Hilbert space consisting of square
integrable functions on $\ab$ with respect to some measure $\mu$. Recall 
that one can define a measure on a locally compact Hausdorff space 
by prescribing a positive linear functional $\chi_\mu$ on the space 
of continuous functions thereon. The particular measure
we choose is given by $\chi_{\mu_0}(\tilde{T}_I)=1$ if $I=\{\{p\},
\vec{0},\vec{1} \}$ and $\chi_{\mu_0}(\tilde{T}_I)=0$ otherwise. Here
$p$ is any point in $\Sigma$, $0$ denotes the 
trivial representation and $1$ the trivial contraction matrix. In other 
words, (Gel'fand transforms of) spin-network functions play the same role 
for $\mu_0$ as 
Wick-polynomials do for Gaussian measures and like those they form
an orthonormal basis in the Hilbert space ${\cal H}:=L_2(\ab,d\mu_0)$ 
obtained by completing their finite linear span $\Phi$.\\
An equivalent definition of $\ab,\mu_0$ is as follows :\\ 
$\ab$ is in one to one correspondence, via the surjective map $H$ defined 
below, with the set $\ab':=\mbox{Hom}({\cal X},G)$
of homomorphisms from the groupoid $\cal X$ of composable, holonomically
independent, analytical paths
into the gauge group. The correspondence is explicitly given by
$\ab\ni\bar{A}\mapsto H_{\bar{A}}\in\mbox{Hom}({\cal X},G)$
where ${\cal X}\ni e\mapsto H_{\bar{A}}(e):=\bar{A}(h_e)=
\tilde{h}_e(\bar{A})\in G$ and $\tilde{h}_e$ is the Gel'fand transform
of the function $\a\ni A\mapsto h_e(A)\in G$. Consider now the restriction
of $\cal X$ to ${\cal X}_\gamma$, the groupoid of composable edges of  
the graph $\gamma$. One can then show that the projective limit of the 
corresponding {\it cylindrical sets} 
$\ab'_\gamma:=\mbox{Hom}({\cal X}_\gamma,G)$ coincides with $\ab'$.
Moreover, we have $\{\{H(e)\}_{e\in E(\gamma)};\;H\in\ab'_\gamma\}=
\{\{H_{\bar{A}}(e)\}_{e\in E(\gamma)};\;\bar{A}\in\ab\}=
G^{|E(\gamma)|}$.
Let now $f\in{\cal B}$ be a function cylindrical over $\gamma$ then 
$$
\chi_{\mu_0}(\tilde{f})=\int_{\ab} d\mu_0(\bar{A}) \tilde{f}(\bar{A})
=\int_{G^{|E(\gamma)|}} \otimes_{e\in E(\gamma)} d\mu_H(h_e)
f_\gamma(\{h_e\}_{e\in E(\gamma)})
$$
where $\mu_H$ is the Haar measure on $G$.
As usual, $\a$ turns out to be contained in a measurable subset of 
$\ab$ which has measure zero with respect to $\mu_0$.

Let $\Phi_\gamma$, as before, be the finite linear span of spin-network 
functions
over $\gamma$ and ${\cal H}_\gamma$ its completion with respect to
$\mu_0$. Clearly, $\cal H$ itself is the completion of the finite linear
span $\Phi$ of vectors from the mutually orthogonal $\tilde{\Phi}_\gamma$. 
Our 
basic coordinates of $M_\gamma$ are promoted to operators on ${\cal H}$ with 
dense domain $\Phi$. As $h_e$ is group-valued and $P^e$ is real-valued
we must check that the adjointness relations coming from these reality 
conditions as well as the Poisson brackets (\ref{2.6}) are implememted on
our ${\cal H}$. This turns out to be precisely the case if we choose
$\hat{h}_e$ to be a multiplication operator and 
$\hat{P}^e_j=i\hbar\kappa X^e_j/2$
where $X^e_j=X_j(h_e)$ and $X^j(h),\;h\in G$ is the vector field on $G$
generating left translations into the $j-th$ coordinate direction of 
$Lie(G)\equiv T_h(G)$ (the tangent space of $G$ at $h$ can be identified 
with the Lie algebra of $G$) and $\kappa$ is the coupling constant of the 
theory. For details see \cite{7,37}.

\section{Ehrenfest Theorems}
\label{s3}

Let us recall the most important facts from \cite{32}.\\
\\
Instead of working with the quantities $P^e_i$ of section \ref{s2}
we use the dimensionless objects $p^e_i=P^e_i/a^{n_D}$. If  
$P^e_i$ is already dimensionless then so is $a$ and we choose
$n_D=1$. Otherwise $a$
is an arbitrary but fixed constant of the dimension $[a]=\mbox{cm}^1$
and the power $n_D$ is so chosen that $p^e_i$ is dimensionless. In both
cases, the numerical value of $a$ is macroscopic, say $a=1$ or $a=$1cm 
respectively. The power $n_D$ depends on the 
dimensionality of $\Sigma$ and the theory, e.g. $n_D=2$ for general 
relativity in $D=3$ spatial dimensions. Also, if $\kappa$ is 
the coupling constant of the theory (the coefficient $1/\kappa$ in front
of the classical action) then $P^e_i$ in (\ref{2.6}) has to be replaced 
by $P^e_i/\kappa$. It follows from the canonical commutation relations
that if $\hat{h}_e$ as before is a multiplication operator in the 
connection representation then $\hat{p}^e_j=it X^e_j/2$ where
\be \label{3.1}
t:=\frac{\alpha}{a^{n_D}} \mbox{ and } \alpha=\kappa\hbar
\ee
define the classicality parameter and the Feinstruktur constant 
respectively. For instance, in four-dimensional general relativity 
$\alpha=\ell_p^2$ is the Planck area and for $a=1$cm we have 
$\sqrt{t}/a=\ell_p/cm\approx 10^{-32}$. All our estimates are based on the 
fact that $t$ is a tiny positive number.  

Consider first only one edge $e$ of a graph $\gamma$, then we define the 
complexifier 
\be \label{3.2}
\hat{C}_e:=\frac{a^{n_D}}{2\kappa} \delta^{ij} \hat{p}^e_i \hat{p}^e_j 
\ee
and the coherent state, labelled by the phase space point 
$g_e=e^{-i p^e_j \tau_j/2}h_e\in G^\Co$ and the classicality parameter $t$, 
by
\be \label{3.3}
\psi^t_{e,g_e}(h_e)
:=[e^{-\frac{\hat{C}_e}{\hbar}} \delta_{h'}(h_e)]_{h'\to g_e}
=[e^{t\Delta_e/2} \delta_{h'}(h_e)]_{h'\to g_e}
\ee
where $\delta_{h'}$ denotes the $\delta$ distribution on $G$ with 
support at $h'$ and in (\ref{3.3}) one is supposed to analytically 
continue $h'$. One can give the following explicit sum formula for 
$\psi^t_{e,g_e}(h_e)$,
\be \label{3.4}
\psi^t_{e,g_e}(h_e)=\sum_\pi d_\pi e^{-t\lambda_\pi/2} 
\chi_\pi(g_e h_e^{-1})
\ee
where the sum is over the equivalence classes of irreducible 
representations $\pi$ of $G$, $\chi_\pi$ is the character of $\pi$ and  
$-\lambda_\pi$ is the eigenvalue of $-\Delta_e$ with eigenfunctions
$\chi_\pi(g_e h_e^{-1})$. The operator $e^{t\Delta_e/2}$ is sometimes 
called the heat kernel operator. Notice that the states $\psi^t_{e,g_e}$
are not normalized.

The generalization to the whole graph $\gamma$ is straightforward,
it is simply given by the product over edges
\be \label{3.5}
\psi^t_{\gamma,g_\gamma}(h_\gamma)=\prod_{e\in E(\gamma)}
\psi^t_{e,g_e}(h_e)
\ee
where $g_\gamma=\{g_e\}_{e\in E(\gamma)},\; 
h_\gamma=\{h_e\}_{e\in E(\gamma)}$. The product states (\ref{3.5})
are obtained by applying the operator $\exp(t\Delta_\gamma/2),\;
\Delta_\gamma=\sum_{e\in E(\gamma)} \Delta_e$ to the product
delta -- distribution 
\be \label{3.6}
\delta_{\gamma,h'_\gamma}(h_\gamma)=\prod_{e\in E(\gamma)}
\delta_{h'_e}(h_e)
\ee
followed by analytical continuation. This formula is meaningful only
if $\gamma\in \Gamma^\omega_0$ is a finite graph, for truly 
infinite graphs in $\Gamma^\omega_\sigma$ we must
work immediately with products of normalized coherent states as otherwise
such states would not be normalizable. For the purpose of this paper
it will be sufficient to stick with $\Gamma^\omega_0$ for the following
reason. Since the Poisson brackets (\ref{2.6}) are promoted to the 
following commutation relations
\ba \label{3.7}
{[}\hat{h}_e,\hat{h}_{e'}] &=& 0\nonumber\\
{[}\hat{p}^e_j,\hat{h}_{e'}] &=&
it \delta^e_{e'} \frac{\tau_j}{2} \hat{h}_e\nonumber\\
{[}\hat{p}^e_i,\hat{p}^{e'}_j] &=&
-it \delta^{ee'}f_{ij}\;^k \hat{p}^e_k
\ea
on the Hilbert space ${\cal H}_\gamma$, the completion of $\Phi_\gamma$,
it follows that due to the commutativity of operators labelled by 
different edges every polynomial of the elementary operators
$\{\hat{h}_e,\hat{p}^e_j\}_{e\in E(\gamma)}$ can in fact be reduced to 
sums of monomials of the form
\be \label{3.8}
\hat{O}_\gamma=\prod_{e\in E(\gamma)} \hat{O}_e
\ee
where for each $e$ the operator $\hat{O}_e=\hat{O}_e(\hat{h}_e,\hat{p}^e_j)$ is a 
certain polynomial of the 
$2\dim(G)$ independent operators $(\hat{h}_e)_{AB},\hat{p}^e_j$ 
for the same $e$ ($A,B,C,...$ are group indices).
Obviously the order of the operators $\hat{O}_e$ is irrelevant but not
the order of the elementary operators appearing in $\hat{O}_e$.
It follows that the expectation value of (infinite) sums of monomials
of the type (\ref{3.8}) is given by the same sum over expectation values 
of monomials and the latter have the following simple product structure
with respect to the state (\ref{3.5})
\be \label{3.9}
\frac{<\psi^t_{\gamma,g_\gamma},
\hat{O}_\gamma \psi^t_{\gamma,g_\gamma}>}{||\psi^t_{\gamma,g_\gamma}||^2}
=\prod_{e\in E(\gamma)} 
\frac{<\psi^t_{e,g_e},\hat{O}_e \psi^t_{e,g_e}>}{||\psi^t_{e,g_e}||^2}
\ee
Also, as far as commutators are concerned, notice the commutator formula
for monomial operators $\hat{O}_\gamma,\hat{O}'_\gamma$ of the type
(\ref{3.8}) given by
\be \label{3.10}
[\hat{O}_\gamma,\hat{O}'_\gamma]=\sum_{e\in E(\gamma)}
[\hat{O}_e,\hat{O}'_e] \prod_{e'\in E(\gamma)-\{e\}} 
(\hat{O}_{e'}\hat{O}'_{e'})
\ee
which can be proved by complete induction over $|E(\gamma)|$. 
Thus, commutators of monomials again reduce to sums over monomials.
We summarize these simple observations by the following theorem.
\begin{Theorem} \label{th3.1}
Let $\gamma\in \Gamma^\omega_0$ be a graph, $g_\gamma\in M_\gamma$
a point in the phase space and $\hat{O}_\gamma,\hat{O}'_\gamma$ 
monomial operators. Suppose that for each $e\in E(\gamma)$ we have 
\ba \label{3.11}
&& \lim_{t\to 0} 
\frac{<\psi^t_{e,g_e},\hat{O}_e \psi^t_{e,g_e}>}{||\psi^t_{e,g_e}||^2}
=O_e(h_e(g_e),p^e_j(g_e))\nonumber\\
&& \lim_{t\to 0} 
\frac{<\psi^t_{e,g_e},\frac{[\hat{O}_e,\hat{O}'_e]}{it}
\psi^t_{e,g_e}>}{||\psi^t_{e,g_e}||^2}
=\{O_e,O'_e\}_e((h_e(g_e),p^e_j(g_e))
\ea
where the polar decomposition $g_e=H_e(g_e) h_e(g_e),\;
H_e(g_e)=\exp(-i\tau_j p^e_j(g_e)/2)$ specifies $h_e(g_e),p^e_j(g_e)$
uniquely. Then
\ba \label{3.12}
&& \lim_{t\to 0} 
\frac{<\psi^t_{\gamma,g_\gamma},\hat{O}_\gamma 
\psi^t_{\gamma,g_\gamma}>}{||\psi^t_{\gamma,g_\gamma}||^2}
=O_\gamma(h_\gamma(g_\gamma),p^\gamma_j(g_\gamma))\nonumber\\
&& \lim_{t\to 0} 
\frac{<\psi^t_{\gamma,g_\gamma},\frac{[\hat{O}_\gamma,\hat{O}'_\gamma]}{it}
\psi^t_{\gamma,g_\gamma}>}{||\psi^t_{\gamma,g_\gamma}||^2}
=\{O_\gamma,O'_\gamma\}_\gamma(h_\gamma(g_\gamma),p^\gamma_j(g_\gamma))
\ea
where $p^\gamma_j=\{p^e_j\}_{e\in E(\gamma)}$.
\end{Theorem}
This theorem shows that in order to establish Ehrenfest theorems it will be 
completely sufficient to consider the problem {\it for one copy of the 
group only}. This is even true in the extension from $\Gamma^\omega_0$ 
to $\Gamma^\omega_\sigma$ because the operators that appear in applications
can be written as infinite sums of monomials {\it each of which depends
on a finite subgraph of an infinite graph only}. However, if 
$\Gamma^\omega_0\ni\gamma\subset\gamma'\in\Gamma^\omega_\sigma$ then we can 
write a given monomial operator
$\hat{O}_\gamma$ also as $\hat{O}_{\gamma'}=
\hat{O}_\gamma\cdot \prod_{e'\in E(\gamma')-\gamma} 1_{e'}$ where 
$1_{e'}$ denotes the unit operator on ${\cal H}_{e'}$. Thus, we get for
the expectation values
\be \label{3.13}
\frac{<\psi^t_{\gamma',g_{\gamma'}},\hat{O}_{\gamma'} 
\psi^t_{\gamma',g_{\gamma'}}>}{||\psi^t_{\gamma',g_{\gamma'}}||^2}
=\frac{<\psi^t_{\gamma,g_\gamma},\hat{O}_\gamma 
\psi^t_{\gamma,g_\gamma}>}{||\psi^t_{\gamma,g_\gamma}||^2}
\ee
so the problem reduces again to one for $\gamma\in \Gamma^\omega_0$.

Equation (\ref{3.13}) seems to indicate that an extension of coherent
states to infinite graphs is not really necessary. However, if $\Sigma$
is non-compact then the only way to approximate, say a classical metric in 
general relativity which is everywhere non-degenerate, by coherent states 
over graphs $\gamma\in \Gamma^\omega_0$ is by using a countably infinite 
superposition of them, say 
$\tilde{\psi}^t_{\gamma,g_\gamma}=
\sum_n z_n\psi^t_{\gamma_n,g_{\gamma_n}},\; z_n\in \Co,n\in \Nl$
where $\gamma=\cup_n \gamma_n$ and the $\psi^t_{\gamma_n,g_{\gamma_n}}$
are the coherent states (\ref{3.5}).
Suppose now that we consider the following
operator $\hat{O}_\gamma=\sum_n \hat{O}_{\gamma_n}$ over $\gamma$.
In applications it is usually true that the $\gamma_n$ are mutually
disjoint but they fill $\Sigma$ everywhere with respect to the metric 
to be approximated, that is, $\gamma$ fills $\Sigma$. Now the 
$\psi^t_{\gamma_n,g_{\gamma_n}}$ are not mutually orthogonal, rather
for $n\not=m$ we have 
$<\psi^t_{\gamma_m,g_{\gamma_m}},\psi^t_{\gamma_n,g_{\gamma_n}}>=1$
while
$||\psi^t_{\gamma_m,g_{\gamma_m}}||>1$ for all $g_\gamma,t$.
Now in applications it turns out that 
$$
<\psi^t_{\gamma_m,g_{\gamma_m}},\hat{O}_{\gamma_n}
\psi^t_{\gamma_p,g_{\gamma_p}}>
=\delta_{m,n}\delta_{n,p}
<\psi^t_{\gamma_n,g_{\gamma_n}},\hat{O}_{\gamma_n}
\psi^t_{\gamma_n,g_{\gamma_n}}>
$$
and thus indeed
$$
<\tilde{\psi}^t_{\gamma,g_\gamma},\hat{O}_\gamma
\tilde{\psi}^t_{\gamma,g_\gamma}>
=\sum_n |z_n|^2
<\psi^t_{\gamma_n,g_{\gamma_n}},\hat{O}_{\gamma_n}
\psi^t_{\gamma_n,g_{\gamma_n}}>
$$
yields the correct expectation value provided that $|z_n|=
1/||\psi^t_{\gamma_n,g_{\gamma_n}}||$. However, then the norm
squared of $\tilde{\psi}^t_{\gamma,g_\gamma}$
is given by
$$
||\tilde{\psi}^t_{\gamma,g_\gamma}||^2
=[\sum_n 1]+2\sum_{m<n} \Re(\bar{z}_m z_n)
$$
which is divergent. Thus,
the only way to deal with semiclassical physics in the case that 
$\Sigma$ is non-compact is to use the extension to 
infinite graphs $\Gamma^\omega_\sigma$.\\
\\
The remainder of this section then is subdivided into two parts. In the 
first one we prove the Ehrenfest Theorem for the polynomial algebra
of operators for one copy of the group only for the case of $G=SU(2)$.
In the second we use these results to extend the theorem to a 
certain class of operators which are not polynomial functions of the 
elementary operators and mix operators labelled by different edges 
by making an appeal to the moment problem by Hamburger for measures.

\subsection{Polynomial Algebra of Operators over One Edge}
\label{s3.1}

As the problem is isomorphic for all the edges of a graph, we drop the 
label $e$ in this subsection and deal only with the operators
$\hat{h}_{AB},\hat{p}_j;\;A,B,C,..\in\{-1/2,1/2\};\;j,k,l..=1,2,3$
which obey the CCR algebra (\ref{3.7}) with the label $e=e'$ dropped.

This subsection is divided into four parts. In the first we reduce the 
computation of expectation values of operator monomials to the
computation of matrix elements of elementary operators between coherent 
states. In the second we estimate the matrix elements for the momentum
operator and in the third for the holonomy operator. As expected the 
matrix elements are concentrated at $g=g'$ and simply given by 
the expectation value of the operator in question times the 
matrix element of the unit operator so that to leading order 
in $t$ the expectation value of the monomial is indeed the monomial
of the expectation values.

The expectation values of operator monomials are 
computed in parts two and three by using the overcompleteness
of the coherent states. This displays them as particularly useful in  
deriving a coherent states path integral formulation of the theory
\cite{45}. In contrast, in the fourth part we use a different method
based on an $SL(2,\Co)$ operator identity and the so-called moment problem
due to Hamburger which we deal with in detail in the second subsection.

\subsubsection{The Completeness Relation}
\label{s3.1.1}

Recall from \cite{36a} for the case of a general compact Lie group or from
\cite{32} for the special case of $G=SU(2)$ that the coherent states
$\psi^t_g$ possess the following ``reproducing property''
\be \label{3.14}
(\hat{U}_t\psi)(g)=<\psi^t_{g^\star},\psi>_{\mu_H}
\ee
where $g^\star =(\bar{g}^T)^{-1}$ is the unique involution on $G^\Co$
that preserves $G$. Here $\psi\in L_2(G,d\mu_H)=:{\cal H}$ is an arbitrary
state and $\hat{U}_t:\;{\cal H}\mapsto {\cal H}^\Co=L_2(G^\Co,d\nu_t)\cap
\mbox{Hol}(G^\Co)$ the coherent state transform of \cite{36a}, that is,
the generalization of (\ref{3.3}) to arbitrary states (heat kernel 
evolution followed by analytic continuation) mapping the $\mu_H$ square
integrable functions on $G$ to holomorphic, $\nu_t$ square integrable 
functions on $G^\Co$. The measure $\nu_t$ of this Bargmann-Segal Hilbert 
space, defined generally in \cite{36a} is computed explicitly in 
\cite{32} for the case of $G=SU(2)$ and is chosen such that $\hat{U}_t$ is 
a unitary operator. Using this unitarity and the reproducing property 
we compute
\ba \label{3.15}
&& <\psi,\psi'>_{\mu_H}=
<\hat{U}_t\psi,\hat{U}_t\psi'>_{\nu_t}
=\int_{G^\Co} d\nu_t(g) 
\overline{(\hat{U}_t\psi)(g)}(\hat{U}_t\psi')(g)\nonumber\\
& =&\int_{G^\Co} d\nu_t(g) 
<\psi,\psi^t_{g^\star}>_{\mu_H}
<\psi^t_{g^\star},\psi'>_{\mu_H}
\ea
and using the involution invariance of the measure $\nu_t$ which is 
essentially a Gaussian in $p_j$ we find the {\it completeness relation}
\be \label{3.16}
\int_{G^\Co} d\nu_t(g) 
|\psi^t_{g}><\psi^t_{g}|=1_{{\cal H}}
\ee
Suppose now that we are given an operator monomial
$\hat{O}=\hat{O}_1 ..\hat{O}_n$ where each of the 
$\hat{O}_k,\;k=1,..,n<\infty$ stands for one of the elementary operators
$\hat{h}_{AB},\hat{p}_j$. Then, using (\ref{3.16}), we can write the 
expectation value of $\hat{O}$ as 
\ba \label{3.17}
&& \frac{<\psi^t_g,\hat{O}\psi^t_g>}{||\psi^t_g||^2} \nonumber\\
&=& 
\frac{1}{||\psi^t_g||^2} \int_{G^\Co} d\nu_t(g_1)..\int_{G^\Co}
d\nu_t(g_{n-1}) \prod_{k=1}^n [<\psi^t_{g_{k-1}},\hat{O}_k\psi^t_{g_k}>]
\nonumber\\
&=& 
\int_{G^\Co} d\Omega(g_1)..\int_{G^\Co} d\Omega(g_{n-1}) 
(\prod_{k=1}^{n-1} [\nu_t(g_k) ||\psi^t_{g_k}||^2])
(\prod_{k=1}^n 
[\frac{<\psi^t_{g_{k-1}},\hat{O}_k
\psi^t_{g_k}>}{||\psi^t_{g_{k-1}}||\;||\psi^t_{g_k}||}])
\ea
where $\Omega$ is the Liouville measure on $G^\Co\cong T^\ast G$ and
we have set $g_0=g_n=g$. Now we recall from \cite{32} that the quantity
\be \label{3.18}
j^t(g,g')=\frac{<\psi^t_g,\psi^t_{g'}>}{||\psi^t_g||\;||\psi^t_{g'}||}
\ee
is exponentially small (at least for $G=SU(2)$) in the sense of a 
Gaussian needle of width $\sqrt{t}$ unless 
$g=g'$ (where it equals unity of course) and that the quantity
$\nu_t(g)||\psi^t_g||^2$ approaches exponentially fast the constant 
$2/(\pi t^3)$. Thus, it is conceivable that
\be \label{3.19}
\frac{<\psi^t_{g_{k-1}},\hat{O}_k
\psi^t_{g_k}>}{||\psi^t_{g_{k-1}}||\;||\psi^t_{g_k}||}
\approx
\frac{<\psi^t_{g_k},\hat{O}_k\psi^t_{g_k}>}{||\psi^t_{g_k}||^2}
j^t(g_{k-1},g_k)
\ee
If that would be the case then the product Liouville measure 
$d\Omega(g_1)..d\Omega(g_{n-1})$ would be essentially supported 
at $g_1=..=g_{n-1}=g$ and we could pull the expectation values 
in (\ref{3.19}) out of the integral (\ref{3.17}) with $g_k$ replaced by $g$
and the remaining integral would then equal unity, of course. Thus we
would have indeed shown that 
\be \label{3.20}
\frac{<\psi^t_g,\hat{O}\psi^t_g>}{||\psi^t_g||^2}
\approx 
\prod_{k=1}^n \frac{<\psi^t_g,\hat{O}_k\psi^t_g>}{||\psi^t_g||^2}
\ee
Thus, in order to prove the desired result (\ref{3.20}) it is sufficient
to prove (\ref{3.19}) together with the precise meaning of $\approx$.
The proof of (\ref{3.19}) will also be the key ingredient for the 
derivation of path integrals based on the coherent states $\psi^t_g$ 
\cite{45}.

\subsubsection{Matrix Elements for the Momentum Operator}
\label{s3.1.2}

Recall from the beginning of this section that $\hat{p}_j=it X_j/2$ 
where $X_j(h)=\mbox{tr}((\tau_j h)^T\partial/\partial h)$ is a basis
of right invariant vector fields on $G$ which generate left 
translations. Thus for any vector $\psi\in {\cal H}$ in the domain
of $\hat{p}_j$ (say smooth square integrable functions) we 
have 
\be \label{3.21}
(\hat{p}_j\psi)(h)=\frac{it}{2} (\frac{d}{ds})_{s=0}\psi(e^{s\tau_j}h)
\ee
Since for our coherent states it holds that 
$\psi^t_g(uh)=\psi^t_{u^{-1}g}(h)$ we have 
\be \label{3.22}
\hat{p}_j\psi^t_g=\frac{it}{2} (\frac{d}{ds})_{s=0}
\psi^t_{e^{-s\tau_j}g}
\ee
It follows by explicit computation of the scalar product (see
\cite{32} for details) that
\be \label{3.23}
<\psi^t_g,\hat{p}_j\psi^t_{g'}>
=\frac{it}{2}(\frac{d}{ds})_{s=0}\psi^{2t}_{e^{-s\tau_j}g'\bar{g}^T}(1_2)
\ee
Upon defining the complex number $z$ uniqely via (see \cite{32} for 
details) 
\be \label{3.24}
\ch(z)=\frac{1}{2}\mbox{tr}(e^{-s\tau_j}g'\bar{g}^T)
=\frac{1}{2}[\mbox{tr}(g'\bar{g}^T)
-s\mbox{tr}(\tau_j g'\bar{g}^T)]+O(s^2)
\ee
and using the Weyl character formula for $G=SU(2)$ we obtain
with $d_j=2j+1,\;j=0,1/2,1,3/2,..$ the spin quantum numbers
\ba \label{3.25}
<\psi^t_g,\hat{p}_j\psi^t_{g'}>
&=&\frac{it}{2}(\frac{d}{ds})_{s=0}
\sum_j d_j e^{-tj(j+1)}\frac{\sh(d_j z)}{\sh(z)}
\nonumber\\
&=&\frac{it}{2}(\frac{d}{ds})_{s=0}
\frac{e^{t/4}}{\sh(z)}
\sum_j d_j e^{-td_j^2/4}\sh(d_j z)
\nonumber\\
&=&\frac{it}{2}(\frac{d}{ds})_{s=0}
\frac{e^{t/4}}{2\sh(z)}
\sum_n n e^{-tn^2/4} e^{nz}
\nonumber\\
&=&\frac{it}{2}(\frac{d}{ds})_{s=0}
\frac{e^{t/4}}{2\sh(z) T}
\sum_n (nT) e^{-(nT)^2} e^{(nT)\xi}
\ea
where $n\in \Zl$ and we have defined $T=\sqrt{t}/2,\;\xi=z/T$. The 
Fourier transform $\tilde{f}(k)=\int_\Rl \frac{dx}{2\pi} e^{-ikx} f(x)$
of the function $f(x):=x e^{-x^2} e^{x\xi}$ is given by 
\be \label{3.26}
\tilde{f}(k)=\frac{\xi-ik}{4\sqrt{\pi}}e^{\frac{(\xi-ik)^2}{4}}
\ee
using a contour argument. An appeal to the Poisson summation formula
(see, e.g., \cite{46,32})  
\be \label{3.27}
\sum_n f(nT)= \frac{2\pi}{T} \sum_n \tilde{f}(\frac{2\pi n}{T})
\ee
then reveals
\be \label{3.28}
<\psi^t_g,\hat{p}_j\psi^t_{g'}>=\frac{it}{2} (\frac{d}{ds})_{s=0}
\frac{\sqrt{\pi}e^{t/4}}{4\sh(z) T^3}
\sum_n (z-2\pi i n) e^{\frac{(z-2\pi i n)^2}{t}}
\ee
By the very same methods we easily obtain
\ba \label{3.29}
||\psi^t_g||^2 &=&
\frac{\sqrt{\pi}e^{t/4}}{4\sh(p) T^3}
\sum_n (p-2\pi i n) e^{\frac{(p-2\pi i n)^2}{t}}
\nonumber\\
||\psi^t_{g'}||^2 &=&
\frac{\sqrt{\pi}e^{t/4}}{4\sh(p') T^3}
\sum_n (p'-2\pi i n) e^{\frac{(p'-2\pi i n)^2}{t}}
\ea
where $\ch(p)=\mbox{tr}(g \bar{g}^T)/2$ with the polar 
decomposition $g=e^{-i p_j\tau_j/2} h,\;p=\sqrt{p_j p_j}$ and similar
for $g'$.

Let now $\ch(z_0):=\frac{1}{2}\mbox{tr}(g'\bar{g}^T)$ and 
$z=z_0+\delta$ where $\delta$ is of first order in $s$. Then comparing
\be \label{3.30}
\ch(z)=\ch(z_0)\ch(\delta)+\sh(z_0)\sh(\delta)=\ch(z_0)+\delta\sh(z_0)
+O(\delta^2)
\ee
with (\ref{3.24}) we conclude that
\be \label{3.31}
\delta=-s\frac{\mbox{tr}(\tau_j g'\bar{g}^T)}{2\sh(z_0)}
\ee
where the sign of $\sh(z_0)$ follows from the formulas given in
\cite{32}. It follows that
\be \label{3.31a}
(\frac{d}{ds})_{s=0}=-\frac{\mbox{tr}(\tau_j g'\bar{g}^T)}{2\sh(z_0)}
(\frac{d}{dz})_{z=z_0}
\ee
and performing the derivative in (\ref{3.28}) we end up with
\ba \label{3.32}
&&<\psi^t_g,\hat{p}_j\psi^t_{g'}>=-\frac{it}{4}
\frac{\mbox{tr}(\tau_j g'\bar{g}^T)}{\sh(z_0)}
\frac{\sqrt{\pi}e^{t/4}}{4 T^3}\times\nonumber\\
&\times& 
\sum_n e^{\frac{(z_0-2\pi i n)^2}{t}}
[2\frac{(z_0-2\pi i n)^2}{t\sh(z_0)}
-(z_0-2\pi i n)\frac{\ch(z_0)}{\sh^2(z_0)}+\frac{1}{\sh(z_0)}]
\ea
Combination with (\ref{3.29}) results in the exact expression
for the momentum matrix element
\ba \label{3.33}
<\hat{p}_j>^t_{gg'} &:=& 
\frac{<\psi^t_g,\hat{p}_j\psi^t_{g'}>}{||\psi^t_g||\;||\psi^t_{g'}||}
\\
&=&
[\frac{-i z_0}{2}\frac{\mbox{tr}(\tau_j g'\bar{g}^T)}{\sh(z_0)}]
\times\nonumber\\
&\times&
\frac{\{\frac{t}{2 z_0}
\sum_n e^{\frac{(z_0-2\pi i n)^2-p^2/2-(p')^2/2}{t}}
[2\frac{(z_0-2\pi i n)^2}{t\sh(z_0)}
-(z_0-2\pi i n)\frac{\ch(z_0)}{\sh^2(z_0)}+\frac{1}{\sh(z_0)}]\}
}{
[\sum_n \frac{p-2\pi i n}{\sh(p)} e^{-\frac{4\pi^2 n^2}{t}}
e^{-i\frac{4\pi n p}{t}}]^{1/2}
[\sum_n \frac{p'-2\pi i n}{\sh(p')} e^{-\frac{4\pi^2 n^2}{t}}
e^{-i\frac{4\pi n p'}{t}}]^{1/2}
}
\nonumber
\ea
The arguments $D^t(p),D^t(p')$ of the square roots in the 
denominator of (\ref{3.33}) were 
already estimated in \cite{32} and we will only recall the result
here without derivation
\be \label{3.34}
\frac{p}{\sh(p)}(1-K_t)\le D^t(p)\le \frac{p}{\sh(p)}(1+K_t)
\mbox{ where }K_t=2\sum_{n=1}^\infty e^{-\frac{4\pi^2 n^2}{t}}
(1+\frac{8\pi^2 n}{t})
\ee
vanishes exponentially fast with $t\to 0$ and similar for $D^t(p')$
with $p$ replaced by $p'$. The term in the curly brackets of 
the numerator in (\ref{3.33}) can be more explicitly written as 
\ba \label{3.35}
&&N^t(g,g') = e^{\frac{z_0^2-p^2/2-(p')^2}{t}} 
\{[\frac{z_0}{\sh(z_0)}
-\frac{t}{2}\frac{\ch(z_0)}{\sh^2(z_0)}+\frac{1}{z_0\sh(z_0)}]
\nonumber\\
&& +\sum_{n=1}^\infty e^{-\frac{4\pi^2 n^2}{t}} 
[
(\frac{2 z_0^2-8\pi^2 n^2}{z_0\sh(z_0)}
-\frac{t \ch(z_0)}{\sh^2(z_0)}+\frac{t}{z_0\sh(z_0)})
\cos(\frac{4\pi n z_0}{t})
\nonumber\\
&& +2\pi n t(\frac{\ch(z_0)}{z_0\sh^2(z_0)}-\frac{4}{t\sh(z_0)})
\sin(\frac{4\pi n z_0}{t})
]\}
\ea
which is superficially divergent at the points $z_0=0,i\pi$
which reminds us, of course, of the singularity structure of the 
overlap function in \cite{32} and we will proceed similarly to 
estimate (\ref{3.35}).
Thus, we will separate the discussion into cases A) and B) respectively,
writing $N^t(g,g')$ in terms of $z_0$ and $z'_0=z_0-i\pi$ respectively
for $0\le\phi\le(1-c)\pi$ and $(1-c)\pi\le\phi\le\pi$ respectively
where $0<c<1$ is a constant. As shown in \cite{32}, $z_0=s+i\phi$
is always uniquely determined with $s\in \Rl,\;\phi\in [0,\pi]$. \\
\\
Case A)\\
Let us pull out a factor of $z_0/\sh(z_0)$ from (\ref{3.35}) since at
$g=g'$ it will cancel against the $p/\sh(p)$ coming from $D^t(p)$
and separate terms into those which are regular and irregular
respectively at $z_0=0$, resulting in
\ba \label{3.36}
&& N^t(g,g')= e^{\frac{z_0^2-p^2/2-(p')^2}{t}} \frac{z_0}{\sh(z_0)}
\{
[1]+[\frac{t}{2}\frac{\frac{\sh(z_0)}{z_0}-\ch(z_0)}{z_0\sh(z_0)}]
\nonumber\\
&& +\sum_{n=1}^\infty e^{-\frac{4\pi^2 n^2}{t}} 
(
[2 \cos(\frac{4\pi n z_0}{t})] 
+[t\frac{\frac{\sh(z_0)}{z_0}-\ch(z_0)}{z_0\sh(z_0)}
\cos(\frac{4\pi n z_0}{t})] 
-[8\pi n\sin(\frac{4\pi n z_0}{t})/z_0]
\nonumber\\
&& +2\pi n t[\frac{\ch(z_0)}{z_0^2\sh(z_0)}\sin(\frac{4\pi n z_0}{t})
-\frac{4\pi n}{t z_0^2}\cos(\frac{4\pi n z_0}{t})]
)
\}
\ea
and obviously all terms in the square brackets, except for the last
one, are regular at $z_0=0$. However, expanding numerator and 
denominator to second order in $z_0$ we see
\ba \label{3.37}
&&\frac{\ch(z_0)}{z_0^2\sh(z_0)}\sin(\frac{4\pi n z_0}{t})
-\frac{4\pi n}{t z_0^2}\cos(\frac{4\pi n z_0}{t})
\nonumber\\
&=&\frac{1}{z_0\sh(z_0)}
[\ch(z_0)\frac{\sin(\frac{4\pi n z_0}{t})}{z_0}
-\frac{4\pi n}{t}\frac{\sh(z_0)}{z_0}\cos(\frac{4\pi n z_0}{t})]
\nonumber\\
&=&\frac{1}{z_0^2(1+O(z_0^2))}
[\frac{4\pi n}{t}(1+O(z_0^2))
-\frac{4\pi n}{t}(1+O(z_0^2))]=O(z_0^2)
\ea
so that (\ref{3.37}) even vanishes at $z_0=0$.

We want to put bounds on all those terms in the square brackets of
(\ref{3.36}) for $0\le \phi\le (1-c)\pi$ except for the first one and, 
in particular, estimate the series. To that end we write (\ref{3.36})
in a yet more suggestive form
\ba \label{3.38}
N^t(g,g') &=& e^{\frac{z_0^2-p^2/2-(p')^2/2}{t}} \frac{z_0}{\sh(z_0)}
\times\nonumber\\
&\times&
\{
[1]+[\frac{t}{2}\frac{\frac{\sh(z_0)}{z_0}-\ch(z_0)}{z_0\sh(z_0)}]
[1+2\sum_{n=1}^\infty e^{-\frac{4\pi^2 n^2}{t}}
\cos(\frac{4\pi n z_0}{t})]
\nonumber\\
&& +2 [\sum_{n=1}^\infty e^{-\frac{4\pi^2 n^2}{t}} 
(\cos(\frac{4\pi n z_0}{t})-\frac{(4\pi n)^2}{t}
\frac{\sin(\frac{4\pi n z_0}{t})}{\frac{4\pi n z_0}{t}})] \nonumber\\
&& +2\pi t[\sum_{n=1}^\infty e^{-\frac{4\pi^2 n^2}{t}} 
n(\frac{\ch(z_0)}{z_0\sh(z_0)}\frac{\sin(\frac{4\pi n z_0}{t})}{z_0}
-\frac{4\pi n}{t z_0^2}\cos(\frac{4\pi n z_0}{t}))]
\}
\nonumber\\
&=:& e^{\frac{z_0^2-p^2/2-(p')^2/2}{t}} \frac{z_0}{\sh(z_0)}
\times\nonumber\\
&\times&
\{
1+\frac{t}{2}\frac{\frac{\sh(z_0)}{z_0}-\ch(z_0)}{z_0\sh(z_0)} I_1
+2 I_2+2\pi t I_3
\}
\ea
and it remains to estimate the sums $I_1,I_2,I_3$ as well as 
the expression 
\be \label{3.39}
I:=\frac{\frac{\sh(z_0)}{z_0}-\ch(z_0)}{z_0\sh(z_0)}
\ee
Notice that for the terms proportional to $t$ in (\ref{3.38}) it will be 
sufficient to estimate them by a function integrable against the 
Gaussian prefactor of (\ref{3.38}).

Focussing first on (\ref{3.39}) we first prove two elementary lemmas.
\begin{Lemma} \label{la3.1}
For any $z=s+i\phi\in \Co$ such that $0\le \phi\le (1-c)\pi$ for some
$0<c<1$ we have 
\ba \label{3.40}
&& \frac{s^2}{\sh^2(s)}\le |\frac{z}{\sh(z)}|^2
\le \frac{\phi^2}{\sin^2(\phi)}\le [\frac{\pi(1-c)}{\sin(\pi(1-c))}]^2=:k_c^2
\\
\label{3.40a}
&& |\frac{z}{\sh(z)}|^2
\le \frac{\phi^2}{\sin^2(\phi)} \frac{s^2}{\sh^2(s)}
\le k_c^2 \frac{s^2}{\sh^2(s)}
\ea
\end{Lemma}
Proof of Lemma \ref{la3.1} :\\
Notice the identity 
\be \label{3.41}
|\frac{z}{\sh(z)}|^2=
\frac{s^2+\phi^2}{\sh^2(s)\cos^2(\phi)+\ch^2(s)\sin^2(\phi)}
=\frac{s^2+\phi^2}{\sh^2(s)+\sin^2(\phi)}
\ee
Both the lower and upper bounds in (\ref{3.40}) turn out to be 
equivalent with the inequality
\be \label{3.41a}
(\frac{\sh(s)}{s})^2\ge(\frac{\sin(\phi)}{\phi})^2
\ee
which is true as the left hand side and right hand side respectively
both take its minimum and maximum respectively at $s=\phi=0$, in fact,
the left hand side and right hand side respectively are strictly 
increasing and decreasing functions respectively.\\
\\
Inequality (\ref{3.40a}) in turn can be transformed into the equivalent
form
\be \label{3.42}
\frac{1}{s^2}-\frac{1}{\sh^2(s)}\le\frac{1}{\sin^2(\phi)}-\frac{1}{\phi^2}
\ee
Both the left and the right hand side of this inequality are always positive
in the range considered and in fact the left hand side approches $0$ as
$s\to\infty$ while the right hand side approaches $+\infty$ as
$\phi\to\pi$. At $\phi=s=0$
both sides equal $1/3$. We will in fact prove that
\be \label{3.41b}
\frac{1}{s^2}-\frac{1}{\sh^2(s)}\le\frac{1}{3}\le 
\frac{1}{\sin^2(\phi)}-\frac{1}{\phi^2}
\ee
Consider first the left hand side. This can be written in the equivalent 
form
\be \label{3.41c}
(3-s^2)\sh^2(s)\le 3s^2
\ee
which is obviously true for $s^2\ge 3$ so that we may restrict examination to
$s^2<3$. In that case we may write (\ref{3.41c}) in the equivalent form
\be \label{3.41d}
\frac{\ch(2s)-1}{2s^2}\le \frac{1}{1-s^2/3}
\ee
As $s^2/3<1$ the right hand side can be expanded into a geometric series.
Introducing $x=2s$ and employing the Taylor series for $\ch$ we can write
(\ref{3.41d}) as 
\be \label{3.41e}
\sum_{n=0}^\infty x^{2n}[\frac{1}{12^n}-\frac{2}{(2(n+1))!}]\ge 0
\ee
and it will be sufficient to establish non-negativity of every coefficient.
Using the basic estimate $\ln(n!)\ge n(\ln(n)-1)+1$ valid for $n\ge 1$
it is easy to see that this is indeed the case for $n>4$ while for the
cases $n=0,1,2,3,4$ this can be checked by direct computation.

Turning to the right hand side of (\ref{3.41b}) we can write it in the 
equivalent form
\be \label{3.41f}
3\phi^2\ge \sin^2(\phi)[3+\phi^2]
\ee
which due to $\sin^2(\phi)\le 1$ is certainly true for $\phi^2\ge 3/2$
so that we can focus attention on the case $\phi^2/3<3/2$. Writing 
(\ref{3.41f})
in the equivalent form
\be \label{3.41g}
\frac{1-\cos(2\phi)}{2\phi^2}\le\frac{1}{1-(-\phi^2/3)}
\ee
exploiting that $0\le \phi^2/3<1/2$ lies in the radius of convergence of the
geometric series, introducing $0\le y=(2\phi)^2/6<1$ and 
\be \label{3.41h}
b_n:=\frac{1}{2^n}-\frac{2 \cdot 6^n}{(2(n+1))!}
\ee
we may write (\ref{3.41g}) in the form 
\be \label{3.41k}
\sum_{n=0}^\infty (-1)^n y^n b_n
=\sum_{n=0}^\infty [y^{2n} b_{2n}-y^{2n+1} b_{2n+1}]\ge 0
\ee
Since $0\le y<1$ this will be the case if $b_{2n}\ge b_{2n+1}$ for
all $n\ge 0$. As one can check, $b_0=b_1=0$ and in (\ref{3.41e}) we have 
already seen that $b_n>0$ for $n\ge 2$. Thus, it is enough to prove
$b_{2n}\ge b_{2n+1}$ for $n\ge 1$. In fact, we will prove more, namely that
$b_n$ is strictly decreasing for $n\ge 2$. This turns out to be equivalent
with $(2(n+1))!-2 \cdot 12^n>1$ for all $n\ge 2$ which in turn would follow from
$(2(n+1))! >(2+\frac{1}{144})12^n$. The latter condition can be demonstrated
to be true by methods similar to those outlined in (\ref{3.41e}).\\
$\Box$\\
\begin{Lemma} \label{la3.1a}
Let
\be \label{3.42b}
S(z):=\frac{\sh(z)-z}{z^2\sh(z)},\; C(z):=\frac{\ch(z)-1}{z\sh(z)} ,\;
k_c':=\sqrt{1+k_c \ch(\pi(1-c))}
\ee
Then for any $z=s+i\phi\in\Co,\;0\le\phi\le(1-c)\pi$ it holds that
\be \label{3.42c}
|S(z)|\le \sqrt{2} k_c' \mbox{ and } |C(z)|\le \sqrt{2} k_c' 
\ee
\end{Lemma}
Proof of Lemma \ref{la3.1a} :\\
Using hyperbolic and trigonometric identities we derive
\ba \label{3.42d}
|S(z)|^2
&=& 
\frac{
(\sh(s)\cos(\phi)-s)^2+(\ch(s)\sin(\phi)-\phi)^2
}{
|z|^4(\ch^2(s)-\cos^2(\phi))
}
\nonumber\\
&=& 
\frac{
(\frac{s^2}{|z|^2}\frac{\sh(s)-s}{s^2}\cos(\phi)+\phi\frac{s\phi}{|z|^2}
\frac{\cos(\phi)-1}{\phi^2})^2
+(\frac{s\phi}{|z|^2}\frac{\ch(s)-1}{s}\frac{\sin(\phi)}{\phi}
+\frac{\phi^2}{|z|^2} \phi\frac{\sin(\phi)-\phi}{\phi^3})^2
}{
\ch^2(s)-\cos^2(\phi)
}
\nonumber\\
&\le& 
\frac{
(|\frac{\sh(s)-s}{s^2}|\;|\cos(\phi)|+\phi |\frac{\cos(\phi)-1}{\phi^2}|)^2
+(|\frac{\ch(s)-1}{s}|\;|\frac{\sin(\phi)}{\phi}|
+\phi |\frac{\sin(\phi)-\phi}{\phi^3}|)^2
}{
\ch^2(s)-\cos^2(\phi)
}
\nonumber\\
&\le& 
\frac{
(|\frac{\sh(s)-s}{s^2}|+\phi |\frac{\cos(\phi)-1}{\phi^2}|)^2
+(|\frac{\ch(s)-1}{s}|+\phi |\frac{\sin(\phi)-\phi}{\phi^3}|)^2
}{
\ch^2(s)-\cos^2(\phi)
}
\nonumber\\
&\le& 
(|\frac{\sh(s)-s}{s^2\sh(s)}|+|\frac{\phi}{\sin(\phi)}|\;
|\frac{\cos(\phi)-1}{\phi^2}|)^2
+(|\frac{\ch(s)-1}{s\sh(s)}|+|\frac{\phi}{\sin(\phi)}|\; 
|\frac{\sin(\phi)-\phi}{\phi^3}|)^2
\nonumber\\
&\le& 
(|S(s)|+k_c |\frac{\cos(\phi)-1}{\phi^2}|)^2
+(|C(s)|+k_c |\frac{\sin(\phi)-\phi}{\phi^3}|)^2
\ea
where in the first inequality we used $|s|,|\phi|\le |z|$, in the second 
that $|\cos(\phi)|,|\sin(\phi)/\phi|\le 1$, in the third that
$\ch^2(s)-\cos^2(\phi)\ge \sh^2(s),\sin^2(\phi)$ and in the fourth
we used Lemma \ref{la3.1} and again the definition of $S(z),C(z)$.

Proceeding similarly with $C(z)$ we arrive at
\ba \label{3.42e}
|C(z)|^2&=& 
\frac{
(\ch(s)\cos(\phi)-1)^2+(\sh(s)\sin(\phi))^2
}{
|z|^2(\ch^2(s)-\cos^2(\phi))
}
\nonumber\\
&=& 
\frac{
(s\frac{\ch(s)-1}{s}\cos(\phi)+\phi^2\frac{\cos(\phi)-1}{\phi^2})^2
+(\phi\sh(s)\frac{\sin(\phi)}{\phi})^2
}{
|z|^2(\ch^2(s)-\cos^2(\phi))
}
\nonumber\\
&\le& 
\frac{
(|\frac{\ch(s)-1}{s}|\;|\cos(\phi)|+\phi |\frac{\cos(\phi)-1}{\phi^2}|)^2
+(\sh(s)|\frac{\sin(\phi)}{\phi}|)^2
}{
(\ch^2(s)-\cos^2(\phi))
}
\nonumber\\
&\le& 
\frac{
(|\frac{\ch(s)-1}{s}|+\phi |\frac{\cos(\phi)-1}{\phi^2}|)^2+\sh^2(s)
}{
(\ch^2(s)-\cos^2(\phi))
}
\nonumber\\
&\le& 
(|\frac{\ch(s)-1}{s\sh(s)}|+|\frac{\phi}{\sin(\phi)}|\; 
|\frac{\cos(\phi)-1}{\phi^2}|)^2+1
\nonumber\\
&\le& 
(|C(s)|+k_c |\frac{\cos(\phi)-1}{\phi^2}|)^2+1
\ea
Using the Taylor series expansion of the trigonometric functions we
easily obtain
\ba \label{3.42f}
&&|\frac{\cos(\phi)-1}{\phi^2}|=
|\sum_{n=1}^\infty \frac{(-1)^n\phi^{2(n-1)}}{(2n)!}|
=|\sum_{n=0}^\infty \frac{(-1)^n\phi^{2n}}{(2n+2)!}|
\le \sum_{n=0}^\infty \frac{\phi^{2n}}{(2(n+1))!}
\nonumber\\
&\le& \ch(\phi)\le 
\ch(\pi(1-c))
\nonumber\\
&&|\frac{\sin(\phi)-\phi}{\phi^3}|=
|\sum_{n=1}^\infty \frac{(-1)^n\phi^{2(n-1)}}{(2n+1)!}|
=|\sum_{n=0}^\infty \frac{(-1)^n\phi^{2n}}{(2n+3)!}|
\le \sum_{n=0}^\infty \frac{\phi^{2n}}{(2n+3)!}\le\ch(\phi)
\nonumber\\
&\le& \ch(\pi(1-c))
\ea
Next notice that $S(s)=S(|s|),\;C(s)=C(|s|)$ so that 
we have reduced our estimate for $|S(z)|,|C(z)|$ to that for 
real non-negative arguments. Finally, using the Taylor series
expression for the hyperbolic functions we find
\ba \label{3.42g}
&&|\frac{\ch(s)-1}{s}|=
|\sum_{n=1}^\infty \frac{s^{2n-1}}{(2n)!}|
=|\sum_{n=0}^\infty \frac{s^{2n+1}}{(2n+2)!}|\le\sh(|s|)
\nonumber\\
&&|\frac{\sh(s)-s}{s^2}|=
|\sum_{n=1}^\infty \frac{\phi^{2n-1}}{(2n+1)!}|
=|\sum_{n=0}^\infty \frac{s^{2n+1}}{(2n+3)!}|\le \sh(|s|)
\ea
so that in fact $|S(s)|,|C(s)|\le 1$. Together with the trivial 
inequality $1\le k_c$ we thus obtain indeed
$|C(z)|^2,|S(z)|^2\le 2[1+k_c \ch(\pi(1-c))]^2=2(k_c')^2$.\\
$\Box$\\
Now we can give a bound on $I$.
\begin{Lemma} \label{la3.2}
For any $z_0=s+i\phi\in \Co$ such that $0\le \phi\le (1-c)\pi$ for some
$0<c<1$ we have            
\be \label{3.43}
|I|\le 2 k_c'
\ee
\end{Lemma}
Proof of Lemma \ref{la3.2} :\\
This follows immediately from the identity 
\be \label{3.44}
I = S(z)-C(z)
\ee
and lemma \ref{la3.1a}.\\
$\Box$\\
Next consider the following expression that appears in $I_3$
\be \label{3.45}
J:=\frac{\ch(z_0)}{z_0\sh(z_0)}\frac{\sin(\frac{4\pi n z_0}{t})}{z_0}
-\frac{4\pi n}{t z_0^2}\cos(\frac{4\pi n z_0}{t})
\ee
which we must estimate in such a way that finally $s$ and $n$ do not appear
in the combination $ns$ inside a hyperbolic function as otherwise we must
worry about convergence of the series $I_3$ as $|s|$ becomes arbitrarily
large in the integrals we are considering. At the same time we must 
bound the superficial singularity at $z_0=0$. To that end we 
introduce $z_0'=4\pi n z_0/t$ and notice the identity, recalling 
the definition (\ref{3.39})
\be \label{3.46}
J = \frac{4\pi n}{t}
\{
-\frac{\sin(z_0')}{z_0'} I+
(\frac{4\pi n}{t})^2
[\frac{\sin(z_0')-z_0'}{(z_0')^3}
-\frac{\cos(z_0')-1}{(z_0')^2}]
\}
\ee
and the task is to estimate the three terms of the form
$(\cos(z)-1)/z^2,\sin(z)/z,(\sin(z)-z)/z^3$ 
for arbitrary $z=x+iy\in\Co$. The inequality 
\be \label{3.47}
|\frac{\sin(z)}{z}| \le 2\ch(y)
\ee
is the content of lemma 4.1 of \cite{32}.
The remaining two estimates are the hardest ones and we have therefore 
devoted the subsequent lemma to them.
\begin{Lemma} \label{la3.3}
Let for any complex number $z$ 
\be \label{3.48b}
c(z):=\frac{\cos(z)-1}{z^2} \mbox{ and }
s(z):=\frac{\sin(z)-z}{z^3}
\ee
Then 
\be \label{3.48a}
|c(z)|\le 4\ch(\Im(z))\mbox{ and } 
|s(z)|\le 4\ch(\Im(z)) 
\ee
\end{Lemma}
Proof of Lemma \ref{la3.3} :\\
i)\\
Splitting $z=x+iy$ we have
\ba \label{3.49}
&& |\frac{\cos(z)-1}{z^2}| = 
|\frac{[\cos(x)\ch(y)-1]-i[\sin(x)\sh(y)]}{z^2}| \nonumber\\
&\le& |\frac{xy}{z^2}|\;|\frac{\sin(x)}{x}|\;|\frac{\sh(y)}{y}|
+|\frac{[\cos(x)-1][\ch(y)-1]+[\cos(x)-1]+[\ch(y)-1]}{z^2}|
\nonumber\\
&\le& \sh(|y|)+|\frac{xy}{z^2}|\;|\frac{\ch(y)-1}{y}|\;|\frac{\cos(x)-1}{x}|
+|\frac{x^2}{z^2}|\;|\frac{\cos(x)-1}{x^2}|
+|\frac{y^2}{z^2}|\;|\frac{\ch(y)-1}{y^2}|
\nonumber\\
&\le& \sh(|y|)+|\frac{\ch(y)-1}{y}|\;|\frac{\cos(x)-1}{x}|
+|\frac{\cos(x)-1}{x^2}|+|\frac{\ch(y)-1}{y^2}|
\ea
using $|x/z|,|y/z|\le 1$. It is easy to see that $|\ch(y)-1|/y^2\le\ch(y)$
by using the Taylor series of $\ch(y)$, see e.g. \cite{32}.
By similar methods it is easy to establish that  
$|\frac{\ch(y)-1}{y}|\le \sh(|y|)$. Furthermore,
we claim that $|\frac{\cos(x)-1}{x}|,|\frac{\cos(x)-1}{x^2}| \le 1$. 

To see the former, notice 
that $|\frac{\cos(x)-1}{x}|=|\frac{\cos(|x|)-1}{|x|}|$ so it 
will be sufficient to demonstrate this for $x\ge 0$. Now for 
$x\ge 0$ 
the inequality $|\frac{\cos(x)-1}{x}|\le 1$ is equivalent with
$1-x\le\cos(x)\le 1+x$, so we claim that 
$f_\pm(x)=x\pm(1-\cos(x))$
are not negative functions for $x\ge 0$. But $f'_\pm(x)=1\pm 
\sin(x)\ge 0$ 
so that $f_\pm$ is never decreasing and takes its minimum at $x=0$ where 
$f_\pm(0)=0$. 

To see the latter, notice 
that $|\frac{\cos(x)-1}{x^2}|=|\frac{\cos(|x|)-1}{|x|^2}|$ so it 
will be sufficient to demonstrate this for $x\ge 0$ as well. Now for 
$x\ge 0$ 
the inequality $|\frac{\cos(x)-1}{x^2}|\le 1$ is equivalent with
$1-x^2\le\cos(x)\le 1+x^2$, so we claim that 
$f_\pm(x)=x^2\pm(1-\cos(x))$
are not negative functions for $x\ge 0$. But $g_\pm=f'_\pm(x)=2x\pm 
\sin(x)$ and $g_\pm'(x)=2\pm\cos(x)>0$. Thus, $g_\pm$ is strictly 
increasing, its minimum at $x=0$ being $g_\pm(0)=0$. Thus 
$f_\pm$ is never decreasing and takes its minimum at $x=0$ where 
$f_\pm(0)=0$. 
It follows that
\be \label{3.50}
|\frac{\cos(z)-1}{z^2}|\le \sh(|y|)+\sh(|y|)+1+\ch(y)
\le 4\ch(y)
\ee
since $1,\sh(|y|)\le\ch(y)$.\\
ii)\\
Proceeding similarly as in i) we have
\ba \label{3.50a}
&& |\frac{\sin(z)-z}{z^3}| =
|\frac{1}{z^3}([\sin(x)\ch(y)-x]+i[\cos(x)\sh(y)-y])|
\nonumber\\
&\le&
\frac{1}{|z^3|}|[xy^2\frac{\sin(x)}{x}\frac{\ch(y)-1}{y^2}]+
[x^3\frac{\sin(x)-x}{x^3}]
+[x^2 y\frac{\cos(x)-1}{x^2}\frac{\sh(y)}{y}]
+[y^3\frac{\sh(y)-y}{y^3}]|
\nonumber\\
&\le&
|\frac{\sin(x)}{x}|\;|\frac{\ch(y)-1}{y^2}|+
|\frac{\sin(x)-x}{x^3}|
+|\frac{\cos(x)-1}{x^2}|\;|\frac{\sh(y)}{y}|
+|\frac{\sh(y)-y}{y^3}|
\nonumber\\
&\le&
\ch(y)+
|\frac{\sin(x)-x}{x^3}|+\ch(y)+\ch(y)
\ea
where we have made use of properties already demonstrated in i)
and the Taylor series of $\sh(y)$. 
We claim that $|\frac{\sin(x)-x}{x^3}|=
|\frac{\sin(|x|)-|x|}{|x|^3}|\le 1$ and it is sufficient to 
prove this for $x\ge 0$. That statement is for $x\ge 0$ equivalent to
$f(x)=x^3+\sin(x)-x\ge 0$ because $\sin(x)\le x$ for $x\ge0$. We have 
$g(x)=f'(x)=3x^2+\cos(x)-1,
h(x)=g'(x)=6x-sin(x),h'(x)=6-\cos(x)>0$. Since $f(0)=g(0)=h(0)=0$ we see
that $h$ is an increasing function whence $h(x)\ge 0$ from which follows 
that $g$ is an increasing function whence $g(x)\ge 0$ from which follows 
that $f$ is an increasing function whence $f(x)\ge 0$ as claimed. 
It follows that  
\be \label{3.50b}             
|\frac{\frac{\sin(z)}{z}-1}{z^2}|\le 1+3\ch(y)\le 4\ch(y)
\ee
as claimed.\\
$\Box$\\
Collecting all the estimates we can now write the estimate for the modulus
of (\ref{3.45}) in the desired form, defining $y'=4\pi n \phi/t$,
\be \label{3.51}
|J| \le 
\frac{4\pi n}{t}
[2\ch(y')|I|+8(\frac{4\pi n}{t})^2 \ch(y')]
\le 
\frac{32\pi n}{t}[k_c'+(\frac{4\pi n}{t})^2]\ch(y')
\ee
which now enables us to estimate the various series $I_1,I_2,I_3$.
We will do this one by one.\\
$I_1$ :\\
The elementary estimate $|\cos(z)| \le |\ch(y)|+|\sh(y)|=e^{|y|}$
applied to $I_1$ reveals
\ba \label{3.53}
|I_1|& \le & 1+2\sum_{n=1}^\infty e^{-\frac{4\pi^2 n^2}{t}}
e^{\frac{4\pi^2 n (1-c)}{t}}
\nonumber\\
& \le & 1+2e^{-\frac{4\pi^2 c}{t}} 
\sum_{n=1}^\infty e^{-\frac{4\pi^2 (n^2-n)}{t}} 
\nonumber\\
& \le & 1+2e^{-\frac{4\pi^2 c}{t}} 
\sum_{n=0}^\infty e^{-\frac{4\pi^2 n^2}{t}} 
=:1+k_t
\ea
where in the last step we have made use of the inequality 
$(n-1)^2\le n^2-n$ valid for $n\ge 1$. The constant $k_t$
is independent of $g,g'$ and vanishes exponentially fast with
$t\to 0$ for any $c>0$.\\
$I_2$ :\\
Using again $|\cos(z)|\le e^{|y|},\;|\sin(z)/z|\le 2\ch(y)\le 2e^{|y|}$
we easily find with $|\phi|\le (1-c)\pi$
\ba \label{3.54}
|I_2| &\le& \sum_{n=1}^\infty e^{-\frac{4\pi^2 n^2}{t}} 
e^{\frac{4\pi^2 n (1-c)}{t}}(1+\frac{32\pi^2 n^2}{t})
\nonumber\\
&\le& e^{-\frac{4\pi^2 c}{t}} 
\sum_{n=1}^\infty e^{-\frac{4\pi^2 (n^2-n)}{t}} 
(1+\frac{32\pi^2 n^2}{t})
\nonumber\\
&\le& e^{-\frac{4\pi^2 c}{t}} 
\sum_{n=0}^\infty e^{-\frac{4\pi^2 n^2}{t}} 
(1+\frac{32\pi^2 (n+1)^2}{t})
=:k'_t
\ea
where again $k'_t$ is a constant, approaching zero exponentially 
fast with $t\to 0$ for any $c>0$.\\
$I_3$ :\\
Invoking our estimate (\ref{3.51}) for the quantity $J$ (\ref{3.45})
that appears in $I_3$ we obtain
\ba \label{3.55}
|I_3| &\le&
\sum_{n=1}^\infty e^{-\frac{4\pi^2 n^2}{t}} e^{\frac{4\pi^2 n (1-c)}{t}}
\frac{32\pi n^2}{t}[k_c'+(\frac{4\pi n}{t})^2]
\nonumber\\
&\le&
32\pi e^{-\frac{4\pi^2 c}{t}}
\sum_{n=1}^\infty e^{-\frac{4\pi^2 (n^2-n)}{t}} 
\frac{n^2}{t}[k_c'+(\frac{4\pi n}{t})^2]
\nonumber\\
&\le&
32\pi e^{-\frac{4\pi^2 c}{t}}
\sum_{n=0}^\infty e^{-\frac{4\pi^2 n^2}{t}} 
\frac{(n+1)^2}{t}[k_c'+(\frac{4\pi (n+1)}{t})^2]
=:\tilde{k}_t
\ea
where again $\tilde{k}_t$
is a constant independent of both $g,g'$ exponentially vanishing with 
$t\to 0$, for any $c>0$.\\
\\
Let us now define 
\ba \label{3.56}
\Delta<\hat{p}_j>^t_{gg'} &:=& <\hat{p}_j>^t_{gg'} 
-\frac{
[\frac{-i}{2}\frac{\mbox{tr}(\tau_j g'\bar{g}^T)}{\sh(z_0)} z_0]
e^{\frac{z_0^2-p^2/2-(p')^2/2}{t}} \frac{z_0}{\sh(z_0)} 
}{
\sqrt{D^t(p) D^t(p')}
}
\\
&=&\frac{
[\frac{-i}{2}\frac{\mbox{tr}(\tau_j g'\bar{g}^T)}{\sh(z_0)} z_0]
e^{\frac{z_0^2-p^2/2-(p')^2/2}{t}} \frac{z_0}{\sh(z_0)} 
\{
\frac{t}{2}\frac{\frac{\sh(z_0)}{z_0}-\ch(z_0)}{z_0\sh(z_0)} I_1
+2 I_2+2\pi t I_3
\}
}{
\sqrt{D^t(p) D^t(p')}
}
\nonumber
\ea
Recall now the relation between the various objects $s,\phi,\tilde{p},
\tilde{\theta},\tilde{\alpha}$ from section 4.2 of \cite{32}. Basically, one 
writes
$g=Hh,g'=H' h'$ in polar decomposed form and defines the polar decomposition
$HH'=\tilde{H}u$ as well as $\tilde{h}=h^{-1}h'u^{-1}$. Then
$\tilde{H}=\exp(-i \tilde{p}_j\tau_j/2),\;
\tilde{h}=\exp(\tilde{\theta}_j\tau_j)$ and 
$\ch(s)\cos(\phi)=\ch(\tilde{p}/2)\cos(\tilde{\theta})$ and
$\sh(s)\sin(\phi)=\sh(\tilde{p}/2)\sin(\tilde{\theta})\cos(\tilde{\alpha})$
with 
$\cos(\tilde{\alpha})=\tilde{p}_j\tilde{\theta}_j/(\tilde{p}\tilde{\theta})$.

Writing $g'\bar{g}^T=e^{-i\tau_j z_0^j}$ we have  
\be \label{3.57}
[\frac{-i}{2}\frac{\mbox{tr}(\tau_j g'\bar{g}^T)}{\sh(z_0)} z_0]
=z_0^j
\ee
and are interested in the relation between $z_0^j$ and $z_0=s+i\phi$.
By definition we have $\ch(z_0)=\mbox{tr}(g'\bar{g}^T)/2$ which reveals
that $z_0^2=\sum_j (z_0^j)^2$, however, it is not true that
$|z_0|^2\ge\sum_j |z_0^j|^2$. In order to estimate the integral
over $\Delta<\hat{p}_j>_{g,g'}$ we thus have to prove one more relation.
\begin{Lemma} \label{la3.4}
For $z_0=s+i\phi,\;0\le\phi\le(1-c)\pi$ and 
$\tilde{g}:=g'\bar{g}^T=e^{-i\tau_j z_0^j}$ with $z_0^2=z_0^j z_0^j$ 
we have 
\be \label{3.58}
|z_0^j|^2\le [k_c \frac{s}{\sh(s)} \sqrt{\ch(\tilde{p})}]^2
\le [2 k_c \frac{s}{\sh(s)} \ch(\frac{p+p'}{2})]^2
\ee
for any $j=1,2,3$.
\end{Lemma}
Proof of Lemma \ref{la3.4} :\\
Using the $SL(2,\Co)$ ``Fierz identity'' $\mbox{tr}(M\tau_j)\tau_j=
\mbox{tr}(M)-2M$ valid for any $2\times 2$ matrix $M$ we find
from (\ref{3.57}) that
\ba \label{3.59}
\sum_j |z_0^j|^2 &=& -\frac{1}{4}|\frac{z_0}{\sh(z_0)}|^2 
\mbox{tr}(\tau_j \tilde{g})
\mbox{tr}(\tau_j \overline{\tilde{g}}^T)
\nonumber\\
& =& -\frac{1}{4}|\frac{z_0}{\sh(z_0)}|^2 (|\mbox{tr}(\tilde{g})|^2
-2\mbox{tr}(\tilde{g}\overline{\tilde{g}}^T))\nonumber\\
&\le& \frac{1}{2}|\frac{z_0}{\sh(z_0)}|^2 
\mbox{tr}(\tilde{H}^2)
= |\frac{z_0}{\sh(z_0)}|^2 \ch(\tilde{p})
\ea
Now recall from \cite{32} that 
\be \label{3.62}
\ch(\tilde{p})=(1+r)\ch^2(\frac{p+p'}{2})+(1-r)\ch^2(\frac{p-p'}{2})-1
\ee
where $r=p_j p'_j/(p p')\in[-1,1]$. Combining (\ref{3.62}) and (\ref{3.59})
yields (notice that $p+p'>|p-p'|,\;|r|\le1$)
\be \label{3.63}
\sum_j |z_0^j|^2 \le 4|\frac{z_0}{\sh(z_0)}|^2 
\ch^2(\frac{p+p'}{2})\le [2 k_c \frac{s}{\sh(s)} \ch(\frac{p+p'}{2})]^2
\ee
$\Box$\\
\\
Next, by estimates established in \cite{32} we have
\be \label{3.65}
\Delta^2(\vec{p},\vec{p}'):=p^2+(p')^2-\frac{\tilde{p}^2}{2}\ge 0 
\mbox{ and } \delta^2(g,g'):=\tilde{p}^2/4-s^2+\phi^2-\tilde{\theta}^2\ge0
\ee
where $\Delta=0$ if and only if $\vec{p}=\vec{p}'$ and $\delta=0$ 
if either a) $\tilde{\alpha}=0,\pi$ and $\tilde{p}, \tilde{\theta}$ are 
arbitrary or b) $\tilde{\alpha}$ is arbitrary and one or both of 
$\tilde{p}=0;\tilde{\theta}=0,\pi$ hold. In both of the cases a),b) we
have $|s|=\tilde{p}/2,\phi=\tilde{\theta}$.
It follows that 
\be \label{3.66}
\Re(z_0^2-p^2/2-(p')^2/2)=
-\{[\frac{p^2}{2}+\frac{(p')^2}{2}-\frac{\tilde{p}^2}{4}]
+\tilde{\theta}^2+
[-s^2+\phi^2-\tilde{\theta}^2+\frac{\tilde{p}^2}{4}]\}
=-[\Delta^2/2+\delta^2+\tilde{\theta}^2]
\ee
Combining the estimates (\ref{3.34}) for $D^t(p)$, (\ref{3.43}) for
$I$, (\ref{3.53}), (\ref{3.54}) and (\ref{3.55}) respectively for 
$I_1,I_2,I_3$ respectively, (\ref{3.58}) for $|z_0^j|$, 
(\ref{3.40}) for $f_c(s)$ and (\ref{3.66}) for the Gaussian prefactor of
$N^t(g,g')$ we conclude 
\ba \label{3.67} 
&& |\Delta<\hat{p}_j>^t_{gg'}| 
\le 
|z_0^j| e^{\frac{\Re(z_0^2-p^2/2-(p')^2/2)}{t}} 
|\frac{z_0}{\sh(z_0)}|
\frac{
\{
\frac{t}{2}|I|\; |I_1|
+2 |I_2|+2\pi t |I_3|
\}
}{
\sqrt{D^t(p) D^t(p')}
}
\\
&\le& 
[2 k_c \frac{s}{\sh(s)}\ch(\frac{p+p'}{2})]\;
[e^{-\frac{\Delta^2/2+\delta^2+\tilde{\theta}^2}{t}}]\;
[\frac{
k_c\frac{s}{\sh(s)}
}{
(1-K_t)\sqrt{\frac{p}{\sh(p)}\frac{p'}{\sh(p')}}
}]
\times\nonumber\\
&\times&
\{
[t k_c' (1+k_t)] + [2 k'_t]
+[2\pi t \tilde{k}_t]
\}
\nonumber\\
&=& 2 k_c^2  
[e^{-\frac{\Delta^2/2+\tilde{\theta}^2}{t}}]\;
[e^{-\delta^2/t}\frac{s^2}{\sqrt{pp'}}
\frac{\ch(\frac{p+p'}{2})\sqrt{\sh(p)\sh(p')}}{\sh^2(s)}]\;
[\frac{t k_c' (1+k_t)+2 k'_t+2\pi t \tilde{k}_t}{1-K_t}]
\nonumber
\ea
Let us discuss this result. The last line of (\ref{3.67}) consists
of three factors corresponding to the three square brackets. The 
first bracket contains a Gaussian with peak of width of order $\sqrt{t}$
at $g=g'$. The third bracket is of the form $t(1+K'_t(c))$ where $K'_t(c)$
is exponentially vanishing with $t$ for any $0<c<1$. These two brackets
are expected from the harmonic oscillator. The remaining second bracket 
\be \label{3.67a}
B_2:=e^{-\delta^2/t}\frac{s^2}{\sqrt{pp'}}
\frac{\ch(\frac{p+p'}{2})\sqrt{\sh(p)\sh(p')}}{\sh^2(s)}
\ee
is unexpected, it is not manifestly bounded from above and thus 
the integral over $g'$ is not obviously converging.
The subsequent paragraph will be devoted to the behaviour of that term.\\ 
\\
Since (\ref{3.67a}) is manifestly regular at $p'=0$, convergence 
problems can arise only for $p'\to\infty$. Formula (\ref{3.62}) 
implies that then also $\tilde{p}\to\infty$, no matter which values  
$p,r$ take, actually $\tilde{p}\to p'$ as $p'\to\infty$ for
fixed $p,r$. By estimate (\ref{3.65}) we then see that either a)
$\delta\to\infty$ or b) $\delta$ stays bounded from above. In the first
case $s$ must stay bounded or grows slowlier than
$\tilde{p}/2$ while in the latter case $s\to\tilde{p}/2\to p/2'$.
Consider first case a). Then for large $p'$ the Gaussian $e^{-\delta^2/t}$
in (\ref{3.67a}) certainly wins over the remaining factor and 
$B_2$ decays exponentially fast. Next consider case b). In that case 
$B_2$ grows as $\sqrt{p'}^3$ at large $p'$. Thus, altogether we have 
shown that $B_2$ grows no worse than polynomially as $p'\to\infty$.
Notice that $s$ stays bounded if and only if $\tilde{\theta}=
\tilde{\alpha}=\pi/2$ which defines a set of $\Omega$ measure zero.
In that case in fact $s=0,\phi=\pi/2$, therefore $z_0=i\pi/2$ 
whence $\tilde{g}=\tau_j\tilde{\theta}_j\ch(\tilde{p}/2)/\tilde{\theta}$ 
and $\sum_j |z_0^j/z_0|^2=\ch^2(\tilde{p}/2)$.\\
\\
We conclude that in the range $0\le\phi\le(1-c)\pi$ 
$\Delta<\hat{p}_j>^t_{gg'}$ can be bounded by a function of $g,g'$ 
which is of the form of a Gaussian with peak at 
$g=g'$ times a function of $g,g'$ bounded by a 
polynomial in $\vec{p},\vec{p}'$ times
$t$ times a constant that approaches unity exponentially fast. Thus 
the integral of that function with respect to $g'$ exists
(since $\Delta^2$ approaches $p'^2/2$ at large $p'$)
and will be of the form of a function of $g$ bounded by a polynomial  
in $p$
times $t$ times a constant that approaches unity exponentially fast.
It should be noted that the six Gaussians in
(\ref{3.67}) of width $\sqrt{t}$ each cancel the $1/t^3$ in the measure 
$d\Omega/t^3$ (recall from (\ref{3.18}), (\ref{3.19}) that 
$||\psi^t_g||^2\nu_t(g)$ approaches $2/(\pi t^3)$ exponentially fast).

It follows that as far as the leading order (in $t$) behaviour of the 
expectation value of any monomial is concerned, we can drop the term
$\Delta<\hat{p}_j>^t_{g,g'}$ from $<\hat{p}_j>^t_{g,g'}$, at least in the 
range $0\phi\le(1-c)\pi$.\\
\\
Case B)\\
Let us now discuss the range $(1-c)\pi\le\phi\le\pi$. We will not be as 
explicit as in case A), the steps to be performed are essentially identical
to the case A) and can be found in more detail in the analogous discussion
of \cite{32}.

The essential point is now to write everything in terms of  
\be \label{3.68}
z_0':=z_0-i\pi=s-i(\pi-\phi)=:s-i\phi' \mbox{ where } 0\le\phi'\le c\pi
\ee
Starting with the expression (\ref{3.33}) we have
\ba \label{3.69}
&& <\hat{p}_j>^t_{gg'}  
=
[\frac{-i}{2}\frac{\mbox{tr}(\tau_j g'\bar{g}^T)}{\sh(z_0)}]
\times\\
&\times&
\frac{\{\frac{t}{2}
\sum_n e^{\frac{(z'_0-i\pi (2n-1))^2-p^2/2-(p')^2/2}{t}}
[2\frac{(z'_0-i\pi(2n-1))^2}{t\sh(z_0)}
-(z'_0-i\pi (2n-1))\frac{\ch(z_0)}{\sh^2(z_0)}+\frac{1}{\sh(z_0)}]\}
}{
[\sum_n \frac{p-2\pi i n}{\sh(p)} e^{-\frac{4\pi^2 n^2}{t}}
e^{-i\frac{4\pi n p}{t}}]^{1/2}
[\sum_n \frac{p'-2\pi i n}{\sh(p')} e^{-\frac{4\pi^2 n^2}{t}}
e^{-i\frac{4\pi n p'}{t}}]^{1/2}
}
\nonumber\\
&=&
[\frac{-i}{2}\frac{\mbox{tr}(\tau_j g'\bar{g}^T)}{\sh(z_0)}]
\times\nonumber\\
&\times& e^{\frac{(z_0')^2-p^2/2-(p')^2/2}{t}}
\frac{\{\frac{t}{2}
\sum_{n=\mbox{odd}} 
e^{-\frac{\pi^2 n^2}{t}} e^{-\frac{2i\pi z_0' n}{t}}
[2\frac{(z'_0-i\pi  n)^2}{t\sh(z'_0)}
-(z'_0-i\pi n)\frac{\ch(z_0)}{\sh^2(z_0)}+\frac{1}{\sh(z_0)}]\}
}{
\sqrt{\frac{p}{\sh(p)}\frac{p'}{\sh(p')}(1-K_t(p))(1-K_t(p'))}
}
\nonumber
\ea
where $|K_t(p)|,|K_t(p')|\le K_t$ are the functions implicit in the 
estimate (\ref{3.34}) bounded from above by exponentially vanishing 
constants. We now pull out a factor of $(z_0')^2/\sh(z_0)$ out of the 
series in (\ref{3.69}), collect terms as to sum over positive, odd
integers $n$ only and observe that
$\sh(z_0)=-\sh(z_0')$ and $\ch(z_0)=-\ch(z_0')$. Then (\ref{3.69})
becomes
\ba \label{3.70}
&& <\hat{p}_j>^t_{gg'}  
=
\frac{
[\frac{-i z_0'}{2}\frac{\mbox{tr}(\tau_j g'\bar{g}^T)}{\sh(z'_0)}]\;
[\frac{z_0'}{\sh(z_0')}]\;e^{\frac{(z_0')^2-p^2/2-(p')^2/2}{t}}
}{
\sqrt{\frac{p}{\sh(p)}\frac{p'}{\sh(p')}(1-K_t(p))(1-K_t(p'))}
}
\times\\
&\times& 
\sum_{n=1,\mbox{odd}}^\infty 
e^{-\frac{\pi^2 n^2}{t}} 
\{
2[\cos(\frac{2\pi n z_0'}{t})
-\frac{(2\pi n)^2}{t}
\frac{\sin(\frac{2\pi n z_0'}{t})}{\frac{2\pi n z_0'}{t}}]
+t[\frac{1}{(z_0')^2}-\frac{\ch(z_0')}{z_0'\sh(z_0')}]
\cos(\frac{2\pi n z_0'}{t})
\nonumber\\
&&
+\pi n t [\frac{\ch(z_0')}{(z_0')^2\sh(z_0')}\sin(\frac{2\pi n z_0'}{t})
-\frac{2\pi n }{(z_0')^2 t}  \cos(\frac{2\pi n z_0'}{t})]
\}
\nonumber
\ea
which can now be estimated essentially as in case A). The most 
important differences are the following : First, lemma \ref{la3.1} has now 
to be replaced by $|z_0'/\sh(z_0')|\le k_{1-c} s/\sh(s)$ which can be seen by 
following the proof given there step by step. Secondly, by the methods
given in \cite{32} the Gaussian is now estimated from above by
\be \label{3.71}
e^{-\frac{\Delta^2+2\delta^2+\tilde{\theta}^2}{2t}}
\ee
where necessarily $\tilde{\theta}\ge\pi/2$ and is therefore exponentially
small for all values of $g'$ in the range $(1-c)\pi\le \phi\le \pi$
provided that we choose $c<1/2$ as we do. Finally, choosing 
as in \cite{32} $c=1/32$ (see specifically formula (4.44) there) and using 
the same estimates of case A) we can display (\ref{3.70}) as a 
function of $g,g'$ bounded by a polynomial in $p,p'$ 
times a Gaussian in $\Delta^2$ multiplied by an overall constant which 
decays exponentially fast to zero as $t\to 0$. Thus, (\ref{3.70}) is  
an $\Omega/t^3$ integrable function with respect to $g'$ and the result
of the integration is a function of $g$ bounded by a polynomial in $p$ 
times a constant which decays exponentially fast as $t\to 0$.

We conclude that the range of integration $(1-c)\pi\le \phi\le \pi$
is irrelevant for the expectation value and all its corrections in
powers of $t$.\\  
\\
\\
We can now finish the estimate of the matrix element. Our discussion 
has demonstrated that to leading order in $t$ we can replace 
$<\hat{p}_j>^t_{gg'}$ by
\be \label{3.72}
\frac{
[\frac{-i}{2}\frac{\mbox{tr}(\tau_j g'\bar{g}^T)}{\sh(z_0)} z_0]
e^{\frac{z_0^2-p^2/2-(p')^2/2}{t}} \frac{z_0}{\sh(z_0)} 
\Theta((1-c)\pi-\phi)
}{
\sqrt{D^t(p) D^t(p')}
}
\ee
where $\Theta$ is the step function. The expression (\ref{3.72})
is Gaussian peaked at $g=g'$ with decay width of order $\sqrt{t}$
at which it equals $p_j$.
In order to perform the integral over $g'$ we will therefore expand 
the square bracket in the numerator of (\ref{3.72}) as 
\be \label{3.73}
p_j+
[\frac{-i}{2}\frac{\mbox{tr}(\tau_j g'\bar{g}^T)}{\sh(z_0)} z_0-p_j]
\ee
and the integral over the $g'$ independent term with respect to 
$\Omega(g')/t^3$ converges exponentially fast to $p_j$ since the 
absolute value squared of
(\ref{3.72}) modulo the square bracket equals precisely the overlap 
function of \cite{32} modulo a multiplicative term whose absolute 
value can be estimated from above by a constant that approaches unity
exponentially fast. The integral over the remaining term can be expanded 
as a function of $\vec{p}-\vec{p}',h (h')^{-1}$ and vanishes at least
linearly in $t$ by standard properties of Gaussian integrals.\\
\\
Collecting all the results we have arrived at the first main theorem
of this paper.
\begin{Theorem} \label{th3.2}
The matrix elements of the momentum operators with respect to coherent 
states can be estimated by 
\be \label{3.74}
|\frac{<\psi^t_g,\hat{p}_j\psi^t_{g'}>}{||\psi^t_g||\;||\psi^t_{g'}||}
-p_j(g)\frac{<\psi^t_g,\psi^t_{g'}>}{||\psi^t_g||\;||\psi^t_{g'}||}|
\le t f(\vec{p},\vec{p}')
\frac{|<\psi^t_g,\psi^t_{g'}>|}{||\psi^t_g||\;||\psi^t_{g'}||}
\ee
where $f$ is a polynomial of 
$p,p'$.
\end{Theorem}
As a corollary to theorem (\ref{th3.2}) we obtain that the expectation value
$<\hat{p}_j>^t_{gg}$ equals $p_j(g)$ up to bounded corrections in $p_j(g)$
and that are proportional to $t$. We will actually
calculate the exact correction in a later section by a different method.

\subsubsection{Matrix Elements for the Holonomy Operator}
\label{s3.1.3}

The computation of the matrix element of the holonomy operator 
\be \label{3.75}
<\hat{h}_{AB}>^t_{gg'}:=
\frac{<\psi^t_g,\hat{h}_{AB}\psi^t_{g'}>}{||\psi^t_g||\;||\psi^t_{g'}||}
\ee
turns out to be rather messy. Let us determine first the following matrix 
element, using the Peter\&Weyl theorem and 
the $SL(2,\Co)$ identity $\chi_j(g)=\chi_j(g^{-1})$ 
\ba \label{3.76}
&&<\psi^t_g,\hat{h}_{A_0 B_0}\psi^t_{g'}>\\
&=&
\sum_{j,j'} d_j d_{j'} e^{-t[\lambda_{j}+\lambda_{j'}]/2}
\overline{\pi_j(g)_{A_1..A_{2j},B_1..B_{2j}}}
\pi_{j'}(g')_{A'_1..A'_{2j'},B'_1..B'_{2j'}}
\times\nonumber\\
&\times&
<\pi_{j'}(.)_{A'_1..A'_{2j},B'_1..B'_{2j}},\hat{h}_{A_0 B_0}
\pi_j(.)_{A_1..A_{2j},B_1..B_{2j}}>
\nonumber\\
&=&
\sum_{j,j'} d_j d_{j'} e^{-t[\lambda_{j}+\lambda_{j'}]/2}
\overline{\pi_j(g)_{A_1..A_{2j},B_1..B_{2j}}}
\pi_{j'}(g')_{A'_1..A'_{2j'},B'_1..B'_{2j'}}\times\nonumber\\
&\times&
[\frac{\delta_{j',j+\frac{1}{2}}}{d_{j'}}
\pi_{j'}(1)_{A'_1..A'_{2j'},A_0 A_1..A_{2j}}
\pi_{j'}(1)_{B'_1..B'_{2j'},B_0 B_1..B_{2j}}
+\frac{\delta_{j',j-\frac{1}{2}}d_{j-\frac{1}{2}}}{d_j d_{j'}}
\times\nonumber\\
&\times& 
\pi_{j'}(1)_{A'_1..A'_{2j'},(A_2..A_{2j}}\epsilon_{A_1) A_0} 
\pi_{j'}(1)_{B'_1..B'_{2j'},(B_2..B_{2j}}\epsilon_{B_1) B_0}]
\nonumber\\
&=&
\sum_j d_j  e^{-t\lambda_j/2} 
\pi_j(\bar{g}^T)_{B_1..B_{2j},A_1..A_{2j}}
\times\nonumber\\
&\times&
[e^{-t\lambda_{j+\frac{1}{2}}/2}
\pi_{j+\frac{1}{2}}(g')_{A_0 A_1..A_{2j}, B_0 B_1..B_{2j}}
-\frac{d_{j-\frac{1}{2}}}{d_j} e^{-t\lambda_{j-\frac{1}{2}}/2}
\epsilon_{A_0 (A_1} 
\pi_{j-\frac{1}{2}}(g')_{A_2..A_{2j}),(B'_2..B'_{2j}}\epsilon_{B_1) B_0}]
\nonumber
\ea
In the second step we have recalled 
the following (Clebsch-Gordan) identity, valid for arbitrary 
$g\in G^\Co=SL(2,\Co)$ and proved in \cite{32}
\ba \label{3.77}
&& g_{A_0 B_0}\pi_j(g)_{A_1..A_{2j},B_1..B_{2j}}
\nonumber\\
&=&\pi_{j+\frac{1}{2}}(g)_{A_0..A_{2j},B_0..B_{2j}}
-\frac{d_{j-\frac{1}{2}}}{d_j}
\epsilon_{A_0 (A_1}\pi_{j-\frac{1}{2}}(g)_{A_2..A_{2j}),(B_2..B_{2j}}
\epsilon_{B_1)B_0}
\ea
with $A,B,..=\pm 1/2$, round brackets around groups of indices denote
total symmetrization taken as an idempotent operation, $\epsilon_{AB}$
is the skew symmetric spinor of rank two, $d_j=\dim(\pi_j)=2j+1$ and the 
relation with the usual
matrix elements of the irreducible representation $\pi_j$ is given by
$\pi_j(g)_{A_1+..+A_{2j},B_1+..+B_{2j}}=\pi_j(g)_{A_1..A_{2j},B_1..B_{2j}}$.

The huge amount of summation indices that appear in (\ref{3.76}) and 
which we cannot nicely contract as irreducible representations of 
different dimension are multiplied with each other make (\ref{3.76}) 
impossible to work with because then we cannot apply the Weyl character --
and Poisson summation formula, our main tools in all the estimates. 
Fortunately, we 
have the following trick at our disposal (it obviously extends to groups of 
higher rank) :\\ 
Let $\Delta_h$ be the 
Laplacian on $G=SU(2)$ acting on $h\in G$. Since 
$\pi_j(hg)_{mn}=\pi_j(h)_{mm'}\pi_j(g)_{m'n}$  is an 
eigenstate of $-\Delta_h$ with eigenvalue $j(j+1)$ we obtain the following
formulas that isolate the irreducible pieces on the right hand side of 
(\ref{3.77})
\ba \label{3.78a}
&& 
\{(j+\frac{1}{4})(hg)_{A_0 B_0}+[-\Delta_h,(hg)_{A_0 B_0}]\}
\pi_j(hg)_{A_1..B_{2j}}\nonumber\\
&=& d_j \pi_{j+\frac{1}{2}}(hg)_{A_0..A_{2j},B_0..B_{2j}}
\\
\label{3.78b}
&& 
\{(j+\frac{3}{4})(hg)_{A_0 B_0}-[-\Delta_h,(hg)_{A_0 B_0}]\}
\pi_j(hg)_{A_1..B_{2j}}\nonumber\\
&=&-d_{j-\frac{1}{2}}
\epsilon_{A_0 (A_1}\pi_{j-\frac{1}{2}}(hg)_{A_2..A_{2j}),(B_2..B_{2j}}
\epsilon_{B_1)B_0}
\ea
Taking the limit $h\to 1$ in (\ref{3.78a}), (\ref{3.78b}) we can cast
(\ref{3.76}) into the following simpler form which allows us to contract 
indices
\ba \label{3.79}
&&<\psi^t_g,\hat{h}_{A_0 B_0}\psi^t_{g'}>\nonumber\\
&=&
\sum_j   e^{-t\lambda_j/2} 
\pi_j(\bar{g}^T)_{B_1..B_{2j},A_1..A_{2j}}
\times\nonumber\\
&\times&
\{
[
e^{-t\lambda_{j+\frac{1}{2}}/2}
((j+\frac{1}{4})(hg')_{A_0 B_0}+[-\Delta_h,(hg')_{A_0 B_0}])
\nonumber\\
&& +e^{-t\lambda_{j-\frac{1}{2}}/2}
((j+\frac{3}{4})(hg')_{A_0 B_0}-[-\Delta_h,(hg')_{A_0 B_0}])
]
\pi_j(hg')_{A_1..A_{2j},B_1..B_{2j}}\}_{|h=1}
\nonumber\\
&=&
\sum_j   e^{-t\lambda_j/2} 
\times\nonumber\\
&\times&
\{
[
e^{-t\lambda_{j+\frac{1}{2}}/2}
((j+\frac{1}{4})(hg')_{A_0 B_0}+[-\Delta_h,(hg')_{A_0 B_0}])
\nonumber\\
&& +e^{-t\lambda_{j-\frac{1}{2}}/2}
((j+\frac{3}{4})(hg')_{A_0 B_0}-[-\Delta_h,(hg')_{A_0 B_0}])
]
\chi_j(hg'\bar{g}^T)\}_{|h=1}
\nonumber\\
&=&
(g')_{A_0 B_0} \sum_j e^{-t\lambda_j/2} 
[
e^{-t\lambda_{j+\frac{1}{2}}/2}(j+\frac{1}{4})
+e^{-t\lambda_{j-\frac{1}{2}}/2} (j+\frac{3}{4})
]\chi_j(g'\bar{g}^T)
\nonumber\\
&+&
\{[-\Delta_h,(hg')_{A_0 B_0}] \sum_j e^{-t\lambda_j/2} 
[e^{-t\lambda_{j+\frac{1}{2}}/2}-e^{-t\lambda_{j-\frac{1}{2}}/2}]
\chi_j(hg'\bar{g}^T)\}_{|h=1}
\ea
Formula (\ref{3.79}) can be further simplified by making use of the 
commutator identity (use $\Delta_h=X^j_h X^j_h$)
\be \label{3.80}
\{[-\Delta_h,(hg')_{A_0 B_0}] f(h)\}_{|h=1}
=\frac{3}{4}g'_{A_0 B_0} f(1)
-\frac{1}{2}(\tau_j g')_{A_0 B_0} (\frac{d}{ds})_{s=0} f(e^{s\tau_j})
\ee
valid for any differentiable function of $h$. Inserting this into 
(\ref{3.79}) results in the final formula
\ba \label{3.81}
&&<\psi^t_g,\hat{h}_{AB}\psi^t_{g'}>\nonumber\\
&=&
g'_{AB} \sum_j e^{-t\lambda_j/2} 
[(j+1)e^{-t\lambda_{j+\frac{1}{2}}/2}
+j e^{-t\lambda_{j-\frac{1}{2}}/2}]\chi_j(g'\bar{g}^T)
\nonumber\\
&-&
\frac{1}{2}(\tau_j g')_{AB} (\frac{d}{ds})_{s=0} 
\sum_j e^{-t\lambda_j/2} 
[e^{-t\lambda_{j+\frac{1}{2}}/2}-e^{-t\lambda_{j-\frac{1}{2}}/2}]
\chi_j(e^{s\tau_j} g'\bar{g}^T)
\ea
to which we can now apply the Weyl character formula.

Let again $\ch(z)=\mbox{tr}(e^{s\tau_j}g'\bar{g}^T)/2$ and 
$\ch(z_0)=\mbox{tr}(g'\bar{g}^T)/2$, then
\ba \label{3.82}
&&<\psi^t_g,\hat{h}_{AB}\psi^t_{g'}>\nonumber\\
&=&
\frac{g'_{AB}}{2\sh(z_0)} \sum_j 
[(d_j+1)e^{-\frac{t}{4}(d_j^2+d_j-1/2)}
+(d_j-1)e^{-\frac{t}{4}(d_j^2-d_j-1/2)}]\sh(d_j z_0)
\nonumber\\
&-&
\frac{1}{2}(\tau_j g')_{AB} (\frac{d}{ds})_{s=0} 
\sum_j  
[e^{-\frac{t}{4}(d_j^2+d_j-1/2)}
-e^{-\frac{t}{4}(d_j^2-d_j-1/2)}]\frac{\sh(d_j z)}{\sh(z)}
\nonumber\\
&=&
\frac{g'_{AB}}{4\sh(z_0)}e^{t/8} \sum_{n=1}^\infty e^{-t n^2/4}
[(n+1)e^{-nt/4}+(n-1)e^{nt/4}][e^{n z_0}-e^{-n z_0}]
\nonumber\\
&-&
\frac{1}{2}(\tau_j g')_{AB} (\frac{d}{ds})_{s=0} \frac{e^{t/8}}{2\sh(z)}
\sum_{n=1}^\infty e^{-t n^2/4}
[e^{-nt/4}-e^{nt/4}][e^{n z}-e^{-n z}]
\nonumber\\
&=&
\frac{g'_{AB}}{4\sh(z_0)}e^{t/8} \sum_{n=1}^\infty e^{-t n^2/4}
\{[n e^{n(z_0-t/4)}+(-n)e^{(-n)(z_0-t/4)}]
+[n e^{n(z_0+t/4)}+(-n)e^{(-n)(z_0+t/4)}]
\nonumber\\
&& +[e^{n(z_0-t/4)}+e^{(-n)(z_0-t/4)}]
-[ e^{n(z_0+t/4)}+e^{(-n)(z_0+t/4)}]\}
\nonumber\\
&-&
\frac{1}{2}(\tau_j g')_{AB} (\frac{d}{ds})_{s=0} \frac{e^{t/8}}{2\sh(z)}
\sum_{n=1}^\infty e^{-t n^2/4}
\{[e^{n(z-t/4)}+e^{(-n)(z-t/4)}]
-[e^{n(z+t/4)}+e^{(-n)(z+t/4)}]\}
\nonumber\\
&=&
\frac{g'_{AB}}{4\sh(z_0)}e^{t/8} \sum_{n=-\infty}^\infty e^{-t n^2/4}
[(n+1) e^{n(z_0-t/4)}+(n-1) e^{n(z_0+t/4)}]
\nonumber\\
&-&
\frac{1}{2}(\tau_j g')_{AB} (\frac{d}{ds})_{s=0} \frac{e^{t/8}}{2\sh(z)}
\sum_{n=-\infty}^\infty e^{-t n^2/4}
[e^{n(z-t/4)}-e^{n(z+t/4)}]
\ea
where in the last step we have recognized that the terms in the curly 
brackets add up to zero at $n=0$ and that the terms with $(-n)$
as argument can be taken care of by extending the series to negative
values of $n$. Introducing
\be \label{3.83}
T:=\sqrt{t}/2,\; z_0^\pm=z_0/T \pm T,\; z^\pm=z/T\pm T
\ee
and remembering (\ref{3.31}) we may write (\ref{3.82}) in the form
(notice that $s$ is to be replaced by $-s$ as compared to (\ref{3.24}))
\ba \label{3.84}
&&<\psi^t_g,\hat{h}_{AB}\psi^t_{g'}>\nonumber\\
&=&
\frac{g'_{AB}}{4\sh(z_0)T}e^{t/8} \sum_{n=-\infty}^\infty e^{-(nT)^2}
[((nT)+T) e^{(nT) z_0^-}+((nT)-T) e^{(nT)z_0^+}]
\nonumber\\
&-&
\frac{1}{2}(\tau_j g')_{AB} 
\frac{\mbox{tr}(\tau_j g'\bar{g}^T)}{2\sh(z_0)}
(\frac{d}{dz})_{z=z_0} \frac{e^{t/8}}{2\sh(z)}
\sum_{n=-\infty}^\infty e^{-(nT)^2}
[e^{(nT)z^-}-e^{(nT)z^+}]
\ea
An appeal to the Poisson summation formula now reveals that
\ba \label{3.85}
&&<\psi^t_g,\hat{h}_{AB}\psi^t_{g'}>\\
&=&
g'_{AB}\frac{e^{t/8}}{4\sh(z_0)T} \frac{2\pi}{T}
\sum_{n=-\infty}^\infty 
[(\frac{z_0^- -\frac{2\pi i n}{T}}{4\sqrt{\pi}}+\frac{T}{2\sqrt{\pi}}) 
e^{(z_0^- -\frac{2\pi i n}{T})^2/4}
+(\frac{z_0^+ -\frac{2\pi i n}{T}}{4\sqrt{\pi}}-\frac{T}{2\sqrt{\pi}}) 
e^{(z_0^+ -\frac{2\pi i n}{T})^2/4}]
\nonumber\\
&-&
\frac{e^{t/8}}{8\sh(z_0)}
(\tau_j g')_{AB} \mbox{tr}(\tau_j g'\bar{g}^T)
(\frac{d}{dz})_{z=z_0} \frac{1}{\sh(z)} \frac{2\pi}{T}
\sum_{n=-\infty}^\infty
[\frac{e^{(z^- -\frac{2\pi i n}{T})^2/4}}{2\sqrt{\pi}}
-\frac{e^{(z^+ -\frac{2\pi i n}{T})^2/4}}{2\sqrt{\pi}}]
\nonumber\\
&=&
\frac{\sqrt{\pi}e^{t/8}}{8\sh(z_0)T^3} [g'_{AB}]
\sum_{n=-\infty}^\infty 
[(z_0+T^2 -2\pi i n) e^{\frac{(z_0-T^2 -2\pi i n)^2}{t}}
+(z_0-T^2 -2\pi i n) e^{\frac{(z_0+T^2 -2\pi i n)^2}{t}}]
\nonumber\\
&-&
\frac{\sqrt{\pi}e^{t/8}}{8\sh(z_0) T^3}
[(\tau_j g')_{AB} \mbox{tr}(\tau_j g'\bar{g}^T)]
(\frac{d}{dz})_{z=z_0} \frac{T^2}{\sh(z)} 
\sum_{n=-\infty}^\infty
[e^{\frac{(z-T^2 -2\pi i n)^2}{t}}-e^{\frac{(z+T^2 -2\pi i n)^2}{t}}]
\nonumber\\
&=&
\frac{\sqrt{\pi}e^{t/8}}{8\sh(z_0)T^3} [g'_{AB}]
\sum_{n=-\infty}^\infty 
[(z_0+T^2 -2\pi i n) e^{\frac{(z_0-T^2 -2\pi i n)^2}{t}}
+(z_0-T^2 -2\pi i n) e^{\frac{(z_0+T^2 -2\pi i n)^2}{t}}]
\nonumber\\
&-&
\frac{\sqrt{\pi}e^{t/8}}{8\sh(z_0) T^3}
[(\tau_j g')_{AB} \mbox{tr}(\tau_j g'\bar{g}^T)] T^2
\{
-\frac{\ch(z_0)}{\sh^2(z_0)} \sum_{n=-\infty}^\infty
[e^{\frac{(z_0-T^2 -2\pi i n)^2}{t}}-e^{\frac{(z_0+T^2 -2\pi i n)^2}{t}}]
\nonumber\\
&& +\frac{1}{2 T^2 \sh(z_0)} \sum_{n=-\infty}^\infty
[(z_0-T^2 -2\pi i n) e^{\frac{(z_0-T^2 -2\pi i n)^2}{t}}
-(z_0+T^2 -2\pi i n) e^{\frac{(z_0+T^2 -2\pi i n)^2}{t}}]\}
\nonumber
\ea
Recalling the norm of the coherent states from (\ref{3.29}), 
\be \label{3.86}
||\psi^t_g||^2=\frac{\sqrt{\pi}e^{t/4}}{4\sh(p) T^3}
\sum_n (p-2\pi i n) e^{\frac{(p-2\pi i n)^2}{t}}
=:\frac{\sqrt{\pi}e^{t/4}}{4 T^3}\frac{p}{\sh(p)}e^{p^2/t}
(1+K_t(p))
\ee
we arrive
at the final exact formula for the matrix element of the holonomy operator
(\ref{3.75}) (all sums run over $n\in\Zl$)
\ba \label{3.87}
&& <\hat{h}_{AB}>^t_{gg'}\nonumber\\
&=& 
\frac{
\frac{e^{-t/8}e^{-\frac{p^2+(p')^2}{2t}}}{2\sh(z_0)}
}{
\sqrt{\frac{p}{\sh(p)}(1+K_t(p))\frac{p'}{\sh(p')}(1+K_t(p'))}
}
\times\nonumber\\
&\times &
\{ 
g'_{AB} 
\sum_n 
[(z_0+T^2 -2\pi i n) e^{\frac{(z_0-T^2 -2\pi i n)^2}{t}}
+(z_0-T^2 -2\pi i n) e^{\frac{(z_0+T^2 -2\pi i n)^2}{t}}]
\nonumber\\
&-&
(\tau_j g')_{AB} \frac{\mbox{tr}(\tau_j g'\bar{g}^T)}{2\sh(z_0)} 
\sum_n
[(z_0-T^2 -2\pi i n-2T^2\frac{\ch(z_0)}{\sh(z_0)}) 
e^{\frac{(z_0-T^2 -2\pi i n)^2}{t}}
\nonumber\\
&& -(z_0+T^2 -2\pi i n-2T^2\frac{\ch(z_0)}{\sh(z_0)}) 
e^{\frac{(z_0+T^2 -2\pi i n)^2}{t}}]
\}
\ea
At this point we must again distinguish between the cases 
A) $0\le \phi\le (1-c)\pi$ and B) $(1-c)\pi\le \phi\le \pi$ for some 
$c<1/2$ where $z_0=s+i\phi$.\\
\\
Case A)\\
We have 
\be \label{3.88}
(z_0\pm T^2-2\pi i n)^2/t=\frac{z_0^2}{t}+\frac{t}{16}-\frac{4\pi^2 n^2}{t}
-\frac{4\pi i n z_0}{t} \pm(z_0/2-i\pi n)
\ee
and thus can write (\ref{3.87}) more explicitely as 
\ba \label{3.89}
&& <\hat{h}_{AB}>^t_{gg'}\nonumber\\
&=& 
\frac{
\frac{e^{-t/16}e^{-\frac{p^2+(p')^2-2 z_0^2}{2t}}}{2\sh(z_0)}
}{
\sqrt{\frac{p}{\sh(p)}(1+K_t(p))\frac{p'}{\sh(p')}(1+K_t(p'))}
}
\times\nonumber\\
&\times &
\{ 
g'_{AB} 
\sum_n (-1)^n e^{-\frac{4\pi^2 n^2}{t}} e^{-\frac{4\pi i n z_0}{t}}
[(z_0+T^2 -2\pi i n) e^{-z_0/2} +(z_0-T^2 -2\pi i n) e^{z_0/2}]
\nonumber\\
&-&
(\tau_j g')_{AB} \frac{\mbox{tr}(\tau_j g'\bar{g}^T)}{2\sh(z_0)} 
\sum_n (-1)^n e^{-\frac{4\pi^2 n^2}{t}} e^{-\frac{4\pi i n z_0}{t}}
[(z_0-T^2 -2\pi i n-2T^2\frac{\ch(z_0)}{\sh(z_0)}) e^{-z_0/2}
\nonumber\\
&& -(z_0+T^2 -2\pi i n-2T^2\frac{\ch(z_0)}{\sh(z_0)}) e^{z_0/2}]
\}
\ea
Let us focus on the curly bracket in (\ref{3.89}) for which we find,
after some considerable amount of algebra,
\ba \label{3.90}
&&\{.\}=2 z_0 \times\nonumber\\
&\times& 
\{ 
g'_{AB}
\{\ch(z_0/2)+[-\frac{T^2}{2}\frac{\sh(z_0/2)}{z_0/2}]
+2\sum_{n=1}^\infty (-1)^n e^{-\frac{4\pi^2 n^2}{t}}
\times\nonumber\\
&\times& 
([(\ch(z_0/2)-\frac{T^2}{2}
\frac{\sh(z_0/2)}{z_0/2})\cos(\frac{4\pi n z_0}{t})]
+[-\frac{8\pi^2 n^2}{t}\ch(z_0/2)
\frac{\sin(\frac{4\pi n z_0}{t})}{\frac{4\pi n z_0}{t}}])
\}
\nonumber\\
&-&
(\tau_j g')_{AB} z_0\frac{\mbox{tr}(\tau_j g'\bar{g}^T)}{2\sh(z_0)} 
\{-\sh(z_0/2)/z_0
+[2T^2\frac{\mbox{coth}(z_0)\sh(z_0/2)-\frac{1}{2}\ch(z_0/2)}{z_0^2}]
\nonumber\\
&&
+2\sum_{n=1}^\infty (-1)^n e^{-\frac{4\pi^2 n^2}{t}} 
\times\nonumber\\
&& \times ([(-\sh(z_0/2)/z_0
+2T^2\frac{\mbox{coth}(z_0)\sh(z_0/2)-\frac{1}{2}\ch(z_0/2)}{z_0^2})
\cos(\frac{4\pi  n z_0}{t})]
\nonumber\\
&&
+[\frac{(2\pi n)^2}{t} \frac{\sh(z_0/2)}{z_0/2}
\frac{\sin(\frac{4\pi n z_0}{t})}{\frac{4\pi n z_0}{t}}])
\}
\}
\nonumber\\
&=:& 2 z_0 
\{ 
g'_{AB}
\{\ch(z_0/2)+ 
[I_1]+2\sum_{n=1}^\infty (-1)^n e^{-\frac{4\pi^2 n^2}{t}} ([I_2]+[I_3])
\}
\nonumber\\
&-&
(\tau_j g')_{AB} z_0\frac{\mbox{tr}(\tau_j g'\bar{g}^T)}{2\sh(z_0)} 
\{-\frac{\sh(z_0/2)}{z_0}+
2T^2 [J_1]+2\sum_{n=1}^\infty (-1)^n e^{-\frac{4\pi^2 n^2}{t}} ([J_2]+[J_3])
\}
\}
\ea
where we have abbreviated the terms in the square brackets in the first 
equality by $I_1,I_2,I_3,J_1,J_2,J_3$ in this order since we wish to estimate 
them separately. 

Combining (\ref{3.89}) with (\ref{3.90}) yields
\ba \label{3.91c}
&& <\hat{h}_{AB}>^t_{gg'}\\
&=& 
\frac{
\frac{e^{-t/16}e^{-\frac{p^2+(p')^2-2 z_0^2}{2t}}z_0}{\sh(z_0)}
}{
\sqrt{\frac{p}{\sh(p)}(1+K_t(p))\frac{p'}{\sh(p')}(1+K_t(p'))}
}
\times\nonumber\\
&\times &
\{ 
g'_{AB}
\{\ch(z_0/2)+ 
[I_1]+2\sum_{n=1}^\infty (-1)^n e^{-\frac{4\pi^2 n^2}{t}} ([I_2]+[I_3])
\}
\nonumber\\
&-&
(\tau_j g')_{AB} z_0\frac{\mbox{tr}(\tau_j g'\bar{g}^T)}{2\sh(z_0)} 
\{-\sh(z_0/2)/z_0+
[J_1]+2\sum_{n=1}^\infty (-1)^n e^{-\frac{4\pi^2 n^2}{t}} ([J_2]+[J_3])
\}
\}
\nonumber
\ea
and proceeding as in section \ref{s3.1.2} we define
\ba \label{3.91}
&& \Delta<\hat{h}_{AB}>^t_{gg'}\nonumber\\
&:=& <\hat{h}_{AB}>^t_{gg'}-
\frac{
\frac{e^{-t/16}e^{-\frac{p^2+(p')^2-2 z_0^2}{2t}}z_0}{\sh(z_0)}
}{
\sqrt{\frac{p}{\sh(p)}(1+K_t(p))\frac{p'}{\sh(p')}(1+K_t(p'))}
}
\times\nonumber\\
&\times & 
[g'_{AB}\{\ch(z_0/2)+ 
(\tau_j g')_{AB} \frac{\mbox{tr}(\tau_j g'\bar{g}^T)}{2\sh(z_0)} 
\sh(z_0/2)]
\nonumber\\
&=&
\frac{
\frac{e^{-t/16}e^{-\frac{p^2+(p')^2-2 z_0^2}{2t}}z_0}{\sh(z_0)}
}{
\sqrt{\frac{p}{\sh(p)}(1+K_t(p))\frac{p'}{\sh(p')}(1+K_t(p'))}
}
\times\nonumber\\
&\times &
\{ 
g'_{AB}
\{ 
[I_1]+2\sum_{n=1}^\infty (-1)^n e^{-\frac{4\pi^2 n^2}{t}} ([I_2]+[I_3])
\}
\nonumber\\
&-&
(\tau_j g')_{AB} z_0\frac{\mbox{tr}(\tau_j g'\bar{g}^T)}{2\sh(z_0)} 
\{2T^2
[J_1]+2\sum_{n=1}^\infty (-1)^n e^{-\frac{4\pi^2 n^2}{t}} ([J_2]+[J_3])
\}
\}
\ea
The tools to estimate these terms have already been laid in section 
\ref{s3.1.2} so that we can be brief here. We just need the following
result.
\begin{Lemma} \label{la3.5}
For any $z_0=s+i\phi,\;0\le \phi\le \pi(1-c)$ we have 
\be \label{3.91a}
|J_1|\le \ch(|z_0|/2)[\sqrt{2}k'_c+\frac{k_c k'_{\frac{1+c}{2}}}{4}]
=:\tilde{k}_c \ch(|z_0|/2)
\ee
\end{Lemma}
Proof of Lemma \ref{la3.5} :\\
We easily establish the following identity
\be \label{3.91b}
J_1=\frac{1}{2}
C(z_0)\frac{\sh(z_0/2)}{z_0/2}-\frac{S(z_0)}{2}\ch(z_0/2)
+\frac{1}{8}\frac{z_0}{\sh(z_0)}\frac{\sh(z_0/2)}{z_0/2}[S(z_0/2)-C(z_0/2)]
\ee
Noticing that $\Im(z_0/2)\le\pi(1-c)/2=\pi(1-\frac{1+c}{2})$ the assertion 
follows from the lemmas of the previous subsection.\\
$\Box$\\
With the help of this lemma one finds
\ba \label{3.92}
|I_1| &=& |-\frac{T^2}{2}\frac{\sh(z_0/2)}{z_0/2}|\le 
\frac{T^2}{2} \ch(|z_0|/2)
\nonumber\\
|I_2| &=& |(\ch(z_0/2)-\frac{T^2}{2}\frac{\sh(z_0/2)}{z_0/2})
\cos(\frac{4\pi n z_0}{t})|
\nonumber\\
&\le & \ch(|z_0|/2)(1+T^2/2)|\cos(\frac{4\pi n z_0}{t})|
\le \ch(|z_0|/2)(1+T^2/) e^{\frac{4\pi^2 n (1-c)}{t}}
\nonumber\\
|I_3| &=& |\frac{8\pi^2 n^2}{t}\ch(z_0/2)
\frac{\sin(\frac{4\pi n z_0}{t})}{\frac{4\pi n z_0}{t}}|
\nonumber\\
&\le&
\frac{8\pi^2 n^2}{t}\ch(|z_0|/2)
|\frac{\sin(\frac{4\pi n z_0}{t})}{\frac{4\pi n z_0}{t}}|
\le \frac{16\pi^2 n^2}{t}\ch(|z_0|/2) e^{\frac{4\pi^2 n (1-c)}{t}}
\nonumber\\
|J_2|&=&|(-\sh(z_0/2)/z_0
+2T^2 J_1)
\cos(\frac{4\pi n z_0}{t})|
\nonumber\\
&\le &| (2\ch(|z_0|/2)+
2T^2 |J_1|)\;|\cos(\frac{4\pi n z_0}{t})|
\nonumber\\
&\le & 2\ch(|z_0|/2)(1+T^2 \tilde{k}_c)
e^{\frac{4\pi^2 n (1-c)}{t}}
\nonumber\\
|J_3| &=& |\frac{(2\pi n)^2}{t} \frac{\sh(z_0/2)}{z_0/2} 
\frac{\sin(\frac{4\pi n z_0}{t})}{\frac{4\pi n z_0}{t}}|
\nonumber\\
&\le&
\frac{(2\pi n)^2}{t} \ch(|z_0|/2) e^{\frac{4\pi^2 n(1-c)}{t}}
\ea
Observing the basic estimates (exploit $(\tau_j)^2=-1$)
\ba \label{3.94}
&& |g'_{AB}|^2\le \sum_{A,B}|g'_{AB}|^2=\mbox{tr}(g'(\bar{g}')^T)
=2\ch(p')\le 4\ch^2(p'/2) \mbox{ and }
\nonumber\\
&& |(\tau_j g')_{AB}|^2\le \mbox{tr}(\tau_j g'(\overline{\tau_j g'})^T)
=-\mbox{tr}(\tau_j^2 g'(\bar{g}')^T)\le 4\ch^2(p'/2) 
\ea
and employing the estimate for the denominator of (\ref{3.91}) from 
the previous section together with (\ref{3.58}) 
we obtain as an estimate (notation the same as in 
section \ref{s3.1.2}) 
\ba \label{3.95}
&& |\Delta<\hat{h}_{AB}>^t_{gg'}|\\
&\le&
\frac{
e^{-t/16}e^{-\frac{p^2+(p')^2-2 \Re(z_0^2)}{2t}} 
k_c|\frac{s}{\sh(s)}|\ch(|z_0|/2)
}{
(1-K_t)\sqrt{\frac{p}{\sh(p)}\frac{p'}{\sh(p')}}
}
\times\nonumber\\
&\times &
\{ 
|g'_{AB}|
\{
\frac{T^2}{2} 
+2\sum_{n=1}^\infty e^{-\frac{4\pi^2 n^2}{t}} 
e^{\frac{4\pi^2 n (1-c)}{t}} 
[1+\frac{T^2}{2}+\frac{16\pi^2 n^2}{t}]
\}
\nonumber\\
&+&
|(\tau_j g')_{AB}|\;|z_0\frac{\mbox{tr}(\tau_j g'\bar{g}^T)}{2\sh(z_0)}|
\{\ 2T^2 tilde{k}_c
+2\sum_{n=1}^\infty e^{-\frac{4\pi^2 n^2}{t}} e^{\frac{4\pi^2 n (1-c)}{t}}
[2(1+T^2 \tilde{k}_c)+\frac{(2\pi n)^2}{t}]
\}
\}
\nonumber\\
&\le&
\frac{
2e^{-t/16}e^{-\frac{\Delta^2+2\delta^2+2\tilde{\theta}^2}{2t}} 
k_c |\frac{s}{\sh(s)}|\ch(|z_0|/2)\ch(p'/2)
}{
(1-K_t)\sqrt{\frac{p}{\sh(p)}\frac{p'}{\sh(p')}}
}
\times\nonumber\\
&\times &
\{ 
\{
\frac{T^2}{2} 
+2 e^{-\frac{4\pi^2 c}{t}}\sum_{n=0}^\infty e^{-\frac{4\pi^2 n^2}{t}} 
[1+\frac{T^2}{2}+\frac{16\pi^2 (n+1)^2}{t}]
\}
\nonumber\\
&+&
4 k_c |\frac{s}{\sh(s)}| \ch(\frac{p+p'}{2})
\{\ 2T^2 tilde{k}_c
+2e^{-\frac{4\pi^2 c}{t}}\sum_{n=0}^\infty e^{-\frac{4\pi^2 n^2}{t}} 
[2(1+T^2\tilde{k}_c)+\frac{(4\pi (n+1))^2}{t}]
\}
\}
\nonumber
\ea
In discussing the behaviour of this function as $p'$ becomes large 
(integrability) we need to 
separate again the regions described by 
a) bounded $\delta$ (i.e. $s\approx\tilde{p}/2\approx p'/2$)
and b) unbounded $\delta$ (i.e. $s/p'$ vanishes as $p'\to\infty$ or $s$ 
is bounded) similar as in the previous section.
In case a) (\ref{3.95}) grows as $\ch(|z_0|/2)\sqrt{p'}^3\approx
\ch(p'/4)\sqrt{p'}^3$ times the Gaussian in $\Delta^2+2\tilde{\theta}^2$.
In case b) it is damped by the Gaussian in $\delta^2$ and is exponentially 
small times the Gaussian in $\Delta^2+2\tilde{\theta}^2$.

Finally looking at the terms inside the two inner curly brackets 
of (\ref{3.95}) we see that they are up to a numerical
factor given by constants of the form $t+K_t(c)$ 
where $K_t(c)$
vanishes exponentially fast as $t\to 0$. We conclude that the integral
over $g'$ of (\ref{3.95}) in the range $0\le \phi\le (1-c)\pi$ results in a 
function of $g$ which is at most exponentially growing with $p$
times a constant that approaches $t$ exponentially fast.\\ 
\\
Case B)\\
Following the by now already familiar trick we will now write
(\ref{3.87}) in terms of $z_0'=z_0-i\pi=s-i(\pi-\phi)=s-i\phi'$
with $0\le\phi'\le c\pi$. This gives (observe that 
$\sh(z_0)=-\sh(z_0'),\ch(z_0)=-\ch(z_0')$)
\ba \label{3.96}
&& <\hat{h}_{AB}>^t_{gg'}\nonumber\\
&=& 
-\frac{
\frac{e^{-t/8}e^{-\frac{p^2+(p')^2}{2t}}}{2\sh(z_0')}
}{
\sqrt{\frac{p}{\sh(p)}(1+K_t(p))\frac{p'}{\sh(p')}(1+K_t(p'))}
}
\times\nonumber\\
&\times &
\{ 
g'_{AB} 
\sum_n 
[(z_0'+T^2 -\pi i (2n-1)) e^{\frac{(z'_0-T^2 -\pi i(2n-1))^2}{t}}
+(z'_0-T^2 -\pi i(2n-1)) e^{\frac{(z'_0+T^2 -\pi i(2n-1))^2}{t}}]
\nonumber\\
&+&
(\tau_j g')_{AB} \frac{\mbox{tr}(\tau_j g'\bar{g}^T)}{2\sh(z'_0)} 
\sum_n
[(z'_0-T^2 -\pi i(2n-1)-2T^2\frac{\ch(z'_0)}{\sh(z'_0)}) 
e^{\frac{(z'_0-T^2 -\pi i (2n-1))^2}{t}}
\nonumber\\
&& -(z'_0+T^2 -\pi i(2n-1)-2T^2\frac{\ch(z'_0)}{\sh(z'_0)}) 
e^{\frac{(z'_0+T^2 -\pi i (2n-1))^2}{t}}]
\}
\nonumber\\
&=& 
-\frac{
\frac{e^{-t/8}e^{-\frac{p^2+(p')^2}{2t}}}{2\sh(z_0')}
}{
\sqrt{\frac{p}{\sh(p)}(1+K_t(p))\frac{p'}{\sh(p')}(1+K_t(p'))}
}
\times\nonumber\\
&\times &
\{ 
g'_{AB} 
\sum_{n=\mbox{odd}} 
[(z_0'+T^2 -\pi i n) e^{\frac{(z'_0-T^2 -\pi i n)^2}{t}}
+(z'_0-T^2 -\pi i n) e^{\frac{(z'_0+T^2 -\pi i n)^2}{t}}]
\nonumber\\
&+&
(\tau_j g')_{AB} \frac{\mbox{tr}(\tau_j g'\bar{g}^T)}{2\sh(z'_0)} 
\sum_{n=\mbox{odd}}
[(z'_0-T^2 -\pi i n-2T^2\frac{\ch(z'_0)}{\sh(z'_0)}) 
e^{\frac{(z'_0-T^2 -\pi i n)^2}{t}}
\nonumber\\
&& -(z'_0+T^2 -\pi i n-2T^2\frac{\ch(z'_0)}{\sh(z'_0)}) 
e^{\frac{(z'_0+T^2 -\pi i n)^2}{t}}]
\}
\ea
With 
\be \label{3.97}
\frac{(z'_0\pm T^2 -\pi i n)^2}{t}=
\frac{(z'_0)^2}{t}+\frac{t}{16}-\frac{\pi^2 n^2}{t}
-\frac{2\pi i n z'_0}{t} \pm(z'_0/2-i\pi n/2)
\ee
and since $i^{-n}=-i^n$ for $n$ odd we find after some pages of algebra
\ba \label{3.98}
&& <\hat{h}_{AB}>^t_{gg'}\nonumber\\
&=& 
-\frac{
\frac{e^{-t/16}e^{-\frac{p^2+(p')^2-2(z_0')^2}{2t}} z_0'}{2\sh(z_0')}
}{
\sqrt{\frac{p}{\sh(p)}(1+K_t(p))\frac{p'}{\sh(p')}(1+K_t(p'))}
}
\times\nonumber\\
&\times &
\{ \{ 
2g'_{AB}\sum_{n=1,\mbox{odd}}^\infty
i^n e^{-\pi^2 n^2/t}
([2i\sh(z_0'/2)\sin(2\pi n z_0'/t)]
\nonumber\\
&&
+[-\pi n \ch(z_0'/2)\frac{\sin(2\pi n z_0'/t)}{2\pi n z_0'/t}]
+[4\pi i n \frac{\sh(z_0'/2)}{z_0'/2}\cos(2\pi n z_0'/t)])
\}
\nonumber\\
&+&
\{
(\tau_j g')_{AB} z_0'\frac{\mbox{tr}(\tau_j g'\bar{g}^T)}{2\sh(z'_0)} 
\sum_{n=1,\mbox{odd}}^\infty i^n e^{-\pi^2 n^2/t}
([-\frac{4\pi i n}{t}(2\ch(z_0'/2)
\nonumber\\
&&
+T^2\frac{\sh(z_0'/2)}{z_0'/2})
\frac{\sin(2\pi n z_0'/t)}{2\pi n z_0'/t}]
\nonumber\\
&& +[2it\frac{\mbox{coth}(z_0')\sin(2\pi n z_0'/t)-\frac{2\pi n}{t} 
\cos(2\pi n z_0'/t)}{(z_0')^2} \ch(z_0'/2)])
\}
\}
\ea
In estimating (\ref{3.98}) the only term that is superficially non-regular
at $z_0'=0$ is the last fraction in the second inner curly 
bracket. However, using the identity
\ba \label{3.99}
K &:=&\frac{\mbox{coth}(z_0')\sin(2\pi n z_0'/t)-\frac{2\pi n}{t} 
\cos(2\pi n z_0'/t)}{(z_0')^2}=\frac{2\pi n}{t}
\times\nonumber\\
&\times& 
\{ \{
[C(z_0')\frac{\sin(\tilde{z}'_0)}{\tilde{z}'_0}-S(z_0')]
+(\frac{2\pi n}{t})^2[s(\tilde{z}'_0)\frac{z'_0}{\sh(z_0')}-c(\tilde{z}_0')]\}
\}
\ea
where $\tilde{z}'_0=2\pi n z_0'/t$
and employing the estimates (\ref{3.48a}), (\ref{3.48b}) 
as well as lemmata 
\ref{la3.1a}, \ref{la3.3}
it is not difficult to show that in the range $0\le\phi'\le c\pi$
\be \label{3.100}
|K|\le 
\frac{16\pi n}{t} [k_{1-c}'+(\frac{2\pi n}{t})^2 k_{1-c}] e^{2\pi^2 n c/t} 
\ee
With these preparations and using previous results we can finish the 
estimate of (\ref{3.98})
\ba \label{3.101}
&& |<\hat{h}_{AB}>^t_{gg'}|\\
&\le& 
\frac{
\frac{e^{-t/16}e^{-\frac{p^2+(p')^2-2\Re((z_0')^2)}{2t}}|z_0'|}{2|\sh(z_0')|} 
}{
(1-K_t)
\sqrt{\frac{p}{\sh(p)}\frac{p'}{\sh(p')}}
}
\times\nonumber\\
&\times &
\{ \{ 
2|g'_{AB}|\sum_{n=1,\mbox{odd}}^\infty
e^{-\pi^2 n^2/t}
([2\sh(|z_0'|/2)|\sin(2\pi n z_0'/t)|]
\nonumber\\
&&
+[\pi n \ch(|z_0'|/2)|\frac{\sin(2\pi n z_0'/t)}{2\pi n z_0'/t}|]
+[4\pi n \frac{\sh(|z_0'|/2)}{|z_0'|/2}|\cos(2\pi n z_0'/t)|])
\}
\nonumber\\
&+&
\{
|(\tau_j g')_{AB}| |z_0'\frac{\mbox{tr}(\tau_j g'\bar{g}^T)}{2\sh(z'_0)}|
\sum_{n=1,\mbox{odd}}^\infty e^{-\pi^2 n^2/t}\times
\nonumber\\
&&\times
([\frac{4\pi n}{t}(2\ch(|z_0'|/2)+T^2\frac{\sh(|z_0'|/2)}{|z_0'|/2})
|\frac{\sin(2\pi n z_0'/t)}{2\pi n z_0'/t}|]
+[2t |K| \ch(|z_0'|/2)])
\}
\}
\nonumber\\
&\le& 
\ch(|z_0'|/2)
\frac{
\frac{e^{-t/16}
e^{-\frac{\Delta^2+2(\tilde{p}^2/4-s^2+(\phi')^2)}{2t}}|z_0'|}{2|\sh(z_0')|} 
}{
(1-K_t)
\sqrt{\frac{p}{\sh(p)}\frac{p'}{\sh(p')}}
}
\times\nonumber\\
&\times &
\{ \{ 
2|g'_{AB}|\sum_{n=1,\mbox{odd}}^\infty
e^{-\pi^2 n^2/t} e^{2\pi^2 n c/t} (2+\pi n + 4\pi n )
\}
\nonumber\\
&+&
\{
|(\tau_j g')_{AB}| |z_0'\frac{\mbox{tr}(\tau_j g'\bar{g}^T)}{2\sh(z'_0)}|
\sum_{n=1,\mbox{odd}}^\infty e^{-\pi^2 n^2/t} e^{2\pi^2 nc/t}
(\frac{4\pi n}{t}[2+T^2]+ 
32\pi n [k_{1-c}'+(\frac{2\pi n}{t})^2 k_{1-c}])
\}
\}
\nonumber\\
&\le& 
\ch(|z_0'|/2)k_{1-c}\frac{s}{\sh(s)}\ch(p'/2)
\frac{
e^{-t/16}
e^{-\frac{\Delta^2+2(\tilde{p}^2/4-s^2+(\phi')^2)}{2t}}
}{
(1-K_t)
\sqrt{\frac{p}{\sh(p)}\frac{p'}{\sh(p')}}
}
\times\nonumber\\
&\times &
\{ \{
2\sum_{n=1,\mbox{odd}}^\infty
e^{-\pi^2 n^2/t} e^{2\pi^2 n c/t} (2+5\pi n)
\}
\nonumber\\
&+&
\{
4 k_{1-c} \frac{s}{\sh(s)} \ch(\frac{p+p'}{2})
\sum_{n=1,\mbox{odd}}^\infty e^{-\pi^2 n^2/t} e^{2\pi^2 nc/t}
(\frac{4\pi n}{t}[2+T^2]+ 
32\pi n [k_{1-c}'+(\frac{2\pi n}{t})^2 k_{1-c}])
\}
\}
\nonumber\\
&\le& 
\ch(|z_0'|/2)k_{1-c}\frac{s}{\sh(s)}\ch(p'/2)
\frac{
e^{-t/16}
e^{-\frac{\Delta^2+2(\tilde{p}^2/4-s^2+(\phi')^2+(1-2c)\pi^2)}{2t}}
}{
(1-K_t)
\sqrt{\frac{p}{\sh(p)}\frac{p'}{\sh(p')}}
}
\times\nonumber\\
&\times &
\{ \{ 
2\sum_{n=0,\mbox{even}}^\infty
e^{-\pi^2 n^2/t}  (2+5\pi (n+1))
\}
\nonumber\\
&+&
\{
4 k_{1-c} \frac{s}{\sh(s)} \ch(\frac{p+p'}{2})
\sum_{n=0,\mbox{even}}^\infty e^{-\pi^2 n^2/t} 
(\frac{4\pi (n+1)}{t}[2+T^2]
\nonumber\\ 
&& +32\pi (n+1) [k_{1-c}'+(\frac{2\pi (n+1)}{t})^2 k_{1-c}])
\}
\}
\nonumber
\ea
where we have used in the last step that for any $n\ge 1$
\be \label{3.102}
-\pi^2 n^2/t+2\pi^2 n c/t=-\pi^2n(n-1)/t-\pi^2n(1-2c)/t
\le -\pi^2(n-1)^2/t-\pi^2(1-2c)/t
\ee
using that $2c<1$ and $(n-1)^2\le n(n-1)$ valid for $n\ge 1$.
Consider now the piece of the argument of the Gaussian given by
\ba \label{3.103}
&&-[\tilde{p}^2/4-s^2+(\phi')^2+(1-2c)\pi^2]
=-[\tilde{p}^2/4-s^2+\phi^2-\tilde{\theta}^2]
-\tilde{\theta}^2+2\pi\phi-2(1-c)\pi^2
\nonumber\\
&\le&
-\delta^2-\tilde{\theta}^2+2\pi\phi-2(1-c)\pi^2
\le -\delta^2-\tilde{\theta}^2+2c\pi^2
=-\delta^2 -(1-d)\tilde{\theta}^2-d\tilde{\theta}^2+2c\pi^2
\nonumber\\
&\le& -\delta^2-(1-d)\tilde{\theta}^2-(d/4-2c)\pi^2
\ea
where in the first inequality we used $\phi'=\pi-\phi$ and (\ref{3.65}),
in the second $\phi\le \pi$ and in the third that $\phi</=/>\pi/2$
iff $\tilde{\theta}</=/>\pi/2$ so that for $2c<1$ we have 
$\tilde{\theta}>\pi/2$. Here $0<d<1$ is an arbitrary real number.
Choosing $c<d/8$, say $c=d/16$ and $d=1/2$ for definiteness we can
complete the estimate (\ref{3.104}) by writing
\ba \label{3.101a}
&& |<\hat{h}_{AB}>^t_{gg'}|\nonumber\\
&\le&
\ch(|z_0'|/2)k_{1-c}\frac{s}{\sh(s)}\ch(p'/2)
\frac{
e^{-t/16}
e^{-\frac{\Delta^2+2\delta^2+\tilde{\theta}^2}{2t}}
}{
(1-K_t)
\sqrt{\frac{p}{\sh(p)}\frac{p'}{\sh(p')}}
}
\times\nonumber\\
&\times &
\{ \{ 
2\sum_{n=0,\mbox{even}}^\infty
e^{-\pi^2 n^2/t}  (2+5\pi (n+1))
\}
\nonumber\\
&+&
\{
4 k_{1-c} \frac{s}{\sh(s)} \ch(\frac{p+p'}{2})
\sum_{n=0,\mbox{even}}^\infty e^{-\pi^2 n^2/t} 
(\frac{4\pi (n+1)}{t}[2+T^2]
\nonumber\\ 
&&+ 32\pi (n+1) [k_{1-c}'+(\frac{2\pi (n+1)}{t})^2 k_{1-c}])
\}
\}
\ea
Comparing (\ref{3.95}) with (\ref{3.101a}) we see that the overall structure
is completely identical, the only essential difference being that 
$2\tilde{\theta}^2$
in the exponent of the Gaussian is replaced by $\tilde{\theta}^2$.
Since now $\tilde{\theta}\ge \pi/2$ we conclude that the integral
of (\ref{3.101}) over $g'$ with respect to $\Omega/t^3$ exists, resulting
in a function of $g$ growing no stronger than exponentially with $p$
times a constant that vanishes exponentially fast with $t\to 0$. \\
\\
\\
Summarizing, as far as the leading order behaviour (in $t$) of the matrix
element of the holonomy operator is concerned, we can replace it by
\ba \label{3.102a}
<\hat{h}_{AB}>^t_{gg'} &\approx& \Theta(\pi(1-c)-\phi)
\frac{
\frac{e^{-t/16}e^{-\frac{p^2+(p')^2-2 z_0^2}{2t}}z_0}{\sh(z_0)}
}{
\sqrt{\frac{p}{\sh(p)}(1+K_t(p))\frac{p'}{\sh(p')}(1+K_t(p'))}
}
\times\nonumber\\
&\times & 
[g'_{AB}\ch(z_0/2)+ 
(\tau_j g')_{AB} \frac{\mbox{tr}(\tau_j g'\bar{g}^T)}{2\sh(z_0)} 
\sh(z_0/2)]
\ea
The Gaussian displays a peak at $g=g'$ where $\tilde{p}/2=p=p',
\tilde{\alpha}=\tilde{\theta}=0$ implying $z_0=p$ whence the prefactor
in front of the square bracket in (\ref{3.102a}) becomes $e^{-t/16}$
and the square bracket itself becomes
\ba \label{3.103a}
[.] &=& g_{AB}\ch(p/2)+ 
(\tau_j g)_{AB} \frac{\mbox{tr}(\tau_j g\bar{g}^T)}{2\sh(p)} 
\sh(p/2)\nonumber\\
&=&
([\ch(p/2)+i\frac{p_j}{p}\sh(p/2)\tau_j]g)_{AB}
=(H(g)^{-1}g)_{AB}=h(g)_{AB}
\ea
as expected. We now write the square bracket as
$$
h(g)_{AB}+
[g'_{AB}\ch(z_0/2)+ (\tau_j g')_{AB} 
\frac{\mbox{tr}(\tau_j g'\bar{g}^T)}{2\sh(z_0)}\sh(z_0/2)-h(g)_{AB}]
$$
and do the Gaussian in $g'$. Notice that the $g'$ independent term is 
given by \\
$h_{AB}(g) <1>^t_{gg'}$ while the integral over the 
additional term will be at 
least of order $t$ since the first contribution to the Gaussian of width
of order $\sqrt{t}$ comes from quadratic terms in $\vec{p}-\vec{p}',
\vec{\theta}-\vec{\theta}'$. This gives rise to the second main theorem.
\begin{Theorem} \label{th3.3}
The matrix elements of the holonomy operators with respect to coherent 
states can be estimated by 
\be \label{3.104}
|\frac{<\psi^t_g,\hat{h}_{AB}\psi^t_{g'}>}{||\psi^t_g||\;||\psi^t_{g'}||}
-h_{AB}(g)\frac{<\psi^t_g,\psi^t_{g'}>}{||\psi^t_g||\;||\psi^t_{g'}||}|
\le t f'(\vec{p},\vec{p}')
\frac{|<\psi^t_g,\psi^t_{g'}>|}{||\psi^t_g||\;||\psi^t_{g'}||}
\ee
where $f'$ is a function of $\vec{p},\vec{p}'$ growing no faster than 
exponentially in either or $p,p'$.
\end{Theorem}
As a corollary to theorem (\ref{th3.2}) we obtain that the expectation value
$<\hat{h}_{AB}>^t_{gg}$ equals $h_{AB}(g)$ up to bounded corrections to 
$h_{AB}(g)$ 
that are proportional to $t$. We will actually
calculate the exact correction in the next section by a different method.
Notice that due to the Gaussian behaviour of the overlap function the 
exponential growth of $f'$ is irrelevant in computing the expectation value
of operator monomials, the corrections of order at least $t$ are always 
integrable.

\subsubsection{Computation of Operator Monomial Expectation Values by a 
Different Method}
\label{s3.1.4}

One can compute the expectation value of operator monomials also 
by a different method which does not rely on the overcompleteness
of the coherent states. To see how this works, notice first of all 
that due to $[\hat{p}_j,\hat{h}_{AB}]=it(\tau_j\hat{h})_{AB}/2$ 
every operator monomial can be reduced to finite linear combinations 
of operator monomials in the following ``standard ordered form'' 
\be \label{3.105}
<(\hat{O}_1..\hat{O}_{m+n})_0>^t_g=
<\hat{p}_{j_1}..\hat{p}_{j_m}\hat{h}_{A_1B_1}..\hat{h}_{A_n B_n}>^t_g
\ee
for some $m,n\ge 0$ and some ordering of the $j_1,..,j_m$.
The idea is now to use the following identity established in 
\cite{32} for general semisimple compact $G$ (we display the case 
$G=SU(2)$), relating the annihilation and holonomy operators, 
\be \label{3.106}
\hat{g}_{AB}=e^{3t/8} (e^{-i\hat{p}_j\tau_j/2}\hat{h})_{AB}
\ee
and to use the eigenvalue property of the coherent states
$\hat{g}_{AB}\psi^t_g=g_{AB}\psi^t_g$. In order to do this we first
have to invert (\ref{3.106}) for $\hat{h}_{AB}$. The naive guess
turns out to be the correct one.
\begin{Theorem} \label{th3.4}
The inversion of (\ref{3.106}) reads
\be \label{3.107}
\hat{h}_{AB}=e^{-3t/8} (e^{i\hat{p}_j\tau_j/2}\hat{g})_{AB}
\ee
\end{Theorem}
The proof of that theorem rests on the following lemma.
\begin{Lemma} \label{la3.6}
Define the strictly positive and self-adjoint operator $\hat{p}$
by
\be \label{3.108}
\hat{p}:=\sqrt{\hat{p}_j\hat{p}_j+\frac{t^2}{4}}
\ee
which commutes with all the $\hat{p}_j$.
Then the following operator identities hold for $G=SU(2)$ (obvious 
generalizations hold for groups of higher rank) :
\ba \label{3.109}
e^{\pm i\hat{p}_j\tau_j/2} &=&
e^{\pm \frac{t}{4}}
\{[\ch(\hat{p}/2) \mp \frac{t}{2}\frac{\sh(\hat{p}/2)}{\hat{p}}]1_2
\pm i\tau_j\hat{p}_j\frac{\sh(\hat{p}/2)}{\hat{p}} \}
\nonumber\\
&& e^{i\hat{p}_j\tau_j/2} e^{-i\hat{p}_j\tau_j/2} =
e^{-i\hat{p}_j\tau_j/2} e^{i\hat{p}_j\tau_j/2} = 1_2 \hat{1}_{{\cal H}}
\ea 
Notice that no operator ordering ambiguities occur in (\ref{3.109}).
\end{Lemma}
Proof of Lemma \ref{la3.6} :\\
By definition (the operator $\hat{p}_j$ is unbounded but the exponential
can be defined by Nelson's analytic vector theorem)
\be \label{3.110}
e^{\pm i\hat{p}_j\tau_j/2}=\sum_{n=0}^\infty \frac{(\pm i/2)^n}{n!}
(\hat{p}_j\tau_j)^n
\ee
Due to the Pauli matrix relation $\tau_j\tau_k=-\delta_{jk} 1_2+
\epsilon_{jkl}\tau_l$ and the commutator relation
\be \label{3.111}
\epsilon_{jkl}\hat{p}_j\hat{p}_k\tau_l=
\frac{1}{2}\epsilon_{jkl}[\hat{p}_j,\hat{p}_k]\tau_l
=-it\frac{1}{2}\epsilon_{jkl}\epsilon_{jkm}\hat{p}_m\tau_l
=-it\hat{p}_j\tau_j
\ee
every power of the matrix valued operator
$\underline{\hat{p}}=\tau_j\hat{p}_j$ can be written as a linear  
combination
\be \label{3.112}
(\underline{\hat{p}})^n=q_n(\hat{x})\underline{\hat{p}} 
+r_n(\hat{x}) 1_2 \hat{1}_{{\cal H}}
\mbox{ where } \hat{x}=-\hat{p}_j\hat{p}_j
\ee
and $q_n,r_n$ are polynomials which are inductively defined by
\be \label{3.113}
(\underline{\hat{p}})^{n+1}=q_{n+1}(\hat{x})\underline{\hat{p}} +
r_{n+1}(\hat{x})1_2 \hat{1}_{{\cal H}}
=[q_n(\hat{x})\underline{\hat{p}}+r_n(\hat{x})1_2 \hat{1}_{{\cal H}}]
\underline{\hat{p}}
\ee
from which we find
\be \label{3.114}
q_{n+1}=-q_n it+r_n,\;r_{n+1}=\hat{x} q_n,\;q_1=1,\;r_1=0
\ee
Notice that no operator ordering problems arise. 

As one can check, the two-dimensional recursion defined in (\ref{3.114}) is 
solved by   
\be \label{3.115}
q_n=\frac{\lambda_+^n-\lambda_-^n}{\lambda_+-\lambda_-},\;
r_n=-\lambda_+ \lambda_- 
\frac{\lambda_+^{n-1}-\lambda_-^{n-1}}{\lambda_+-\lambda_-}
\ee
where 
\be \label{3.116}
\lambda_\pm=-i(\frac{t}{2}\pm\hat{p})
\ee
Inserted back into (\ref{3.110}) gives
\be \label{3.117}
e^{\pm i\underline{\hat{p}}/2}
=\frac{1}{\lambda_+-\lambda_-}
[\underline{\hat{p}}(e^{\pm i \lambda_+/2}-e^{\pm i \lambda_-/2})
-1_2 \hat{1}_{{\cal H}} 
(\lambda_- e^{\pm i \lambda_+/2} - \lambda_+ e^{\pm i \lambda_-/2})]
\ee
and writing out $\lambda_\pm$ results in the first line of (\ref{3.109}).

Using this result and that $[\hat{p}_j,\hat{p}]=0$ we 
easily verify the second line in (\ref{3.109}).\\
$\Box$\\
Remark :\\
In the form (\ref{3.109}) the exponential of $\underline{\hat{p}}$
can directly be defined through the spectral theorem without recourse 
to Nelson's theorem.\\
\\
Proof of Theorem \ref{th3.4} :\\
The proof follows trivially from the second line of (\ref{3.109}).\\
$\Box$\\
Inserting formula (\ref{3.107}) into (\ref{3.105}) does not directly
help us as not all the $\hat{g}_{A_k B_k}$ stand to the right.
However, making use again of a finite number of commutation relations and 
the eigenvalue property we see that we have to leading order in $t$
\be \label{3.118}
<\hat{p}_{j_1}..\hat{p}_{j_m}\hat{h}_{A_1 B_1}..\hat{h}_{A_n B_n}>^t_g
=g_{C_1 B_1}..g_{C_n B_n}
<\hat{p}_{j_1}..\hat{p}_{j_m}
(e^{i\underline{\hat{p}}/2})_{A_1 C_1}..
(e^{i\underline{\hat{p}}/2})_{A_n C_n}>^t_g [1+O(t)]
\ee
Using the explicit expression (\ref{3.109}) for $e^{i\underline{\hat{p}}/2}$ 
and $[\hat{p}_j,\hat{p}]=0$ we see that we can compute the leading order of 
(\ref{3.118}) once we know all expectation values of the form
\be \label{3.119}
<\hat{p}_{j_1}..\hat{p}_{j_m} f(\hat{p}^2)>^t_g
=<f(\hat{p}^2)\hat{p}_{j_1}..\hat{p}_{j_m}>^t_g
\ee
where $f$ is an arbitrary analytical function of $\hat{p}^2$.
If $f$ would be at most a polynomial in $\hat{p}^2$ then it would suffice to 
know all the matrix elements of the form 
\be \label{3.120}
<\hat{p}_{j_1}..\hat{p}_{j_m}>^t_g
\ee
however, the functions of $\hat{p}^2$ that appear in (\ref{3.109}) are not
simply polynomials. In what follows we will show that (\ref{3.119})
can be computed once we know $<f(\hat{p}^2)>^t_g$. 
The latter can then be computed by an appeal to the solution of the moment 
problem by Hamburger.

Recall that 
\be \label{3.121}
\hat{p}_j\psi^t_g(h)=it/2(d/ds)_{s=0}\psi^t_g(e^{s\tau_j}h)
=it/2(d/ds)_{s=0}\psi^t_{e^{-s\tau_j}g}(h)
\ee
and since $\hat{p}_j$ is self-adjoint 
\be \label{3.122}
<\hat{p}_{j_1}..\hat{p}_{j_m} f(\hat{p}^2)>^t_g
=(-\frac{it}{2})^m[(\frac{d}{ds_m})_{s_m=0}..[(\frac{d}{ds_1})_{s_1=0}
\frac{<\psi^t_{e^{-s_1\tau_{j_1}}..e^{-s_m\tau_{j_m}}g},
f(\hat{p}^2)\psi^t_g>}{||\psi^t_g||^2}]..]
\ee
Let $g'=e^{-s_1\tau_{j_1}}..e^{-s_m\tau_{j_m}}g$, then since 
$\hat{p}\pi_{jmn}(h)=t(j+1/2)\pi_{jmn}(h)$ we have with 
$f(.)=F\circ\sqrt{(.)}$
\ba \label{3.123}
<\psi^t_{g'},F(\hat{p})\psi^t_g>
&=& \sum_j d_j e^{-t(d_j^2-1)/4} F(t d_j/2) \chi_j(g'\bar{g}^T) 
\nonumber\\
&=& \frac{1}{2\sh(z)}
\sum_j d_j e^{-t(d_j^2-1)/4} F(t d_j/2) [e^{z d_j}-e^{-z d_j}] 
\nonumber\\
&=& \frac{e^{t/4}}{2\sh(z)T}
\sum_{n\in\Zl}  e^{-(nT)^2)/4} F((nT)T/2) e^{(nT)\frac{z}{T}} 
\ea
where $\ch(z)=\mbox{tr}(g'\bar{g}^T)/2$ and $T=\sqrt{t}$. Applying the 
Poisson summation formula to (\ref{3.123}) we find
\be \label{3.124}
<\psi^t_{g'},F(\hat{p})\psi^t_g>
=\frac{e^{t/4}{2\sh(z}T^2}
\sum_{n\in\Zl}  g_n(z)
\ee
where 
\be \label{3.125}
g_n(z)=[\int_\Rl dx e^{-ikx} g(x)]_{k=2\pi n/T}
=\int_\Rl dx \; e^{-x^2/4}\; F(xT/2)\; x\; e^{x\frac{z-2\pi i n}{T}} 
\ee
Let $\ch(z_0)=\mbox{tr}(g\bar{g}^T)/2$ and define  
$z=z(s_1,..,s_m),\;z_k=z(s_1,.,s_k,0,..,0),\;k=0,..,m$. Let $G(z)$ be any 
function of $z$, then
\ba \label{3.126}
&& [(\frac{d}{ds_m})_{s_m=0}..[(\frac{d}{ds_1})_{s_1=0} G(z)]..]
=[(\frac{d}{ds_m})_{s_m=0}..[(\frac{d}{ds_2})_{s_2=0} 
(\frac{dz}{ds_1})_{s_1=0} G^{(1)}(z_{m-1})]..]
\nonumber\\
&=& 
[(\frac{d}{ds_m})_{s_m=0}..[(\frac{d}{ds_3})_{s_3=0} 
\{(\frac{d^2z}{ds_1 ds_2})_{s_1=s_2=0} G^{(1)}(z_{m-2})
\nonumber\\
&& +(\frac{dz}{ds_1})_{s_1=s_2=0} (\frac{dz}{ds_2})_{s_1=s_2=0} 
G^{(2)}(z_{m-2})\}]..]
\nonumber\\
&=&...\nonumber\\ 
&=&
(\frac{d^m z}{ds_1.. ds_m})_{s_1=..=s_m=0} G^{(1)}(z_0)
+...+(\frac{dz}{ds_1})_{s_1=..=s_m=0}..(\frac{dz}{ds_m})_{s_1=..=s_m=0} 
G^{(m)}(z_0)
\ea
Applied to $G(z)=<\psi^t_{g'},F(\hat{p})\psi^t_g>$ we infer that the  
derivatives of $g_n(z)/\sh(z)$ at $z_0$ of all orders $k$ between $1$ and $m$ 
appear in (\ref{3.122}) with coefficients that involve sums, 
over all partitions of $m=l_1+..+l_k,\;l_j\ge 1$, of products of 
$k$ factors of the form
\be \label{3.127}
(\frac{d^l z}{ds_{I_1}..ds_{I_l}})_{s_1=..=s_m=0},\;1\le I_1,..,I_l\le m
\mbox{ mutually disjoint}
\ee
Performing the $k-$th derivative of $g_n(z)$ we obtain
\be \label{3.128}
g^{(k)}_n(z_0)=\frac{1}{T^k}
\int_\Rl dx \; e^{-x^2/4}\; F(xT/2)\; x^{k+1}\; e^{x\frac{z_0-2\pi i n}{T}} 
\ee
The idea is now to do integrations by parts until only $x$ appears instead 
of $x^{k+1}$ using $x e^{-x^2/4}=-2 (e^{-x^2/4})'$. There are no boundary 
terms due to the Gaussian. Each time the derivative hits 
$e^{x\frac{z-2\pi i n}{T}}$ it brings down a factor of 
$\frac{z-2\pi i n}{T}$ while hitting $x^l F(xT/2)$ produces
a non-negative power of $T$. We conclude that 
\ba \label{3.128a}
g^{(k)}_n(z_0)&=&\frac{(2[z_0-2\pi i n])^k}{T^{2k}}
[\int_\Rl dx \; e^{-x^2/4}\; F(xT/2)\; x\; e^{x\frac{z_0-2\pi i n}{T}}]
(1+O(T))\nonumber\\
& =& \frac{(2[z_0-2\pi i n])^k}{T^{2k}}g_n(z_0)
(1+O(T))
\ea
Since in (\ref{3.122}) we multiply with $T^{2m}$ and 
since the derivatives of $z$ at $s_1=..=s_m=0$ are independent
of $t$ we see that to leading order in $t$ we only need to keep the term
with $k=m$. The same argument reveals that we do not need to take into
account the derivatives of $1/\sh(z)$.

Next, by a substitution and a contour argument we obtain 
(at least for functions $F$ integrable against the Gaussian)
\be \label{3.129}
g_n(z_0)=e^{4\frac{(z_0-2\pi i n)^2}{t}}
\int_\Rl dx \; e^{-x^2/4}\; F(T(x+\frac{z_0-2\pi i n}{T})/2)\; 
(x+\frac{z_0-2\pi i n}{T})
\ee
By our assumption on $F$ the integral exists and 
by the already familiar argument, the $e^{z_0^2}$ in (\ref{3.129})
is controlled by
the $e^{-p^2/t}$ coming from the denominator $||\psi^t_g||^2$ in
(\ref{3.119}) so that (\ref{3.129}) is exponentially suppressed 
with $t\to 0$ for any $n\not=0$. 

Putting everything together we therefore have to leading order in $T$
\ba \label{3.130}
<\hat{p}_{j_1}..\hat{p}_{j_m} f(\hat{p}^2)>^t_g
&=&(-it/2)^m (2z_0/t)^m [\prod_{l=1}^m (\frac{dz}{ds_l})_{s_1=..=s_m=0}]
\frac{\frac{e^{t/4}}{2\sh(z_0)T^2} \sum_{n\in\Zl}  
g_n(z_0)](1+O(T))}{||\psi^t_g||^2}
\nonumber\\
&=& [\prod_{l=1}^m (-iz_0 \frac{dz}{ds_l})_{s_1=..=s_m=0}]
<f(\hat{p}^2)>^t_g(1+O(T))
\ea
Now by the method of section \ref{s3.1.2} we find 
\be \label{3.131}
z(s_1,..,s_m)=p+i\frac{1}{p}\sum_{l=1}^m s_l p_{j_l}+O(s^2)
\ee
and we obtain the desired result
\be \label{3.131a}
<\hat{p}_{j_1}..\hat{p}_{j_m} f(\hat{p}^2)>^t_g
= p_{j_1}(g)..p_{j_m}(g) <f(\hat{p}^2)>^t_g(1+O(T))
\ee
It therefore remains to compute the expectation value 
$<f(\hat{p}^2)>^t_g$ where $f$ for the purpose of computing
(\ref{3.105}) can be chosen analytical in $\hat{O}:=\hat{p}^2$ and at most 
exponentially growing with $\hat{p}$. Consider first the expectation values 
of the powers $\hat{O}^n$. Since they are of the form (\ref{3.131}) with
the choice $f=1$ we immediately find
\be \label{3.131b}
\lim_{t\to 0} <\hat{O}^n>^t_g = O(g)^n=(p_j(g)p_j(g))^n
\ee
The assertion $\lim_{t\to 0} <f(\hat{p}^2)>^t_g=f(p_j(g)p_j(g))$ follows
therefore immediately from the solution of the moment problem due to
Hamburger which is the subject of the next subsection.\\
\\
Remark :\\
Obviously, since we can compute commutators of polynomial operators
and express it in terms of elementary operators again, the Ehrenfest
theorem to first order in $t$ for such operators is trivially satisfied 
because the operator algebra of elementary operators precisely mirrors 
the classical Poisson algebra.

\subsection{Expectation Values of Non-Polynomial Operators and the Moment 
Problem due to Hamburger}
\label{s3.2}

Recall the following theorem (see, e.g. \cite{47})
\begin{Theorem}[Hamburger] \label{th3.5}
Let be given a sequence of real numbers $a_n\in \Rl,\,n=0,1,2,..$. 
A necessary and sufficient criterion for the existence of a positive,
finite measure $d\rho(x)$ on $\Rl$ such that the $a_n$ are its moments, 
that is,
\be \label{3.132}
a_n=\int_\Rl d\rho(x) x^n
\ee
is that for any natural number $0\le N<\infty$ and arbitrary complex numbers
$z_k,\;k=0,..,N$ it holds that
\be \label{3.133}
\sum_{k,l=0}^N \bar{z}_k z_l a_{n+m}\ge 0
\ee
The measure is faithful if equality in (\ref{3.133}) occurs only
for $z_k=0$. Moreover, if there exist
 constants $\alpha,\beta>0$ such that
$|a_n|\le \alpha \beta^n (n!)$ for all $n$, then the measure $\rho$ is 
unique.
\end{Theorem}
Necessity is easy to see by considering the $L_2$ norm of the functions
$\sum_{k=0}^N z_k x^k$. Sufficiency follows from the spectral theorem
and uniqueness can be established by an appeal to Nelson's analytic vector
theorem.

In this section we assume that all operators under consideration are  
densely defined on a common domain which they together with arbitrary powers
leave invariant. We are then able to prove the following theorem. 
\begin{Theorem} \label{th3.6}
Let $\hat{O}$ be a self-adjoint operator $\hat{O}$ 
on ${\cal H}_\gamma$ for some $\gamma\in \Gamma^\omega_0$ built from
$\hat{p}^e_j,\hat{h}_e,\;e\in E(\gamma)$, that is, 
$\hat{O}=O(\{\hat{p}_e,\hat{h}_e\}_{e\in E(\gamma)}$).
Let $O(\vec{g})=O(\{p_e(g_e),h_e(g_e)\}_{e\in E(\gamma)})$ be its 
real valued classical counterpart and suppose 
that for every $n\in \Nl$
\be \label{3.134}
\lim_{t\to 0}<\hat{O}^n>^t_{\gamma,\vec{g}}=O(\vec{g})^n 
\ee
Then for any Borel measurable function $f$ on $\Rl$ such that 
$<f(\hat{O})^\dagger f(\hat{O})>^t_g<\infty$ we have 
\be \label{3.135}
\lim_{t\to 0}<f(\hat{O})>^t_{\gamma,\vec{g}}=f(O(\vec{g}))
\ee
\end{Theorem}
Proof of Theorem \ref{th3.6} :\\
Let $E(x),\;x\in \Rl$ be the spectral projections of 
$\hat{O}$. Then, by assumption and the spectral theorem 
\be \label{3.136}
\lim_{t\to 0} \int_\Rl 
d<\xi^t_{\gamma,\vec{g}},E(x)\xi^t_{\gamma,\vec{g}}> x^n
=O(\vec{g})^n
\ee
where $\xi^t_{\vec{g}}=\psi^t_{\vec{g}}/||\psi^t_{\vec{g}}||$.
Now $a_n:=O(\vec{g})^n$ obviously satisfies all the criteria of theorem
\ref{th3.5} and we conclude that there exists a measure 
$d\rho_{\vec{g}}(x)$ on $\Rl$ such that 
\be \label{3.137}
\int_\Rl d\rho_{\vec{g}}(x) x^n=O(\vec{g})^n
\ee
The Dirac measure $d\rho_{\vec{g}}(x)=\delta_\Rl(x,O(\vec{g})) dx$
obviously satisfies (\ref{3.137}) and choosing $\alpha=1,
\beta=|O(\vec{g})|$ in theorem \ref{th3.5} obviously satisfies the 
uniqueness part of the criterion. Thus, the Dirac measure is in fact
the unique solution to our moment problem. Thus the spectral measure
$d\rho^t_{\vec{g}}(x):=
d<\xi^t_{\gamma,\vec{g}},E(x)\xi^t_{\gamma,\vec{g}}>$
approaches the Dirac $\delta$ distribution when evaluated on monomials
$x^n$. It follows that the support of $\rho^t_{\vec{g}}$ gets confined
to $\{x_0\}$ as $t\to 0$ by definiton of the Lebesgue integral.

Now for any function $f$ satisfying the assumptions of the theorem
the spectral theorem applies and we have 
\be \label{3.138}
<f(\hat{O})>^t_{\vec{g}}=\int_\Rl d\rho^t_{\vec{g}}(x) f(x)
\ee
Thus, since $f$ is in particular measurable, the limit of both sides 
of (\ref{3.138}) turns into (\ref{3.135}). \\
$\Box$\\
\begin{Corollary} \label{col3.1}
Let $\hat{O}_1,..,\hat{O}_m$ be self-adjoint, not necessarily commuting,
operators such that
\be \label{3.140}
\lim_{t\to 0} <\prod_{k=1}^m \hat{O}_k^{n_k}>^t_{\gamma,\vec{g}}
=\prod_{k=1}^m O_k(\vec{g})^{n_k}
\ee
Then for any Borel measurable function $f$ on $\Rl^m$
\be \label{3.141}
\lim_{t\to 0} <f(\{\hat{O}_k\}_{k=1}^m)>^t_{\gamma,\vec{g}}
=f(\{O_k(\vec{g})\}_{k=1}^m\})
\ee
\end{Corollary}
Proof of Corollary \ref{col3.1} :\\
By the spectral theorem
\be \label{3.142}
\lim_{t\to 0} \int_{\Rl^m} d^m<E_1(x_1)..E(x_m)>^t_{\vec{g}}
\prod_{k=1}^m x_k^{n_k}=\prod_{k=1}^m O_k(\vec{g})^{n_k}
\ee
where $E_k(x)$ is the family of spectral projections of $\hat{O}_k$. 
Thus, by the unique solution to the moment problem the measure 
in (\ref{3.142}) approaches the product Dirac measure  
$d^m x \prod_k \delta_{\Rl}(x_k,O_k(\vec{g}))$ similar as in 
theorem \ref{th3.6}.\\
$\Box$\\
Next we turn to commutators.
\begin{Theorem} \label{th3.7}
Suppose that $\hat{O}_1,\hat{O}_2$ are self-adjoint
operators satisfying the assumptions of corollary \ref{col3.1}.
Suppose, moreover, that $\hat{O}_1$ is positive semi-definite and that
\be \label{3.143}
\lim_{t\to 0} \frac{<[\hat{O}_1,\hat{O}_2]>^t_g}{it}=\{O_1,O_2\}(\vec{g})
\ee
Then for any real number $r$
\be \label{3.144}
\lim_{t\to 0} \frac{<[(\hat{O}_1)^r,\hat{O}_2]>^t_g}{it}=
\{(O_1)^r,O_2\}(\vec{g})
\ee
\end{Theorem}
Proof of Theorem \ref{th3.7} :\\
It suffices to prove the theorem for rational $r=m/n$ with $m,n$ integers 
and $n>0$. We have the identity 
\ba \label{3.145}
\frac{[\hat{O}_1^m,\hat{O}_2]}{it}
&=&\sum_{k=1}^{m}
\hat{O}_1^{k-1}\frac{[\hat{O}_1,\hat{O}_2]}{it}\hat{O}_1^{m-k}
\nonumber\\
&=&\sum_{k=1}^{n}
\hat{O}_1^{m(k-1)/n}\frac{[\hat{O}_1^r,\hat{O}_2]}{it}\hat{O}_1^{m(n-k)/n}
\ea
Now for any measurable function $f$ we have by assumption and completeness 
relation 
\ba \label{3.146}
\lim_{t\to 0} <f(\hat{O}_1)f(\hat{O}_1)>^t_g
& = & \lim_{t\to 0} 
(\frac{2}{\pi t^3})^N\int_{(G^\Co)^N} d^N\Omega(\vec{g}') 
<f(\hat{O}_1)>^t_{\vec{g}\vec{g}'} <f(\hat{O}_1)>^t_{\vec{g}'\vec{g}}
\nonumber\\
& = &\lim_{t\to 0} 
<f(\hat{O}_1)>^t_{\vec{g}\vec{g}} <f(\hat{O}_1)>^t_{\vec{g}\vec{g}}
\ea
meaning that $(\frac{2}{\pi t^3})^N |<f(\hat{O}_1)>^t_{\vec{g}\vec{g}'}|^2$
approaches a delta distribution times $(<f(\hat{O}_1)>^t_{\vec{g}})^2$, 
for any $f$, with respect to $\Omega^N$ as $t\to 0$ where $N=|E(\gamma)|$.
It follows that $<f(\hat{O}_1)>^t_{gg'}$ is concentrated at $g=g'$
as explicitly displayed in sections \ref{s3.1.2}, \ref{s3.1.3} and we 
therefore find for the expectation value of (\ref{3.145}) by using 
again the completeness relation in a similar fashion
\be \label{3.147}
m\lim_{t\to 0}
<\hat{O}_1^{m-1}>^t_{\vec{g}} <\frac{[\hat{O}_1,\hat{O}_2]}{it}>^t_{\vec{g}}
=n\lim_{t\to 0}
<\hat{O}_1^{m(n-1)/n}>^t_{\vec{g}} 
<\frac{[\hat{O}_1^r,\hat{O}_2]}{it}>^t_{\vec{g}}
\ee
Using the assumptions of the theorem we thus find
\be \label{3.148}
\lim_{t\to 0}
<\frac{[\hat{O}_1^r,\hat{O}_2]}{it}>^t_g
=\frac{m}{n}             
O_1(\vec{g})^{\frac{m}{n}-1} \{O_1,O_2\}(\vec{g})=\{O_1^r,O_2\}(\vec{g})
\ee
as claimed.\\
$\Box$\\
\\
The application of these theorems concerns operators which are 
not polynomials of the elementary ones. Such operators occur in quantum 
general relativity
where diffeomorphism invariance requires that Hamiltonian constraint 
operators
are density one valued and therefore free of UV singularities. This enforces 
that non-analytic functions, specifically roots of the volume operator 
\cite{12,14,15,16}, appear. The spectral measure of this operator 
is not explicitly known and therefore a direct computation of its 
expectation values and its commutators with holonomy operators 
that appear in \cite{17,18,19,20,22} is a hopeless task. Theorems 
\ref{th3.6}, \ref{th3.7} and corollary \ref{col3.1} circumvent this 
problem at least as far as the leading order behaviour of expectation
values is concerned by using the following trick : The fourth power of the 
volume operator $\hat{O}:=\hat{V}^4$ is in fact a polynomial of the 
$p^e_j$ and thus the expectation values of $\hat{O}^n$ and the commutators  
with holonomy operators can be straightforwardly computed, leading to 
the expected result. Defining then $\hat{V}:=\hat{O}^{1/4}$ and using 
the above results shows that the Hamiltonian constraint indeed has the 
correct classical limit. Details will appear elsewhere \cite{36f}.\\
\\
\\
\\
{\large Acknowledgements}\\
\\
O. W. thanks the Studienstiftung des Deutschen Volkes for financial 
support.

\begin{appendix}

\section{The $U(1)$ case}
\label{sa}

In this appendix we will apply the results of this paper to the case of 
$U(1)$ as the gauge group. As will become clear, the much simpler 
structure of $U(1)$ leads to a considerable simplification of the 
derivation of all the results. The main reason for this is, of course, 
the fact that $U(1)$ is Abelian and as a consequence of this that all 
its irreducible representations are one-dimensional. This means that 
one has to deal with numbers only, instead of matrices.
\\

\subsection{Expectation Values of the Momentum Operator}
\label{sa.1}

We will first show that the expectation value of the momentum operator 
with respect to the $U(1)$ coherent states has the proper 
(semi-)classical limit. Recall from \cite{32} the form of the 
coherent states for $U(1)$:
\be \label{a1.1}
 \psi^t_g (h) = \sum_n e^{- \frac{t}{2} n^2} (g h^{-1} )^n 
\ee
with $g= e^p e^{i\theta_0} $ and $h= e^{i \theta }$. 
By the same token as in section \ref{s3.1.4}, the expectation value of 
$\hat{p}$ with respect to these states is given by :
\be \label{a1.2}
 \frac{\langle \psi^t_g , \hat{p} \, \psi^t_g \rangle }{ \| \psi^t_g
\|^2 } 
= it        (\frac{d}{dr})_{r=0} \frac{\sum_n e^{-t n^2} (e^{-ir} e^{2p} 
)^n}{\sum_n e^{-t n^2} e^{2np}} = \frac{\sum_n e^{-t n^2} nt
\, e^{2np}}
{\sum_n e^{-t n^2} e^{2np}} 
\ee

To see the behaviour of this expression for $t \to 0$, we have to perform 
a Poisson transformation. For the denominator this has already been done 
in the appendix of \cite{32}, the result being
\be \label{a1.3}
\sum_n e^{-t n^2} e^{2np} = \sqrt{\pi} e^{\frac{p^2}{t}} 
\sum_n e^{-\frac{\pi^2 n^2 + 2i\pi np}{t}}.
\ee
The transformation for the numerator, however, has to be calculated
anew. 
With $s=\sqrt{t}$, as usual, we get
\ba \label{a1.4}
\tilde{f} (k) &=& s \int dx \: x \, e^{-x^2} e^{(\frac{2p}{s} - ik)x}
\nonumber \\
	     &=& s \, e^{\frac{(\frac{2p}{s} -ik)}{4}} \int dx \: 
(x - \frac{ik- 2p/s }{2}) \, e^{-x^2} \nonumber\\
	     &=& \sqrt{\pi} s \, \frac{(2p/s -ik)}{2} \,
e^{\frac{(\frac{2p}{s} 
-ik)}{4}}.
\ea
Inserting this into (\ref{a1.2}) yields
\ba \label{a1.5}
    \frac{\langle \psi^t_g , \hat{p} \, \psi^t_g \rangle }{ \|
\psi^t_g \|^2 }
      & =& \frac{s \sum_n (p - \pi n i)/s \: e^{\frac{(2p - 2\pi n    
     i)^2}{4t}}}{e^{p^2/t} \sum_n e^{-\frac{\pi^2 n^2 + 2                 
	    \pi i n p}{t}}} \nonumber \\
	     &=& \frac{p  \sum_n e^{-\frac{\pi^2 n^2 + 2\pi i n p}{t}}    
			 - 2\pi i  \sum_n n \, e^{-\frac{\pi^2 n^2 +
2\pi i  n p}{t}}}{ \sum_n e^{-\frac{\pi^2 n^2 + 2\pi i n p }{t}}}
\nonumber \\
		      &=& p - \frac{2\pi i  \sum_n n \,  
e^{-\frac{\pi^2 n^2 + 2\pi i n p}{t}}}{ \sum_n 
e^{-\frac{\pi^2 n^2 + 2\pi i n p}{t}}}.
\ea
and this result makes it immediately obvious that
\be \label{a1.6}
\lim{t\to 0} \frac{\langle \psi^t_g , \hat{p} \, \psi^t_g \rangle }
     { \| \psi^t_g \|^2 }  = p
\ee   

Our next task is to generalize this calculation to an arbitrary integer 
power of the momentum operator. We thus have
\be \label{a1.7}
 \frac{\langle \psi^t_g , \hat{p}^m \psi^t_g \rangle }{ \| \psi^t_g \|^2 }
 = \frac{\sum_n e^{-t n^2}(nt)^m e^{2np}}{\sum_n e^{-t n^2} e^{2np}}
\ee
which now has to be Poisson transformed again. The transform for the 
denominator stays the same, so we can concentrate on the numerator.
As the 
general steps are the same as above we will be more concise here. The 
Poisson transform is given by
\ba \label{a1.8}
\tilde{f}(k) &=& s^m e^{\frac{(ik - 2p/s)^2}{4}} \int dy \: e^{-y^2} 
      (y+(p/s - ik/2    ))^m \nonumber \\
	     &=& s^m e^{\frac{(ik - 2p/s)^2}{4}} \int dy \: e^{-y^2} 
  \sum_{l=0}^m y^{m-l} (p/s - ik/2)^l {m \choose l}
\ea
As we are only interested in the limit $t \to 0$ and due to the prefactor 
$s^m$ the only surviving term will be the $l=m$ term. We therefore have
\ba \label{a1.9}
\lim_{t \to 0} \tilde{f}(k) &=& s^m e^{\frac{(ik - 2p/s)^2}{4}} \int
dy \,  e^{-y^2} (p/s - ik/2)^m \nonumber \\
		 &=& \sqrt{\pi} (p - i \pi n)^m e^{ \frac{(2i\pi n -2p)^2}
	 {4t}}
\ea
where we already substituted the $k$ variable by $2 \pi n/\sqrt{t}$. This 
expression now has to be put back into (\ref{a1.7}) which yields
\ba \label{a1.10} 
	\lim_{t\to 0}  \frac{\langle \psi^t_g , \hat{p}^m \psi^t_g 
\rangle }{ \| \psi^t_g \|^2 } & =&  \lim_{t\to 0} \frac{ 
\frac{2\pi}{\sqrt{t}} \sqrt{\pi} e^{p^2/t} \sum_n (p - i\pi n)^m 
e^{- \frac{\pi^2 n^2 + 2\pi i n p}{t}}}{\frac{2\pi}{\sqrt{t}} \sqrt{\pi}  
	     e^{p^2/t} \sum_n e^{- \frac{\pi^2 n^2 + 2\pi i n p}{t}}}
\nonumber \\
	     &=&  p^m \: ,
\ea
as the only term in the sum, which survives in the limit, is the one
with 
$n=0$.   

Although these results are quite satisfying, one often encounters other 
powers of the momentum operators, especially square and higher roots, so 
it would be reassuring to know that they, too, have the expected 
semiclassical behaviour. A direct calculation as performed above becomes 
quite difficult for roots of arbitrary polynomials of $\hat{p}_e$ 
where $e$ labels the edges of a graph (for one edge and, say, 
$\sqrt{|\hat{p}|}$ the computational effort is still low and is left to the 
reader as an exercise) so we have to resort to other methods. A clue comes 
from 
reformulating the expectation value for integer powers of $\hat{p}$:
\ba \label{a1.11}
\lim_{t\to 0}  \frac{\langle \psi^t_g , \hat{p}^m \psi^t_g \rangle }{ \|  
      \psi^t_g \|^2 } & =&  \lim_{t\to 0} \frac{\sum_n \langle \psi^t_g , 
  \hat{p}^m n\rangle \langle n, \psi^t_g \rangle }{ \| \psi^t_g \|^2}
\nonumber \\
		     &=&  \lim_{t\to 0} \frac{\sum_n (nt)^m | \langle n, 
	 \psi^t_g \rangle |^2 }{ \| \psi^t_g \|^2} \nonumber \\
		  &=& p^m
\ea
where the first two lines are an expansion in terms of $| n\rangle$, the 
basis consisting of eigenvectors of $\hat{p}$ - recall, that $U(1)$ 
momenta have discrete spectrum - , and the last line follows from our 
calculations above. This suggests that $\langle n,
\psi^t_g \rangle |^2/ \| \psi^t_g \|^2 $ approaches - in the sense of 
distributions - just $\delta_{n,p/t} $, an observation that receives 
additional support from the explicit form of  $\lim_{t\to 0} | \langle n,
\psi^t_g \rangle |^2/ \| \psi^t_g \|^2 $ that was calculated in
\cite{35}. 
That this also holds in a rigorous sense is guaranteed by the solution to 
the moment problem by Hamburger as quoted in the main text. 
In our case the 
$a_n$ are given by the $p^n$, therefore $a_{n+m} = a_n a_m $ and 
thus the condition of the theorem is obviously satisfied. We can 
therefore conclude that our results for the integer powers of the 
momentum operator indeed determine  $| \langle n,\psi^t_g 
\rangle |^2/ \| \psi^t_g \|^2 $ to approach $\delta_{n,p/t} $. 
This important 
result will considerably simplify the calculations in the following 
subsections. We now come back to the problem of the roots of the momentum 
operator. Let $m$ be an odd integer. Then
\ba \label{a1.13}
   \lim_{t\to 0}  \frac{\langle \psi^t_g , \hat{p}^{\frac{m}{2}} \psi^t_g 
\rangle }{ \| \psi^t_g \|^2 } & =&  \lim_{t\to 0} \frac{\sum_n \langle 
\psi^t_g , \hat{p}^{\frac{m}{2}} n\rangle \langle n, \psi^t_g \rangle }
{ \| \psi^t_g \|^2} \nonumber \\
		     &=&  \lim_{t\to 0} \frac{\sum_n (nt)^{\frac{m}{2}} | 
\langle n, \psi^t_g \rangle |^2 }{ \| \psi^t_g \|^2} \nonumber \\
		  &=& p^{\frac{m}{2}}
\ea
where the last equality is now justified by the aforementioned theorem.

\subsection{Expectation Values of the Holonomy Operator}
\label{sa.2}

In this subsection we will compute the semiclassical limit of expectation 
values of (powers of) the configuration operator $\hat{h}$. We can 
basically reduce this case to the one in the last subsection by the 
useful observation that $\hat{h} = e^{- 1/2 t} e^{- \hat{p}} \hat{g} $, 
see \cite{35}. For $m$ integer or half-integer we get
\ba \label{a2.1}
      \lim_{t\to 0}  \frac{\langle \psi^t_g , \hat{h}^m \psi^t_g \rangle }
{ \|\psi^t_g \|^2 } & =&  \lim_{t\to 0} \frac{e^{-m/2 t}  \langle         
	   \psi^t_g , e^{-\hat{p}} \hat{g} \ldots  e^{-\hat{p}}
\hat{g} \,  \psi^t_g \rangle }{ \| \psi^t_g \|^2} \nonumber \\
	      &=&  \lim_{t\to 0} \frac{e^{-m/2 t} \langle \psi^t_g , 
e^{- m\hat{p}}\hat{g}^m  \psi^t_g \rangle }{ \| \psi^t_g \|^2}
\nonumber \\
	      &=&  \lim_{t\to 0} \frac{e^{-m/2 t} \langle \psi^t_g , 
e^{- m\hat{p}}  \psi^t_g \rangle g^m}{ \| \psi^t_g \|^2} \nonumber \\
	      &=& \lim_{t\to 0} e^{-m/2 t} e^{-mp} g^m \nonumber \\
	     &=& h^m
\ea
where we used in line two that all remaining commutator terms are at 
least of order $t$ and therefore vanish in the limit $t\to 0$, and in  
line three that our coherent states are eigenstates of $\hat{g}$. It 
should be  obvious from this that arbitrary mixed polynomials in
$\hat{p}$ and $\hat{h}$ can be treated equivalently, to leading
order in $t$.

\subsection{Expectation Values of Commutators}
\label{sa.3}

In this subsection we intend to obtain the semiclassical limit of 
expectation values of commutator terms by direct computation. The main 
example we have in mind 
here is $\hat{h}^{-1} [ \sqrt{\hat{V}} , \hat{h} ]$ which plays an 
important role in the Hamiltonian constraint operator constructed in 
\cite{17,18}. Here $\hat{V}$ denotes the volume operator. As this
requires  rather tedious calculations due to its structure, requiring
at least a 
graph with three-valent vertices (in the gauge-variant case), we will
restrict ourselves to the following case : we would like to check that 
$\hat{h}^{-1} [ \sqrt{\hat{p}} , \hat{h} ] /(it)$ has the right
semiclassical
limit, i.e. reproduces $h^{-1} \{ \sqrt{ | p|} , h\} $ which is 
$i \mbox{sgn}(p)/ (2 \sqrt{ |p|} )$.
We start with the observation that
\be \label{a3.1}
\hat{h} \, \psi^t_g (h) = \sum_n e^{-n^2t/2} g^n h^{-n+1}
\ee
We then have
\ba \label{a3.2}
 \lefteqn{   \frac{\langle \psi^t_g , \hat{h}^{-1} [ \sqrt{|\hat{p}
|} , \hat{h} ]
/(it)  \, \psi^t_g \rangle }{ \| \psi^t_g \|^2 }  } \nonumber \\
 &=& -i\frac{\sum_n e^{- n^2 t} e^{2pn} ( \sqrt{| (n-1) \, t|} -
\sqrt{|nt|} )/t }{\sum_n e^{-n^2 t} e^{2pn}} \nonumber \\
    &=& -i\frac{\frac{1}{\sqrt{\pi}} \sum_n \int_{-\infty}^{\infty} dx 
\frac{( \sqrt{| x-s|} - \sqrt{|x|})}{s^{3/2}} e^{-x^2} 
e^{(2p/s -2i\pi n/s )x} }{e^{\frac{p^2}{s^2}} \sum_n 
e^{-\frac{\pi^2 n^2 + 2i\pi n p}{s^2}}} \nonumber \\
	 &=&  \frac{\frac{1}{\sqrt{\pi}} \sum_n 
e^{(2p/s -2i\pi n/s )^2 } \int_{-\infty}^{\infty} dx 
\frac{( \sqrt{| x + p/s - i\pi n/s - s|} - \sqrt{|x + p/s - i\pi n/s |})}
{s^{3/2}} e^{-x^2} }{e^{\frac{p^2}{s^2}} \sum_n 
e^{-\frac{\pi^2 n^2 + 2i\pi n p}{s^2}}}
\ea
where the last integral can involves the choice of a branch cut
for $n\not=0$. Since the integral certainly converges for any $n$ and 
is multiplied by the exponentially fast vanishing function 
$e^{-4\pi^2n^2/s}$, by the argument already familiar from \cite{32} 
for the limit $t\to 0$ it will be sufficient to keep the term
$n=0$ for what follows.
We thus obtain for the expectation value
\be \label{a3.3}
\lim_{t\to 0}  
\frac{\langle \psi^t_g , \hat{h}^{-1} [ \sqrt{|\hat{p} |} , \hat{h} ]
/t  \, \psi^t_g \rangle }{ \| \psi^t_g \|^2 }  = -i
\frac{1}{\sqrt{\pi}} \int_{-\infty}^{\infty} dx                    
	  \frac{( \sqrt{| x + p/s - s|} - \sqrt{|x + p/s  |})}{s^{3/2}} 
	    e^{-x^2}
\ee
As we are ultimately only interested in the limit $t\to 0$, and
therefore $s\to 0$, we aim at putting the integrand into a form that
allows taking the $s\to 0$ limit inside: 
\ba \label{a3.4}
\lefteqn{\frac{-i}{\sqrt{\pi}} \int_{-\infty}^{\infty} dx
\frac{( \sqrt{| x + p/s - s|} - \sqrt{|x + p/s  |})}{s^{3/2}} e^{-x^2}
} \nonumber \\
&= &  \frac{-i}{\sqrt{\pi}} \int_{-\infty}^{\infty} dx 
\frac{( \sqrt{| xs + p- s^2|} - \sqrt{|xs + p  |})}{s^2} e^{-x^2}  
\nonumber \\
&= &  \frac{-i}{\sqrt{\pi}} \int_{-\infty}^{\infty} dx
\frac{(xs + p -s^2)^2 - ( xs + p )^2 }{s^2 (| xs + p -s^2|^{1/2} +
|xs + p  |^{1/2}) (|xs + p - s^   2| + |xs + p|) } e^{-x^2} 
\nonumber \\
    &= &  \frac{-i}{\sqrt{\pi}} \int_{-\infty}^{\infty} dx
 \frac{s^2 - 2(xs+p) }{ (| xs + p -s^2|^{1/2} + |xs + p  |^{1/2}) 
 (|xs + p - s^   2| + |xs + p|)} e^{-x^2}
\ea
It is easy to see that the limit $f_p(x)$ as $s\to 0$ of 
the integrand $f^s_p(x)$ exists pointwise. 
Furthermore it is clear that the modulus of the 
integrand is $L^1$-integrable. Next, we write the integrand of 
(\ref{a3.4}) as $e^{-x^2}f^s_p(x)=e^{-x^2/2} g^s_p(x)$ and we 
seek to give a bound on $g^s_p(x)$ independent of $s,x$ for $s$
smaller than some $s_0$.
To that end we estimate $e^{-x^2/2}\le 1$ for $|x|\le 1$ and 
$e^{-x^2/}\le e^{-|x|/2}$ for $|x|\ge 1$ when estimating $g^s_p(x)$. 
Consider first the region $|x|\ge 1$. The first derivative 
of the estimated $|g^s_p(x)|$ then leads to a quadratic equation whose roots 
depend on the signs of both $p,x$. The local maxima turn out to
lie at $x=\pm 1$ and $x\approx -p/s$. Only the former one is an absolute
maximum. The value of $|g^s_p(x)|$ can then be estimated roughly by 
$1/\sqrt{|p|}$ up to a multiplicative, numerical constant and the same is 
true for the region $|x|\le 1$. Altogether we have found, up to a 
numerical factor 
the following $L_1$ function, independent of $s$ that dominates $f^s_p$
\be \label{a3.4a}
|f^s_p(x)|\stackrel{<}{\sim} \frac{e^{-x^2/2}}{\sqrt{|p|}}
\ee
so that all conditions of the dominated 
convergence theorem are satisfied, and the $s\to 0$ limit can be taken 
inside the integral. We thus obtain
\ba \label{a3.5}
 \lefteqn{ \lim_{t\to 0} \frac{\langle \psi^t_g , \hat{h}^{-1} [
\sqrt{|\hat{p} |} ,
	     \hat{h} ]/t \psi^t_g \rangle }{ \| \psi^t_g \|^2 }}
\nonumber \\
     &=& \frac{-i}{\sqrt{\pi}} \int_{-\infty}^{\infty} dx \lim_{s\to 0} 
\frac{s^2 - 2(xs+p) }{ (| xs + p -s^2|^{1/2} + 
|xs + p  |^{1/2}) (|xs + p^2 s^2| + |xs + p|)} e^{-x^2}
\nonumber \\
     &=&  \frac{-i}{\sqrt{\pi}} \int_{-\infty}^{\infty} dx \, 
( - \frac{1}{2 \sqrt{|p|}} \mbox{sgn}(p) ) e^{-x^2} \nonumber \\
    &=& - \frac{1}{2 \sqrt{|p|}} \mbox{sgn}(p).
\ea

\end{appendix}

\end{document}